\documentclass[iop,apj,tighten]{emulateapj}

\usepackage{apjfonts}
\usepackage{graphicx}
\usepackage{dcolumn}
\usepackage{bm}
\usepackage{epsfig}

\usepackage{float}
\usepackage{amsmath,amstext}
\usepackage{floatflt}
\usepackage{subfigure}
\usepackage[usenames]{color}
\usepackage{ulem}
\usepackage{ragged2e}

\usepackage[breaklinks,colorlinks,citecolor=blue,linkcolor=magenta]{hyperref}
\usepackage[all]{hypcap} 

\shorttitle{Tilted flat and untilted non-flat $\Lambda\textrm{CDM}$ models}
\shortauthors{Park \& Ratra}

\begin{document}


\title{Using the tilted flat-$\Lambda$CDM and the untilted non-flat $\Lambda$CDM inflation models to measure cosmological parameters from a compilation of observational data}

\author{
Chan-Gyung Park\altaffilmark{1} and
Bharat Ratra\altaffilmark{2}
}

\altaffiltext{1}{Division of Science Education and Institute of Fusion
                 Science, Chonbuk National University, Jeonju 54896, South Korea;
                 e-mail: park.chan.gyung@gmail.com}
\altaffiltext{2}{Department of Physics, Kansas State University, 116 Cardwell Hall,
                 Manhattan, KS 66506, USA;
                 e-mail: ratra@phys.ksu.edu}

\date{\today}



%
%
\begin{abstract}
We use the physically-consistent tilted spatially-flat and untilted 
non-flat 
$\Lambda\textrm{CDM}$ inflation models to constrain cosmological 
parameter values with the Planck 2015 cosmic microwave background (CMB) 
anisotropy data and recent Type Ia supernovae measurements, baryonic 
acoustic oscillations (BAO) data, growth rate observations, and 
Hubble parameter measurements. The most dramatic consequence of 
including the four non-CMB data sets is the significant strengthening 
of the evidence for non-flatness in the non-flat $\Lambda$CDM model, 
from 1.8$\sigma$ for the CMB data alone to 5.1$\sigma$ for the full 
data combination. The BAO data is the  most powerful of the 
non-CMB data sets in more tightly constraining model parameter values 
and in favoring a spatially-closed Universe in which spatial curvature 
contributes about a percent to the current cosmological energy budget. 
The untilted non-flat $\Lambda$CDM model better fits the large-angle CMB 
temperature anisotropy angular spectrum and is more consistent with 
the Dark Energy Survey constraints on the current value of the rms 
amplitude of mass fluctuations ($\sigma_8$) as a function of the 
current value of the nonrelativistic matter density parameter
($\Omega_m$) but does not provide as good a fit to the smaller-angle CMB
temperature anisotropy data as does the tilted flat-$\Lambda$CDM model. 
Some measured cosmological parameter values differ 
significantly between the two models, including the reionization optical 
depth and the baryonic matter density parameter, both of whose 2$\sigma$ 
ranges (in the two models) are disjoint or almost so.       
\end{abstract}

\maketitle

%
%

\section{Introduction}

In the standard spatially-flat $\Lambda$CDM cosmogony
\citep{Peebles1984} the cosmological constant $\Lambda$ dominates the current
energy budget, cold dark matter (CDM) and baryonic matter are the second 
and third biggest contributors to the cosmological energy budget now, followed 
by small contributions from neutrinos and photons. For reviews of this model, 
see \citet{RatraVogeley2008}, \citet{Martin2012}, and 
\citet{HutererShafer2017}. Many different observations are largely 
consistent with the standard picture, including CMB anisotropies data
\citep{PlanckCollaboration2016}, BAO distance measurements 
\citep{Alametal2017}, Hubble parameter observations \citep{Farooqetal2017}, 
and Type Ia supernova (SNIa) apparent magnitude data \citep{Betouleetal2014}. 
However, there still is room for mild dark 
energy dynamics or a bit of spatial curvature, among other possibilities.

The standard model is characterized by six cosmological parameters that are 
conventionally taken to be:
$\Omega_{\rm b} h^2$ and $\Omega_{\rm c} h^2$, the current values of the baryonic
and cold dark matter density parameters multiplied by the square of the 
Hubble constant $H_0$ (in units of 100 km s$^{-1}$  Mpc$^{-1}$); 
$\theta_{\rm MC}$, the 
angular diameter distance as a multiple of the sound horizon at recombination; 
$\tau$, the reionization optical depth; and $A_{\rm s}$ and $n_{\rm s}$, the 
amplitude and spectral index of the (assumed) power-law primordial scalar 
energy density inhomogeneity power spectrum \citep{PlanckCollaboration2016}. 
The standard model assumes a flat spatial geometry \citep{PlanckCollaboration2016}. 

However, using a physically consistent non-flat inflation model power 
spectrum of energy density inhomogeneities \citep{RatraPeebles1995, Ratra2017},
\citet{Oobaetal2018a} recently found that Planck 2015 CMB anisotropy 
measurements \citep{PlanckCollaboration2016} do not require flat spatial 
geometry in the six 
parameter non-flat $\Lambda$CDM model. In the non-flat $\Lambda$CDM model, 
compared to the flat-$\Lambda$CDM model, there is no simple tilt option so 
$n_s$ is no longer a free parameter and is replaced by the current 
value of the spatial curvature density parameter 
$\Omega_k$.\footnote{The CMB anisotropy data also do not require flat 
spatial geometry in the seven parameter 
non-flat XCDM inflation model \citep{Oobaetal2018b,ParkRatra2019}. Here the equation of state 
relating the pressure and energy density of the dark energy fluid is 
$p_X = w_0 \rho_X$ and $w_0$ is the additional, seventh, 
parameter. XCDM is often used to model dynamical dark energy but is not a 
physically consistent model as it cannot describe the evolution of 
energy density inhomogeneities. Also, XCDM does not accurately model $\phi$CDM 
\citep{PeeblesRatra1988,RatraPeebles1988} dark energy dynamics 
\citep{PodariuRatra2001}. In the simplest, physically consistent, seven 
parameter non-flat $\phi$CDM inflation model \citep{Pavlovetal2013} --- in 
which a scalar
field $\phi$ with potential energy density $V(\phi) \propto \phi^{-\alpha}$ is 
the dynamical dark energy and $\alpha > 0$ is the seventh parameter that
governs dark energy evolution --- \citet{Oobaetal2018c} again found that
CMB anisotropy data do not require flat spatial hypersurfaces 
\cite[also see][]{ParkRatra2018a}. (In both the non-flat XCDM and $\phi$CDM 
cases, $n_s$ is again replaced by $\Omega_k$.)}

In non-flat models non-zero spatial curvature sets the second, new length 
scale. This is in addition to the Hubble length scale. Inflation provides 
the only known way to 
define a physically consistent non-flat model power spectrum. For open 
spatial geometry the open-bubble inflation model of \cite{Gott1982} is used 
to compute the non-power-law power spectrum 
\citep{RatraPeebles1994,RatraPeebles1995}.\footnote{For early discussions 
of observational consequences of the open inflation model, see
\citet{Kamionkowskietal1994}, \citet{Gorskietal1995}, and \citet{Gorskietal1998}.}
For closed spatial geometry Hawking's prescription for the quantum state 
of the universe \citep{Hawking1984,Ratra1985} can be used to construct a closed 
inflation model that can be used to compute the non-power-law power spectrum
of energy density inhomogeneities \citep{Ratra2017}. Both these open and closed inflation models are slow-roll inflation models \citep{Gott1982,Hawking1984,Ratra1985} so the resulting energy density inhomogeneity power spectra are untilted \citep{RatraPeebles1995,Ratra2017}.

Non-CMB observations, even combinations thereof to date, do not rule out 
non-flat dark energy models 
\citep[see, e.g.,][]{Farooqetal2015, Caietal2016, Chenetal2016, YuWang2016, LHuillierShafieloo2017, Farooqetal2017, Lietal2016, WeiWu2017, Ranaetal2017, Yuetal2018, Mitraetal2018, Mitraetal2019, Ryanetal2018, Ryanetal2019, ParkRatra2018b}. The most restrictive
constraints on spatial curvature come from CMB anisotropy measurements,
but, as shown by \citet{Oobaetal2018a}, when the correct non-power-law power
spectrum for energy density inhomogeneities is used for the CMB anisotropy
analyses, a spatial curvature density parameter contribution of magnitude a 
percent or two is 
still allowed, with the CMB anisotropy data \citep{PlanckCollaboration2016} 
favoring a mildly closed model. \citet{Oobaetal2018a} also added a few BAO 
distance measurements to the mix and found that a mildly closed model was 
still favored. Moreover, the mildly closed model better fits the observed 
low-$\ell$ CMB temperature anisotropy multipole number ($\ell$) power spectrum $C_\ell$ 
and was more consistent with rms fractional energy density inhomogeneity 
averaged over 8$h^{-1}$ Mpc radius spheres, $\sigma_8$, current values
determined from weak lensing observations, although the flat-$\Lambda$CDM model 
better fits the observed higher-$\ell$ $C_\ell$'s.

In this paper we examine the constraints on the non-flat $\Lambda$CDM inflation
model that result from a joint analysis of the Planck 2015 CMB anisotropy 
data \citep{PlanckCollaboration2016}, the Joint Light-curve Analysis (JLA) 
SNIa apparent magnitude measurements \citep{Betouleetal2014}, and all reliable 
BAO distance, growth factor, and Hubble parameter measurements to date. We also
perform a similar analysis for the tilted flat-$\Lambda$CDM inflation model.

The main purposes of our analyses here are, firstly, to examine the
effect that the inclusion of a significant amount of reliable, recent, non-CMB
data has on the finding of \citet{Oobaetal2018a} that the Planck
2015 CMB anisotropy observations and a handful of BAO distance measurements are
not inconsistent with the untilted closed-$\Lambda$CDM inflation model, and, secondly, 
to use this large new compilation of reliable non-CMB data to examine the 
consistency between the cosmological constraints of each type of data and
to more tightly measure cosmological parameters than has been done to date.

Our main findings here are that our carefully gathered compilation of 
cosmological observations, the largest to date, does not require flat 
spatial hypersurfaces, with the untilted non-flat $\Lambda$CDM 
inflation model 
in which spatial curvature contributes about a percent to the current
cosmological energy budget being more than 5$\sigma$ away from flatness;
the untilted non-flat model better fits the low-$\ell$ CMB 
temperature anisotropy $C_\ell$'s as well as
the weak lensing constraints in the $\sigma_8$--$\Omega_m$ plane, while 
the tilted flat-$\Lambda$CDM model is more consistent with the higher-$\ell$
$C_\ell$'s; $H_0$ is robustly measured in an almost model-independent 
manner and the value is consistent with most other measurements; and some
measured cosmological parameter values, including those of $\Omega_{\rm b} h^2$,
$\tau$, and $\Omega_{\rm c} h^2$, differ significantly between the two 
models and so care must be exercised when utilizing cosmological measurements
of such parameters. 

This paper is organized as follows. In Sec.\ 2 we describe the cosmological 
data sets we use in our analyses. In Sec.\ 3 we summarize the methods we 
use for our analyses here. The observational constraints resulting 
from these data for the tilted flat-$\Lambda$CDM and the non-flat 
$\Lambda$CDM inflation models are presented in Sec.\ 4. We summarize our 
results in Sec.\ 5.

\section{Data}

\subsection{Planck 2015 CMB anisotropy data}

We use the Planck 2015 TT + lowP and TT + lowP + lensing CMB anisotropy
data \citep{PlanckCollaboration2016}. Here TT represents the low-$\ell$
($2 \le \ell \le 29$) and high-$\ell$ ($30 \le \ell \le 2508$; PlikTT)
Planck temperature-only $C_\ell^{TT}$ data and lowP denotes low-$\ell$
polarization $C_\ell^{TE}$, $C_\ell^{EE}$, and $C_\ell^{BB}$ power
spectra measurements at $2 \le \ell \le 29$. The collection of low-$\ell$
temperature and polarization measurements is denoted as lowTEB.
For CMB lensing data we use the power spectrum of the lensing potential
measured by Planck.

\subsection{JLA SNIa data}

We use the JLA compilation of 740 SNIa apparent magnitude measurements 
released by the SDSS-II and 
SNLS collaborations \citep{Betouleetal2014}. The JLA data set is composed of 
several low-redshift SNIa ($z < 0.1$) and higher redshift samples from the 
SDSS-II ($0.05 < z < 0.4$) and SNLS ($0.2 < z < 1$).

\subsection{BAO data}

The anisotropy of BAO features in the line-of-sight and the transverse
directions enable us to constrain both the Hubble parameter $H(z)$ and
the comoving angular diameter distance
\begin{equation}
   D_M (z) = (1+z) D_A (z)
\end{equation}
where $D_A$ is the physical angular diameter distance at redshift $z$.
The radius of the sound horizon at the drag epoch $z_d$ is
\begin{equation}
   r_d = \int_{z_d}^{\infty} \frac{c_s (z)}{H(z)} dz
\end{equation}
where $c_s(z)$ is the sound speed of the photon-baryon fluid. Because the size 
of the sound horizon $r_d$ depends on the cosmological model and the energy 
contents, the BAO features in the large-scale structure actually 
constrain $D_M(z)/r_d$ and $H(z) r_d$.

\begin{table*}
\caption{BAO measurements.}
\begin{ruledtabular}
\begin{tabular}{lccccc}
  Data Set     & LSS tracers    &  $z_\textrm{eff}$  &  Observable                          &  Measurement        &   Reference    \\[+0mm]
 \hline \\[-2mm]
  BOSS DR12    & galaxies       &  $0.38$     & $D_M (r_{d,\textrm{fid}} / r_d)$ [Mpc]      & $1518\pm22$  &  \cite{Alametal2017} \\[+1mm]
               &                &  $0.51$     & $D_M (r_{d,\textrm{fid}} / r_d)$ [Mpc]      & $1977\pm27$  &  \cite{Alametal2017} \\[+1mm]
               &                &  $0.61$     & $D_M (r_{d,\textrm{fid}} / r_d)$ [Mpc]      & $2283\pm32$  &  \cite{Alametal2017} \\[+1mm]
               &                &  $0.38$     & $H (r_d / r_{d,\textrm{fid}})$ [km s$^{-1}$ Mpc$^{-1}$]  & $81.5\pm1.9$  &  \cite{Alametal2017} \\[+1mm]
               &                &  $0.51$     & $H (r_d / r_{d,\textrm{fid}})$ [km s$^{-1}$ Mpc$^{-1}$]  & $90.4\pm1.9$  &  \cite{Alametal2017} \\[+1mm]
               &                &  $0.61$     & $H (r_d / r_{d,\textrm{fid}})$ [km s$^{-1}$ Mpc$^{-1}$]  & $97.3\pm2.1$  &  \cite{Alametal2017} \\[+1mm]
               &                &  $0.38$     & $f \sigma_8$                                & $0.497\pm0.045$  &  \cite{Alametal2017} \\[+1mm]
               &                &  $0.51$     & $f \sigma_8$                                & $0.458\pm0.038$  &  \cite{Alametal2017} \\[+1mm]
               &                &  $0.61$     & $f \sigma_8$                                & $0.436\pm0.034$  &  \cite{Alametal2017} \\[+1mm]
 \hline \\[-2mm]
  6dF          & galaxies       &  $0.106$    & $r_d / D_V$                                 & $0.327\pm0.015$  &  \cite{Beutleretal2011} \\[+1mm]
   \hline \\[-2mm]
  SDSS DR7 MGS & galaxies       &  $0.15$     & $D_V (r_{d,\textrm{fid}} / r_d)$ [Mpc]      & $664\pm25$       & \cite{Rossetal2015} \\[+1mm]
   \hline \\[-2mm]
  eBOSS DR14   & QSOs           &  $1.52$     & $D_V (r_{d,\textrm{fid}} / r_d)$ [Mpc]      & $3855\pm170$     & \cite{Ataetal2018} \\[+1mm]
   \hline \\[-2mm]
  BOSS DR12 Ly$\alpha$ forest  & Ly$\alpha$   &  $2.33$    &   $D_H^{0.7} D_M^{0.3} / r_d$  & $13.94\pm0.35$   &  \cite{Bautistaetal2017} \\[+1mm]
   \hline \\[-2mm]
  BOSS DR11 Ly$\alpha$ forest  & QSO \& Ly$\alpha$    &  $2.36$    & $D_H / r_d$            & $9.0\pm0.3$      &  \cite{Font-Riberaetal2014} \\[+1mm] 
                               &                      &  $2.36$    & $D_A / r_d$            & $10.8\pm0.4$     &  \cite{Font-Riberaetal2014} \\[+0mm]
\end{tabular}
\\[+1mm]
Note: The sound horizon size of the fiducial model is $r_{d,\textrm{fid}}=147.78~\textrm{Mpc}$ in \cite{Alametal2017} and \cite{Ataetal2018}, and $r_{d,\textrm{fid}}=148.69~\textrm{Mpc}$ in \cite{Rossetal2015}. 
\end{ruledtabular}
\label{tab:bao}
\end{table*}

We use the recent, more reliable BAO distance measurements from the 6dF 
Galaxy Survey (6dFGS) \citep{Beutleretal2011}, the Sloan Digital Sky 
Survey (SDSS) Data Release 7 (DR7) main galaxy sample (MGS) 
\citep{Rossetal2015}, the Baryon Oscillation Spectroscopic Survey (BOSS) 
DR12 galaxies \citep{Alametal2017}, the eBOSS DR14 QSO's \citep{Ataetal2018},
and the BOSS DR11 and DR12 Ly$\alpha$ forest \citep{Font-Riberaetal2014,Bautistaetal2017}, 15 points in total, which are summarized in Table 
\ref{tab:bao}.\footnote{For the BAO data point of
\cite{Ataetal2018} we use the value presented in arXiv:1705.06373v1. In the revised published
version (arXiv:1705.06373v2) they updated the data point to $D_V (r_{d,\textrm{fid}} / r_d)=3843\pm147$ Mpc where $r_{d,\textrm{fid}}$ is the value of $r_d$ for the fiducial model used in the analysis.}
We call this collection of BAO measurements `NewBAO' to distinguish it from
the earlier BAO data compilation (which we call `BAO') of 6dFGS 
\citep{Beutleretal2011}, BOSS LOWZ and CMASS \citep{Andersonetal2014}, 
and SDSS MGS \citep{Rossetal2015} BAO distance measurements, used in 
the analyses of \citet{PlanckCollaboration2016} and 
\citet{Oobaetal2018a,Oobaetal2018b,Oobaetal2018c}.

For BAO data provided by \cite{Alametal2017}, we include the growth rate ($f\sigma_8$)
data in our BAO (and not in our growth rate) analyses here, to be able to properly account
for the correlations in the \citet{Alametal2017} measurements.
For the SDSS DR7 MGS \citep{Rossetal2015} and BOSS DR11 Ly$\alpha$ 
forest \citep{Font-Riberaetal2014} measurements, we use the probability 
distributions of the BAO data points, instead of using the Gaussian 
approximation constraints. \cite{Bautistaetal2017} provide one BAO parameter $D_H^{0.7} D_M^{0.3} / r_d$ at $z=2.33$ measured from
BOSS DR12 Ly$\alpha$ forest observations, where $D_H$ is defined as  
\begin{equation}
   D_H (z) = c / H(z)
\end{equation} 
where $c$ is the speed of light. \cite{Font-Riberaetal2014} provide BAO
parameters ($D_H/r_d$ and $D_A / r_d$) measured from the cross-correlation
between QSO and Ly$\alpha$ forest data. They actually provide the probability
distribution of parameters that describe shifts of the BAO peak position with
respect to the fiducial cosmology in perpendicular and parallel directions
to the line-of-sight,
\begin{equation}
   \alpha_{\perp}=\frac{D_M(z) r_{d,\textrm{fid}}}{D_M^{\textrm{fid}}(z) r_d}, \quad
   \alpha_{\parallel}=\frac{H^{\textrm{fid}} (z) r_{d,\textrm{fid}}}{H(z) r_d}.
\end{equation}
The angle-averaged shift and the ratio of the two $\alpha$ parameters can be converted into
the angle-averaged version of the distance scale
\begin{equation}
   D_V (z) = \left[ cz D_M^2 (z) / H(z) \right]^{1/3},
\end{equation} 
and the Alcock-Paczynski parameter
\begin{equation}
   F_\textrm{AP} (z) = D_M (z) H(z) / c.
\end{equation}
For the BAO data of \citet{Alametal2017}, instead of using 
$D_M (r_{d,\textrm{fid}} / r_d)$ and $H (r_d / r_{d,\textrm{fid}})$, we 
actually transform these into $D_V / r_d$ and $F_\textrm{AP}$ and also
use their growth rate $f\sigma_8$ measurements and account for
correlations (data publicly available at the BOSS website).

\subsection{Hubble parameter data}

Hubble parameter measurements can be used to constrain dark energy
parameters, as well as other cosmological parameters, including the 
spatial curvature of the Universe 
(see e.g., \citealt{Farooqetal2017}).\footnote{Early developments include
\citet{SamushiaRatra2006}, \citet{Samushiaetal2007}, and 
\citet{ChenRatra2011b}; recent work includes \citet{Tripathietal2017}, 
\citet{Lonappanetal2017}, \citet{Rezaeietal2017}, \citet{Maganaetal2017},
\citet{AnagnostopoulosBasilakos2017}, \citet{Yuetal2018}, and 
\citet{Caoetal2018}.
We note that there are many different $H(z)$ compilations discussed
in the literature. Unfortunately a significant fraction of these
include non-independent or unreliable measurements.}
Here we adapt and use a recent Hubble parameter measurement 
compilation to constrain both 
the tilted flat-$\Lambda$CDM inflation model and the non-flat $\Lambda$CDM 
inflation model.
Table \ref{tab:hubble} lists all more reliable recent measurements of the 
Hubble 
parameter at various redshifts (with 31 data points in total).\footnote{Table \ref{tab:hubble} does not list radial or line-of-sight BAO $H(z)$ measurements; 
these are instead listed in Table \ref{tab:bao}.} 
See \citet{Farooqetal2017} and 
\citet{Yuetal2018} for discussions about how these data were 
selected.\footnote{The redshift range over which the Hubble parameter has been 
measured encompasses the redshift of the cosmological deceleration-acceleration 
transition in the standard cosmological model. This transition is between the 
earlier nonrelativistic-matter-powered decelerating cosmological expansion and 
the more recent dark-energy-driven accelerating cosmological expansion. This 
transition redshift has recently been measured and is at roughly the value 
expected in the standard $\Lambda$CDM and other dark energy models 
\citep{FarooqRatra2013,Morescoetal2016,Farooqetal2017}.}

\begin{table}
\caption{Hubble parameter data.}
\begin{ruledtabular}
\begin{tabular}{lccc}
  $z$              & $H(z)$                     &  $\sigma_H$                 &  Reference  \\[+0mm]
                   & (km s$^{-1}$ Mpc$^{-1}$)   &  (km s$^{-1}$ Mpc$^{-1}$)   &             \\[+0mm]
  \hline \\[-2mm]
  $0.070$         & $69$              &  $19.6$           &  \cite{Zhangetal2014}   \\[+1mm]
  $0.090$         & $69$              &  $12$ 	          &  \cite{Simonetal2005}  \\[+1mm]
  $0.120$         & $68.6$            &  $26.2$           &  \cite{Zhangetal2014}  \\[+1mm]
  $0.170$         & $83$              &  $8$              &  \cite{Simonetal2005}  \\[+1mm]
  $0.179$         & $75$              &  $4$              &  \cite{Morescoetal2012} \\[+1mm]
  $0.199$         & $75$              &  $5$              &  \cite{Morescoetal2012}   \\[+1mm]
  $0.200$         & $72.9$            &  $29.6$           &  \cite{Zhangetal2014}   \\[+1mm]
  $0.270$         & $77$              &  $14$             &  \cite{Simonetal2005} \\[+1mm]
  $0.280$         & $88.8$            &  $36.6$           &  \cite{Zhangetal2014} \\[+1mm]
  $0.352$         & $83$              &  $14$             &  \cite{Morescoetal2012} \\[+1mm]
  $0.3802$        & $83$              &  $13.5$           &  \cite{Morescoetal2016} \\[+1mm]
  $0.400$         & $95$              &  $17$             &  \cite{Simonetal2005}  \\[+1mm]
  $0.4004$        & $77$              &	 $10.2$           & \cite{Morescoetal2016} \\[+1mm]
  $0.4247$        & $87.1$            &  $11.2$           &  \cite{Morescoetal2016} \\[+1mm] 
  $0.4497$        & $92.8$            &  $12.9$           & \cite{Morescoetal2016} \\[+1mm]
  $0.47$          & $89$              &  $50$             & \cite{Ratsimbazafyetal2017} \\[+1mm]
  $0.4783$        & $80.9$            &  $9$              & \cite{Morescoetal2016} \\[+1mm]
  $0.480$         & $97$              &	 $62$             & \cite{Sternetal2010}  \\[+1mm]
  $0.593$         & $104$             &  $13$             & \cite{Morescoetal2012}  \\[+1mm]
  $0.680$         & $92$              &  $8$              & \cite{Morescoetal2012} \\[+1mm]
  $0.781$         & $105$             &  $12$             & \cite{Morescoetal2012} \\[+1mm]
  $0.875$         & $125$             &  $17$	          & \cite{Morescoetal2012} \\[+1mm]
  $0.880$         & $90$              &  $40$             & \cite{Sternetal2010} \\[+1mm]
  $0.900$         & $117$             &  $23$             & \cite{Simonetal2005} \\[+1mm]
  $1.037$         & $154$             &  $20$             & \cite{Morescoetal2012} \\[+1mm]
  $1.300$         & $168$             &  $17$             & \cite{Simonetal2005}   \\[+1mm]
  $1.363$         & $160$             &  $33.6$           & \cite{Moresco2015} \\[+1mm]
  $1.430$         & $177$             &  $18$             & \cite{Simonetal2005} \\[+1mm]
  $1.530$         & $140$             &	 $14$             & \cite{Simonetal2005} \\[+1mm]
  $1.750$         & $202$             &  $40$             & \cite{Simonetal2005} \\[+1mm]
  $1.965$         & $186.5$           &  $50.4$           & \cite{Moresco2015} \\[+0mm]
\end{tabular}
\end{ruledtabular}
\label{tab:hubble}
\end{table}

\subsection{Growth rate data}

\begin{table*}
\caption{Growth rate data.}
\begin{ruledtabular}
\begin{tabular}{lcccc}
 Data set       &   $z_{\textrm{eff}}$   & $f(z) \sigma_8 (z)$  &  References        & Notes \\[+0mm]
  \hline \\[-2mm]
 SNIa+IRAS PSCz &   $0.02$           & $0.398 \pm 0.065$      &  \cite{Turnbulletal2012,HudsonTurnbull2012}  & $(\Omega_m,\Omega_k)=(0.3,0)$ \\[+1mm]
 2MASS          &   $0.02$           & $0.32 \pm 0.04$        &  \cite{Springobetal2016}    & 	$(\Omega_m,\Omega_k)=(0.308,0)$ \\[+1mm]
 6dFGS          &   $0.067$          & $0.423 \pm 0.055$      &  \cite{Beutleretal2012}     & 	$(\Omega_m,\Omega_k)=(0.27,0)$   \\[+1mm]
 SDSS MGS       &   $0.1$            & $0.37 \pm 0.13$        &  \cite{Feixetal2015}        &  $(\Omega_m,\Omega_k)=(0.3,0)$   \\[+1mm]
 SDSS MGS       &   $0.15$           & $0.49 \pm 0.145$       &  \cite{Howlettetal2015}     &  $(\Omega_m,h,\sigma_{8,0})=(0.31,0.67,0.83)$ \\[+1mm]
 GAMA           &   $0.18$           & $0.29 \pm 0.10$        &  \cite{Simpsonetal2016}     &  $(\Omega_m,\Omega_k)=(0.27,0)$      \\[+1mm]
 GAMA           &   $0.38$           & $0.44 \pm 0.06$        &  \cite{Blakeetal2013}       &                       \\[+1mm]
 VIPERS PDR-2   &   $0.6$            & $0.55 \pm 0.12$        &  \cite{Pezzottaetal2017}    &  $(\Omega_m,\Omega_b)=(0.3,0.045)$ \\[+1mm]
 VIPERS PDR-2   &   $0.86$           & $0.40 \pm 0.11$        &  \cite{Pezzottaetal2017}    &                                     \\[+1mm]
 FastSound      &   $1.4$            & $0.482 \pm 0.116$      &  \cite{Okumuraetal2016}     &  $(\Omega_m,\Omega_k)=(0.27,0)$   \\[+0mm]
\end{tabular}
\\[+1mm]
The fiducial models assumed in the analyses are listed in the notes.
\end{ruledtabular}
\label{tab:growth}
\end{table*}

The growth rate is defined as
\begin{equation}
   f(a)=\frac{d \ln D(a)}{d \ln a}
\end{equation}
where $a$ is the scale factor, $D(a)$ is the amplitude of the matter
density perturbation, and $f\approx \Omega_m^{0.55}$ for the
$\Lambda\textrm{CDM}$ model.
Information on the growth rate is derived from the peculiar velocities
of galaxies. The peculiar velocities can be obtained from the redshift
space distortion information imprinted in the large-scale structures of
galaxy redshift surveys.
The growth rate measurement is sometimes given in terms of $\beta=f/b$,
where $b$ is the bias parameter that relates the galaxy density
perturbation to the matter one via $\delta_g = b \delta_m$.
Since the $\beta$ parameter strongly depends on the bias parameter,
the combination $f(z)\sigma_8 (z)$ is more widely used to quantify
the growth rate of the matter density perturbation.  
Here the rms of density fluctuations within a sphere of
$8~h^{-1}\textrm{Mpc}$ radius is represented by $\sigma_8$ for the
mass and $\sigma_{8,g}$ for the galaxy distributions. These are
related through $\sigma_8=\sigma_{8,g}/b$ and $f\sigma_8 = \beta \sigma_{8,g}$.
The rms mass fluctuation at epoch $a$ is
\begin{equation}
   \sigma_8 (a) = \sigma_{8,0} \frac{D(a)}{D_0},
\end{equation}
where the subscript 0 indicates the present epoch. In the following we denote 
the present value $\sigma_{8,0}$ as $\sigma_8$ for simplicity.

Table \ref{tab:growth} lists all more reliable recent measurements
of growth rate $f(z)\sigma_8 (z)$ at various redshifts, 10 points in total. 
As already noted, the three growth rate data points of \citet{Alametal2017}
are included in the collection of BAO data points in order to properly account 
for correlations between these BAO and growth rate data points.

\section{Methods}

\subsection{Model computations}

We use the publicly available CAMB/COSMOMC package (version of Nov.\ 2016)
\citep{ChallinorLasenby1999,Lewisetal2000,LewisBridle2002}
to constrain the tilted flat and the non-flat $\Lambda$CDM inflation models 
with Planck 2015 CMB and other non-CMB data sets. The Boltzmann code CAMB 
computes the CMB angular power spectra for temperature fluctuations, 
polarization, and lensing potential, and COSMOMC applies the Markov chain 
Monte Carlo (MCMC) method to explore and determine model-parameter space 
that is favored by the data used. We use the COSMOMC settings 
adopted in the Planck team's analysis \citep{PlanckCollaboration2016}. We 
set the present CMB temperature to $T_0=2.7255$ K \citep{Fixsen2009} and 
the effective number of neutrino
species to $N_\textrm{eff}=3.046$. We assume the existence of a single species
of massive neutrinos with mass $m_\nu = 0.06$ eV. The primordial Helium
fraction $Y_\textrm{He}$ is set from the Big Bang nucleosynthesis prediction.
In the parameter estimation the lensed CMB power spectra for each model
are compared with observations. When the Planck lensing data are included
in the analysis, we also need to consider the non-linear lensing effect that
is important in the lensing potential reconstruction
\citep{PlanckCollaboration2014}.
As needed, we turn on the options for CMB lensing and nonlinear lensing
in every case, regardless of whether the Planck lensing data are used or not.

The primordial power spectrum in the spatially-flat tilted $\Lambda$CDM 
inflation model \citep{LucchinMatarrese1985,Ratra1992,Ratra1989} is
\begin{equation}
   P(k)=A_s \left(\frac{k}{k_0} \right)^{n_s},
\label{eq:Pk}
\end{equation}
where $k$ is wavenumber and $A_s$ is the amplitude at the pivot scale 
$k_0=0.05~\textrm{Mpc}^{-1}$. On the other 
hand, the primordial power spectrum in the non-flat $\Lambda$CDM inflation
model \citep{RatraPeebles1995,Ratra2017} is
\begin{equation}
   P(q) \propto \frac{(q^2-4K)^2}{q(q^2-K)},
\label{eq:Pq}
\end{equation}
which goes over to the $n_s=1$ spectrum in the spatially-flat limit ($K=0$).
For scalar perturbations, $q=\sqrt{k^2 + K}$ is the wavenumber 
where $K=-(H_0^2 / c^2 ) \Omega_k$ is the spatial curvature.
For the spatially-closed model, with negative $\Omega_k$, the normal modes 
are characterized by the 
positive integers $\nu=q K^{-1/2}=3,4,5,\cdots$, and the eigenvalue of the 
spatial Laplacian is $ -(q^2 - K)/K \equiv - {\bar k}^2/K$. 
We use $P(q)$ as the
initial power spectrum of perturbations for the non-flat model by
normalizing its amplitude at the pivot scale $k_0$ to the value of $A_s$. 

The Planck 2015 non-flat model analyses \citep{PlanckCollaboration2016} are not based on either of the above power spectra, instead they assume
\begin{equation}
   P_{\rm Planck}(q) \propto \frac{(q^2-4K)^2}{q(q^2-K)} \left(\frac{\bar k}{k_0} \right)^{n_s-1},
\label{eq:PqP} 
\end{equation}
where in addition to the non-flat space wavenumber $q$, the wavenumber 
$\bar k$ is also used to define and tilt the non-flat model 
$P_{\rm Planck}(q)$. The ${\bar k}^{n_s-1}$ tilt factor in $P_{\rm Planck}(q)$ 
assumes that tilt in non-flat space works somewhat as it does in flat space, 
which seems unlikely since spatial curvature sets an additional length scale in 
non-flat space (i.e., in addition to the Hubble length). It is not known if the 
power spectrum of Eq.~(\ref{eq:PqP}) can be the consequence of quantum 
fluctuations during an early epoch of inflation. This power spectrum is 
physically sensible if $K= 0$ or if $n_s = 1$, when it reduces to the 
power spectra in Eqs.~(\ref{eq:Pk}) and (\ref{eq:Pq}), both of which are 
consequences of quantum fluctuations during inflation.\footnote{Using this expression for the power spectrum (in the seven parameter non-flat $\Lambda$CDM model) in analyses of the Planck 2015 data, \citet{PlanckCollaboration2016} finds $n_s = 0.9717 \pm 0.0066$ in the TT + lowP case \citep[][Table 19.1]{PlanckParameterTables2015}; since $n_s \neq 1$ it is unclear what significance this result has.} 

\subsection{Constraining model parameters}

We explore the parameter space of the tilted flat-$\Lambda$CDM model with 
six cosmological parameters ($\Omega_b h^2$, $\Omega_c h^2$,
$\theta_\textrm{MC}$, $\tau$, $A_s$, and $n_s$) and
the untilted non-flat $\Lambda$CDM model with six parameters ($\Omega_b h^2$,
$\Omega_c h^2$, $\Omega_k$, $\theta_\textrm{MC}$, $\tau$, and
$A_s$). $\theta_{\textrm{MC}}$ is the approximate angular
size of the sound horizon ($r_* / D_A$) at redshift $z_*$ for which
the optical depth equals unity \citep{PlanckCollaboration2014}.
Unresolved extragalactic foregrounds due to point sources, cosmic
infrared background, and thermal and kinetic Sunyaev-Zeldovich components
contribute to the temperature power spectrum. Thus the foreground model
parameters are also constrained as nuisance parameters by the MCMC method.
We also compute three derived parameters, $H_0$, $\Omega_m$, and $\sigma_8$.

For each model (and set of six parameter values), we compare
the lensed CMB power spectra obtained from the CAMB Boltzmann code with
the Planck 2015 TT + lowP data and TT + lowP + lensing data, excluding and
including the power spectrum of the lensing potential, respectively.
For BAO, SNIa, and Hubble parameter data, the prediction determined from
the spatially homogeneous background evolution equations solution for each 
set of model
parameters is compared with the observations.\footnote{For parameter
estimation using the JLA SNIa data set we need to consider hidden
nuisance parameters, $\alpha_\textrm{JLA}$ and $\beta_{\textrm{JLA}}$
related to the stretch and color correction of the SNIa light curves,
$B$-band absolute magnitude $M_B$, and the offset of the absolute magnitude
due to the environment (host stellar mass) $\Delta_M$. Thus, the number
of degrees of freedom for the JLA data is less than the total number of SNIa
($N=740$). For example, for the flat-$\Lambda\textrm{CDM}$ model that fits
the matter density parameter $\Omega_m$, $\alpha_\textrm{JLA}$,
$\beta_\textrm{JLA}$, $M_B$, and $\Delta_M$, the number of degrees
of freedom becomes 735 ($=740-5$). In our analysis, we assume flat priors for
these parameters ($0.01 \le \alpha_\textrm{JLA} \le 2$ and
$0.9 \le \beta_\textrm{JLA} \le 4.6$) during parameter estimation.}
For growth rate data, the matter density perturbation evolved by the CAMB
code is used to compute $f\sigma_8$ at the needed redshifts.

We set priors for some parameters. The Hubble constant is restricted to the 
range $20 \le H_0 \le 100$, in units of km s$^{-1}$ Mpc$^{-1}$.
The reionization optical depth is explored only in the range $\tau > 0.005$.
The other basic parameters have flat priors that are sufficiently wide
such that the final constraints are within the prior ranges.
For every model considered here sufficient MCMC chains are generated
in order that the Gelman and Rubin $R$ statistics satisfy the condition
$R \lesssim 0.01$.

\section{Observational Constraints}

We constrain the spatially-flat tilted and the untilted 
non-flat $\Lambda$CDM 
inflation models using the Planck 2015 TT + lowP (excluding and including the 
CMB lensing) data and other non-CMB data sets.

\begin{figure*}
\centering
\mbox{\includegraphics[width=87mm]{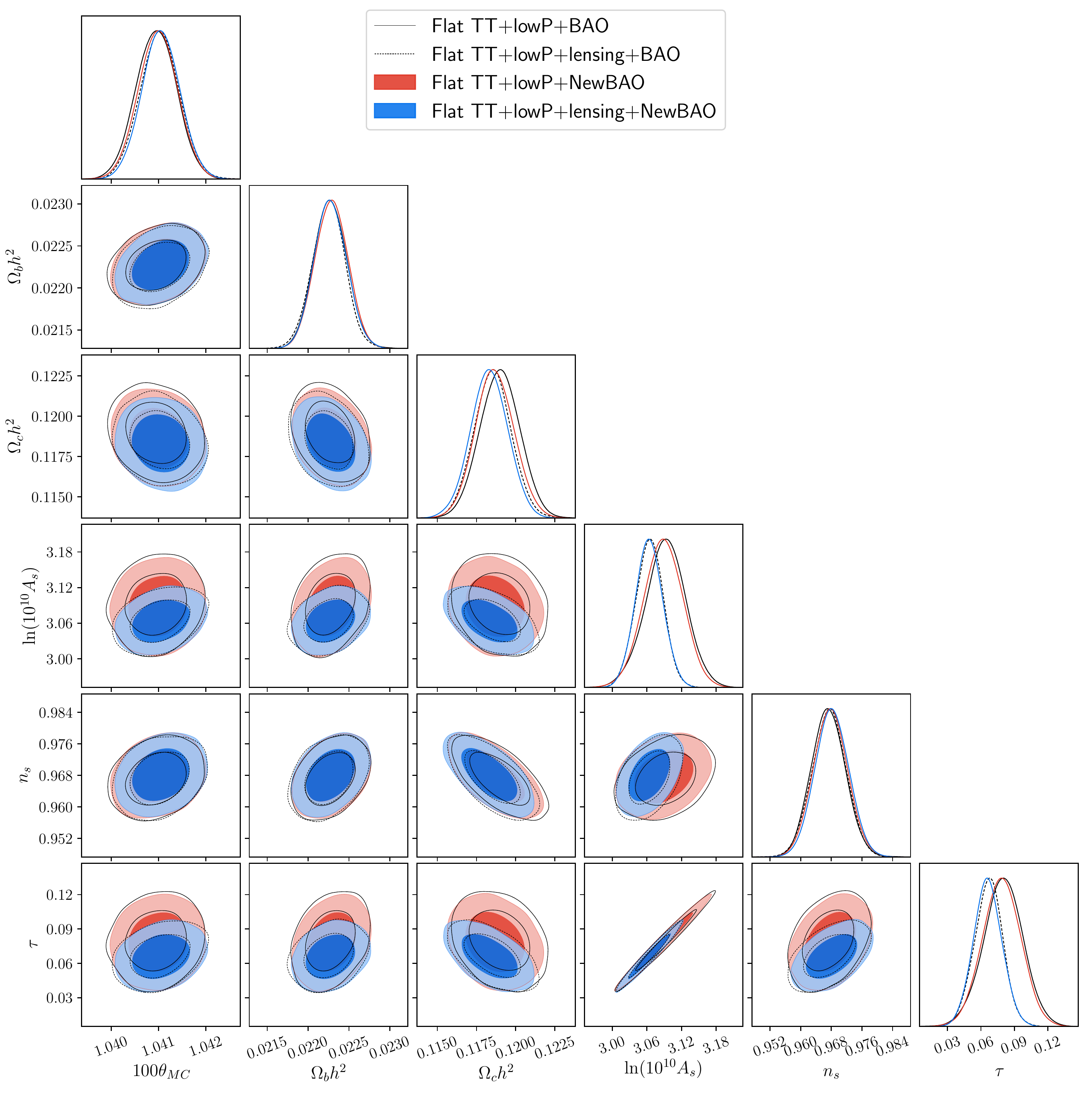}}
\mbox{\includegraphics[width=87mm]{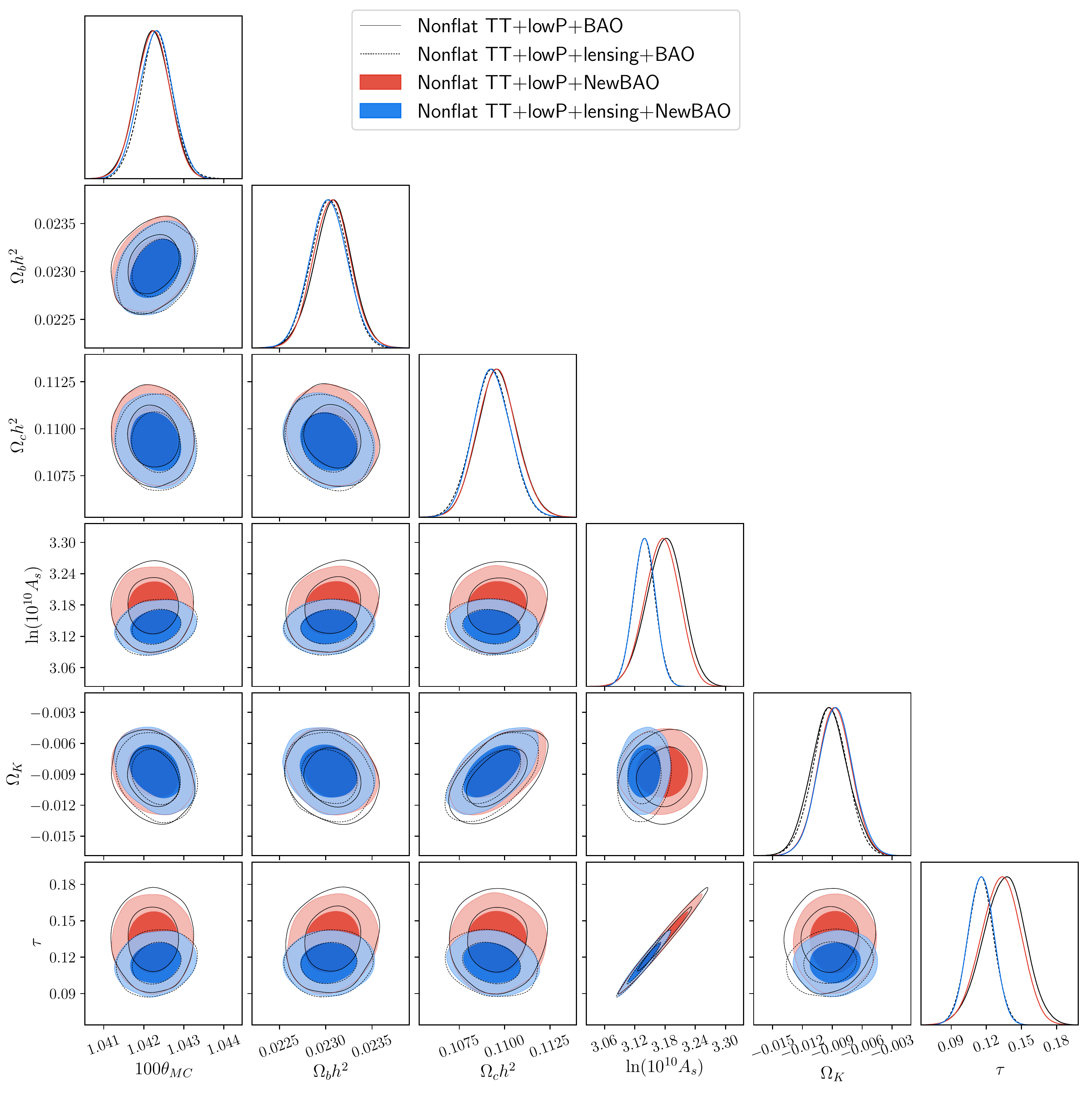}}
\caption{Likelihood distributions of the tilted flat (left) and untilted 
non-flat
(right) $\Lambda\textrm{CDM}$ model parameters favored by the Planck 2015
CMB TT + lowP (+ lensing) and BAO data. Here the parameter constraints are
compared for the old BAO data and the new (NewBAO) data summarized
in Table \ref{tab:bao}.
Two-dimensional marginalized likelihood distributions of all possible
combinations of model parameters together with one-dimensional likelihoods
are shown as solid and dashed black curves for BAO and filled contours and colored
curves for NewBAO data.
}
\label{fig:bao_newbao}
\end{figure*}

We first examine how efficient the new BAO data are in constraining
parameters, relative to the old BAO data. Figure \ref{fig:bao_newbao} 
compares the likelihood distributions of the model parameters for the old 
(`BAO') and new BAO (`NewBAO') data sets, in conjunction with the CMB 
observations.
The mean and 68.3\% confidence limits of model parameters are presented in
Table \ref{tab:para_bao}. We see that adding CMB lensing data results in 
a reduction of $\ln (10^{10} A_s)$ and $\tau$ in both models and that the 
NewBAO data improve parameter estimation with slightly narrower
parameter constraints (more so for the cases when the lensing data are
excluded).

The entries in the TT + lowP + BAO and TT + lowP + lensing + BAO columns
for the non-flat $\Lambda$CDM model in Table \ref{tab:para_bao} agree well 
with the corresponding entries in Table 2 of \citet{Oobaetal2018a}.
\citet{Oobaetal2018a} used CLASS \citep{Blasetal2011} to compute the
$C_\ell$'s and Monte Python \citep{Audrenetal2013} for the MCMC analyses,
so it is gratifying and reassuring that our results agree well with those of
\citet{Oobaetal2018a}.\footnote{ Note that \citet{Oobaetal2018a} use CLASS $\theta$ that is defined as the ratio of comoving sound horizon to the angular diameter distance at decoupling while here we use CAMB $\theta_\textrm{MC}$ that is an approximate version of $\theta$.}

\begin{table*}
\caption{Mean and 68.3\% confidence limits of tilted flat and untilted non-flat $\Lambda\textrm{CDM}$ model parameters. BAO versus NewBAO.}
\begin{ruledtabular}
\begin{tabular}{lcccc}
 \multicolumn{5}{c}{Tilted flat-$\Lambda\textrm{CDM}$ model} \\
 \hline \\[-2mm]
  Parameter               & TT+lowP+BAO            &  TT+lowP+lensing+BAO    &  TT+lowP+NewBAO         &  TT+lowP+lensing+NewBAO  \\[+0mm]
 \hline \\[-2mm]
  $\Omega_b h^2$          & $0.02227 \pm 0.00020$  &  $0.02225 \pm 0.00020$  &  $0.02229 \pm 0.00020$  &  $0.02227 \pm 0.00020$   \\[+1mm]
  $\Omega_c h^2$          & $0.1190  \pm 0.0013$   &  $0.1185  \pm 0.0012$   &  $0.1187  \pm 0.0012$   &  $0.1183  \pm 0.0012$    \\[+1mm]
  $100\theta_\textrm{MC}$ & $1.04095 \pm 0.00042$  &  $1.04103 \pm 0.00041$  &  $1.04099 \pm 0.00041$  &  $1.04105 \pm 0.00040$   \\[+1mm]
  $\tau$                  & $0.080   \pm 0.018$    &  $0.067   \pm 0.013$    &  $0.079   \pm 0.017$    &  $0.066   \pm 0.013$     \\[+1mm]
  $\ln(10^{10} A_s)$      & $3.092   \pm 0.035$    &  $3.065   \pm 0.024$    &  $3.088   \pm 0.034$    &  $3.064   \pm 0.024$     \\[+1mm]
  $n_s$                   & $0.9673  \pm 0.0044$   &  $0.9674  \pm 0.0044$   &  $0.9678  \pm 0.0044$   &  $0.9682  \pm 0.0044$    \\[+1mm]  
 \hline \\[-2mm]
  $H_0$ [km s$^{-1}$ Mpc$^{-1}$] & $67.65 \pm0.57$ &  $67.81 \pm 0.54$       &  $67.78   \pm 0.55$     &  $67.92   \pm 0.54$      \\[+1mm]  
  $\Omega_m$              & $0.3102 \pm0.0076$     &  $0.3077  \pm 0.0072$   &  $0.3083  \pm 0.0074$   &  $0.3063  \pm 0.0071$    \\[+1mm]  
  $\sigma_8$              & $0.829  \pm 0.014$     &  $0.8158  \pm 0.0089$   &  $0.826   \pm 0.014$    &  $0.8150  \pm 0.0089$    \\[+1mm]  
    \hline \hline \\[-2mm]
   \multicolumn{5}{c}{Untilted non-flat $\Lambda\textrm{CDM}$ model} \\
   \hline \\[-2mm]
  Parameter               & TT+lowP+BAO            &  TT+lowP+lensing+BAO    &  TT+lowP+NewBAO         &  TT+lowP+lensing+NewBAO  \\[+0mm]
 \hline \\[-2mm]
  $\Omega_b h^2$          & $0.02307 \pm0.00020$  &  $0.02304  \pm 0.00020$  &  $0.02307 \pm 0.00020$  &  $0.02303 \pm 0.00020$   \\[+1mm]
  $\Omega_c h^2$          & $0.1096  \pm0.0011$   &  $0.1093   \pm 0.0011$   &  $0.1096  \pm 0.0011$   &  $0.1093  \pm 0.0010$    \\[+1mm]
  $100\theta_\textrm{MC}$ & $1.04222 \pm0.00042$  &  $1.04232  \pm 0.00041$  &  $1.04223 \pm 0.00042$  &  $1.04230 \pm 0.00042$   \\[+1mm]
  $\tau$                  & $0.135   \pm0.018$    &  $0.115    \pm 0.011$    &  $0.132   \pm 0.017$    &  $0.115   \pm 0.011$     \\[+1mm]
  $\ln(10^{10} A_s)$      & $3.179   \pm0.036$    &  $3.138    \pm 0.022$    &  $3.174   \pm 0.034$    &  $3.139   \pm 0.022$     \\[+1mm]
  $\Omega_k$              & $-0.0093 \pm0.0019$   &  $-0.0093  \pm 0.0018$   &  $-0.0088 \pm 0.0017$   &  $-0.0087 \pm 0.0017$    \\[+1mm]  
 \hline \\[-2mm]
  $H_0$ [km s$^{-1}$ Mpc$^{-1}$] & $67.46\pm0.72$ &  $67.56 \pm 0.67$        &  $67.69   \pm 0.66$     &  $67.81   \pm 0.66$      \\[+1mm]  
  $\Omega_m$              & $0.2931 \pm0.0064$    &  $0.2914   \pm 0.0059$   &  $0.2910  \pm 0.0059$   &  $0.2893  \pm 0.0058$    \\[+1mm]  
  $\sigma_8$              & $0.832  \pm 0.016$    &  $0.814    \pm 0.010$    &  $0.830   \pm 0.015$    &  $0.8148  \pm 0.0097$    \\[+0mm]  
\end{tabular}
\end{ruledtabular}
\label{tab:para_bao}
\end{table*}

\begin{figure*}
\centering
\mbox{\includegraphics[width=87mm]{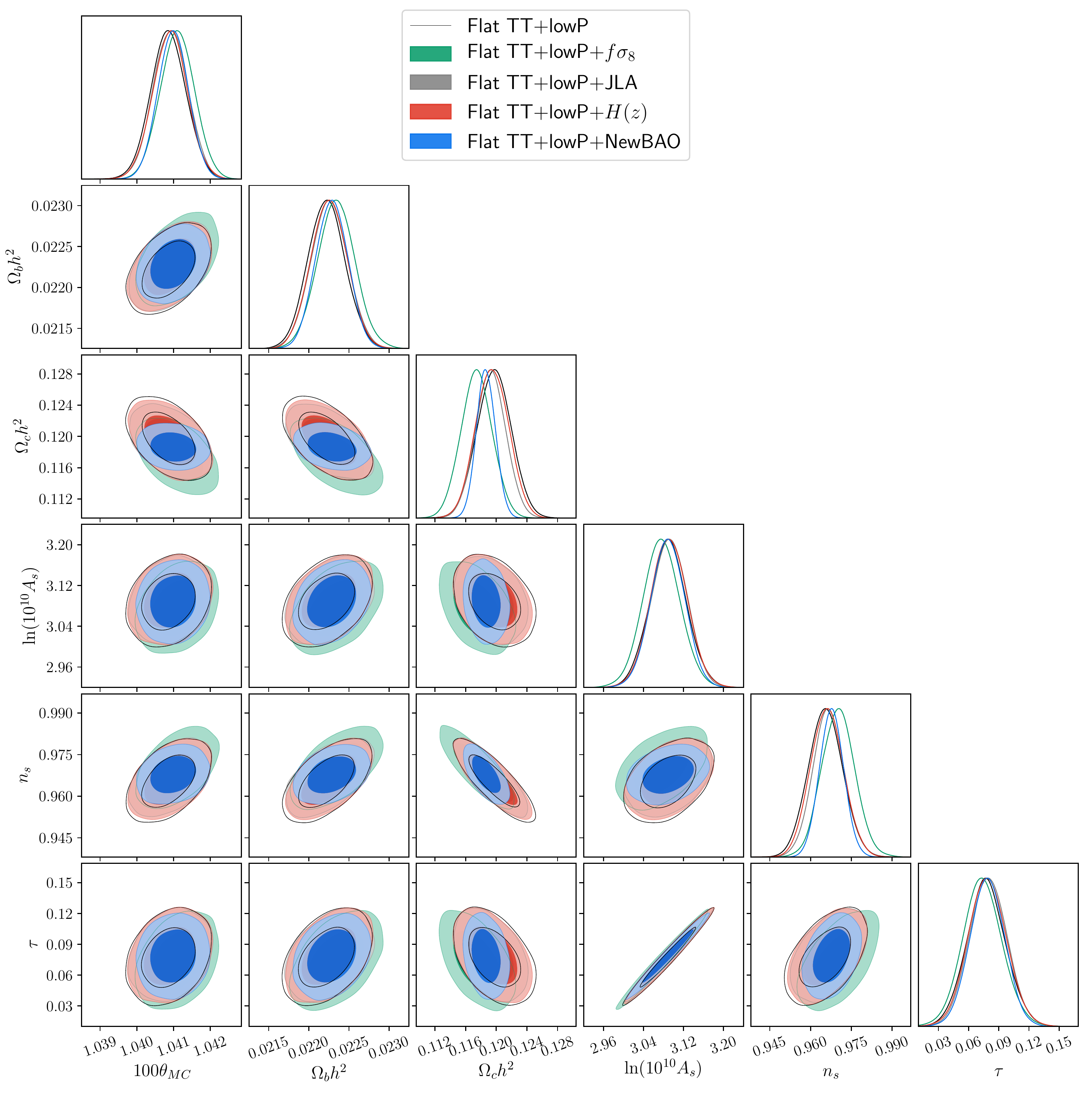}} 
\mbox{\includegraphics[width=87mm]{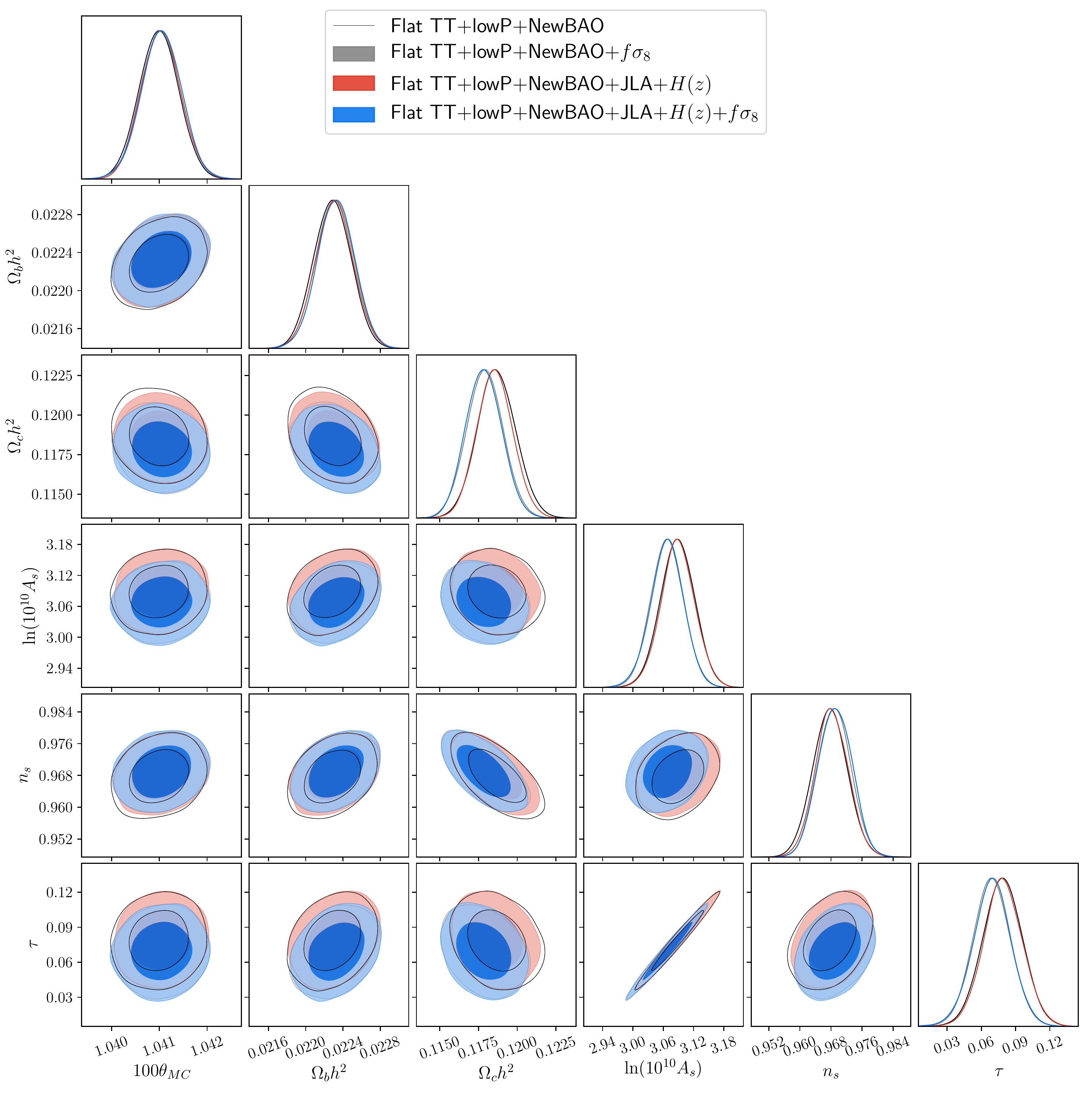}} 
\caption{Likelihood distributions of the tilted flat-$\Lambda\textrm{CDM}$
model parameters constrained by Planck CMB TT + lowP, JLA SNIa, NewBAO, $H(z)$,
and $f\sigma_8$ data.
Two-dimensional marginalized likelihood distributions of all possible
combinations of model parameters together with one-dimensional likelihoods are
shown for cases when each non-CMB data set is added to the Planck TT + lowP data
(left panel) and when the growth rate, JLA SNIa, Hubble parameter data, and
the combination of them, are added to TT + lowP + NewBAO data (right panel).  
For ease of viewing, the cases of TT + lowP (left) and TT + lowP + NewBAO
(right panel) are shown as solid black curves.
}
\label{fig:para_flat}
\end{figure*}

\begin{figure*}
\centering
\mbox{\includegraphics[width=87mm]{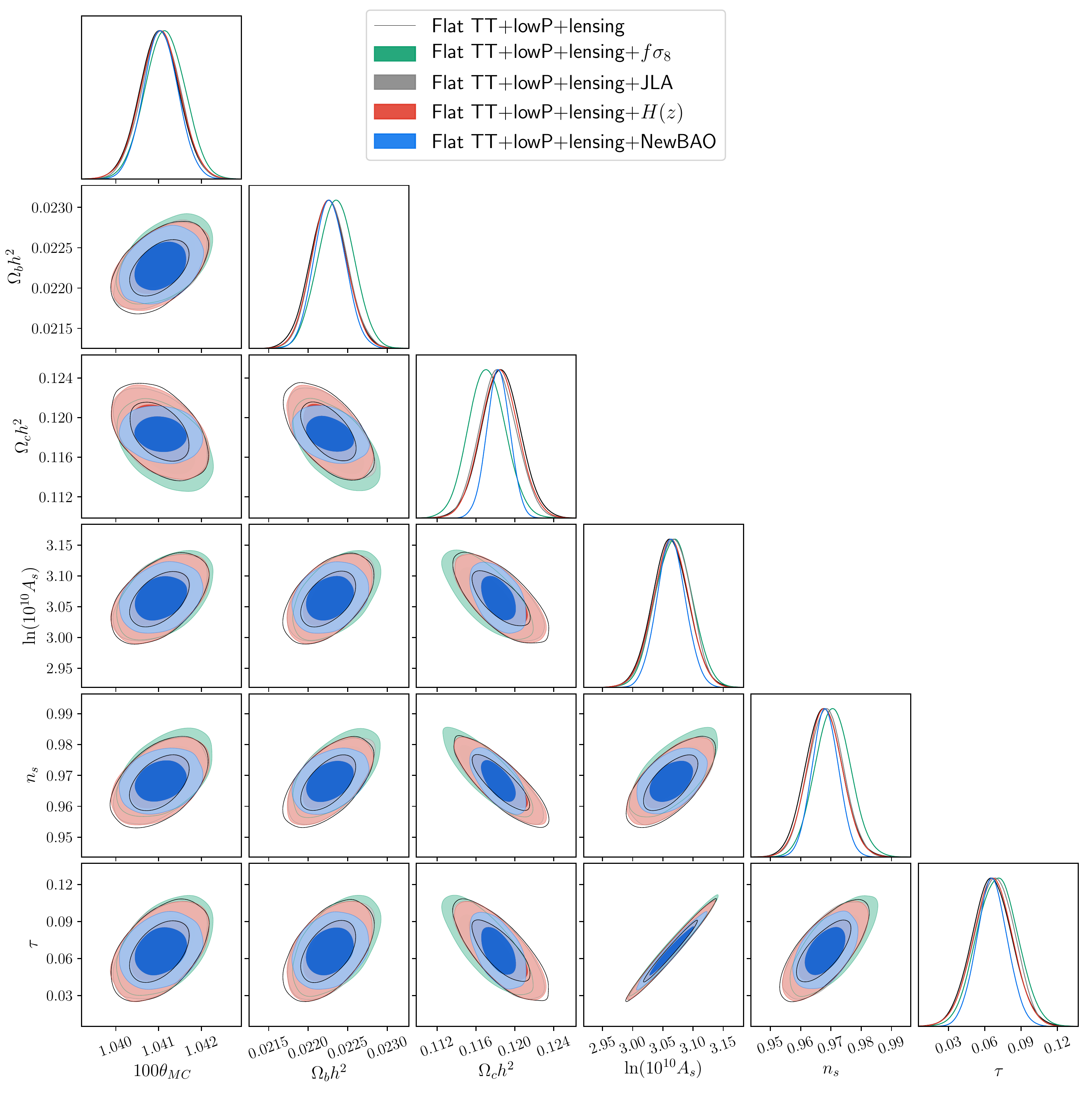}} 
\mbox{\includegraphics[width=87mm]{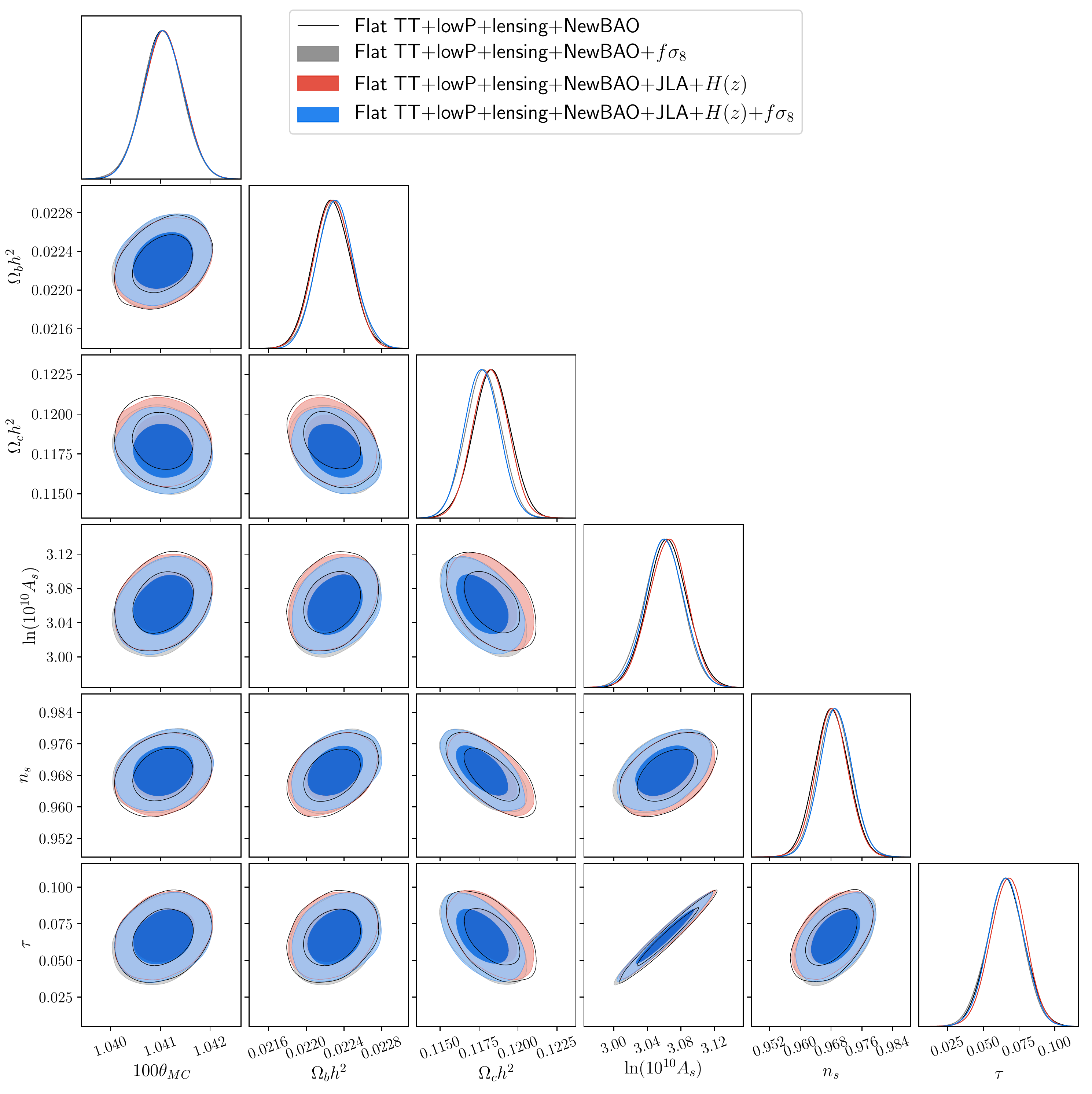}} 
\caption{Same as Fig.\ \ref{fig:para_flat} but now including
the Planck CMB lensing data.
}
\label{fig:para_flat_lensing}
\end{figure*}

\begin{figure*}
\centering
\mbox{\includegraphics[width=87mm]{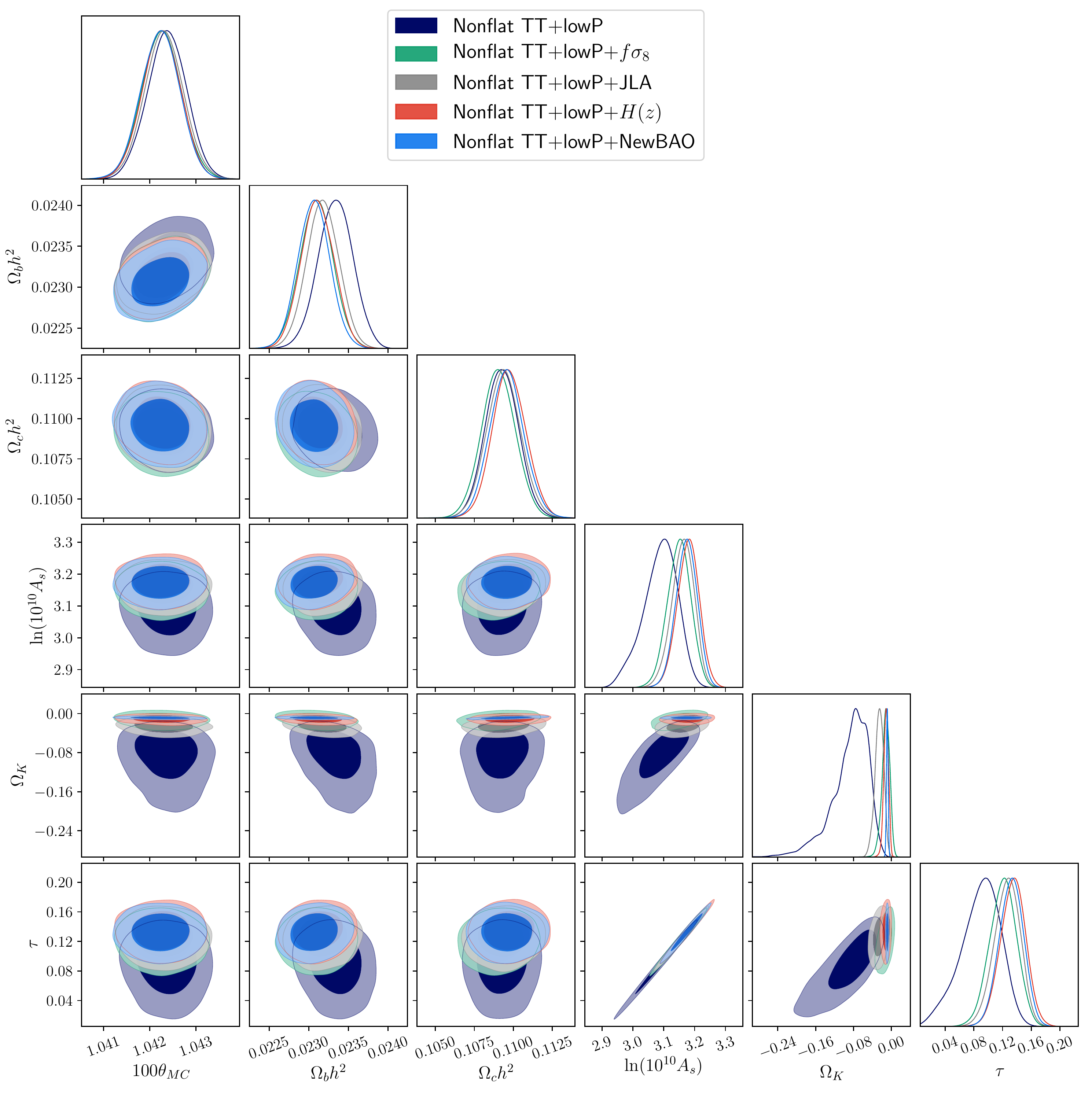}} 
\mbox{\includegraphics[width=87mm]{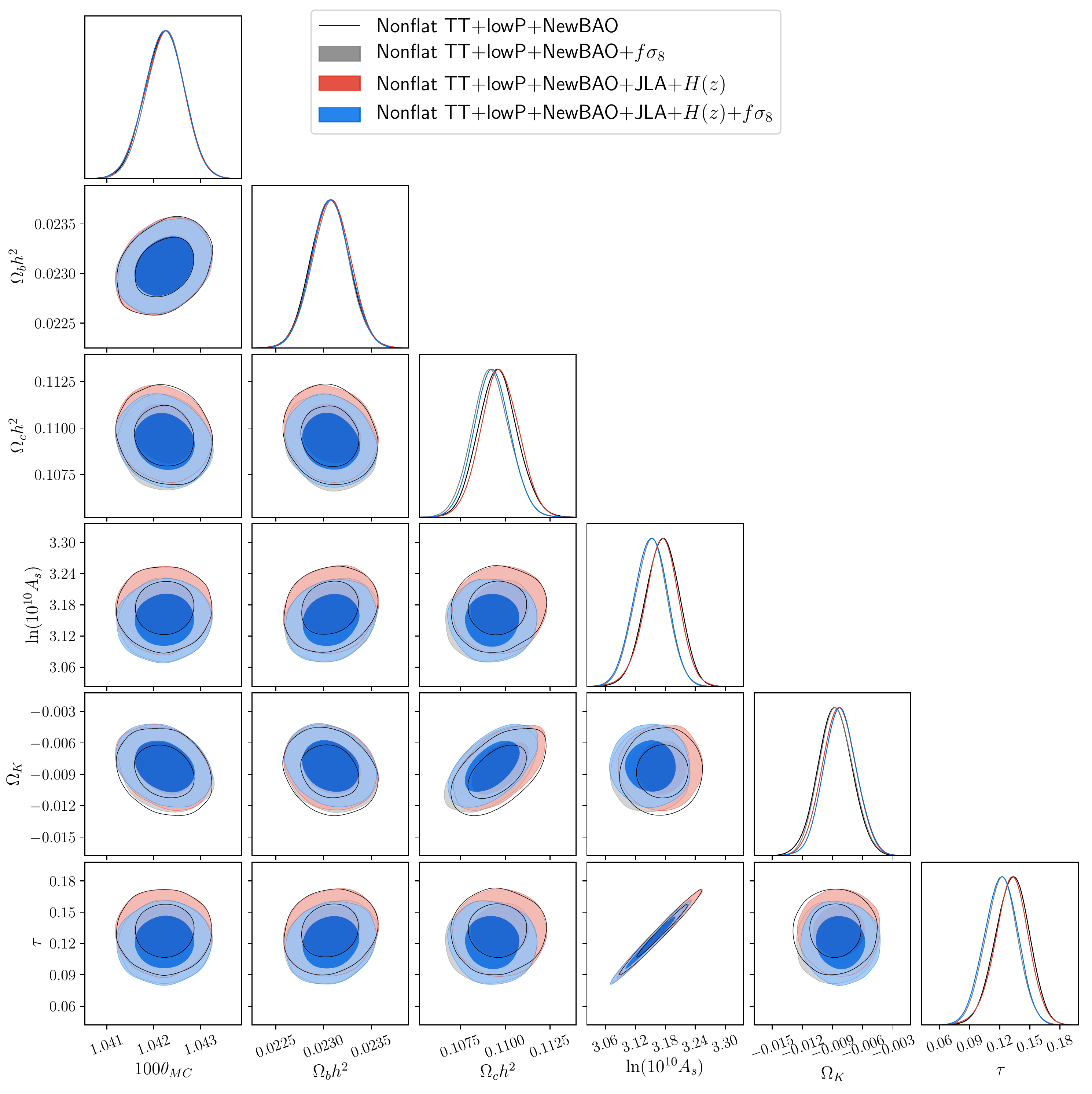}}
\caption{
Likelihood distributions of the untilted non-flat $\Lambda\textrm{CDM}$ model
parameters constrained by Planck CMB TT + lowP, JLA SNIa, NewBAO, $H(z)$, and
$f\sigma_8$ data.
Two-dimensional marginalized likelihood distributions of all possible
combinations of model parameters together with one-dimensional likelihoods are
shown for cases when each non-CMB data set is added to the Planck TT + lowP data
(left panel) and when the growth rate, JLA SNIa, Hubble parameter data,
and the combination of them, are added to TT + lowP + NewBAO data
(right panel). For ease of viewing, the result of TT + lowP + NewBAO
is shown as solid black curves in the right panel.
}
\label{fig:para_nonflat}
\end{figure*}

\begin{figure*}
\centering
\mbox{\includegraphics[width=87mm]{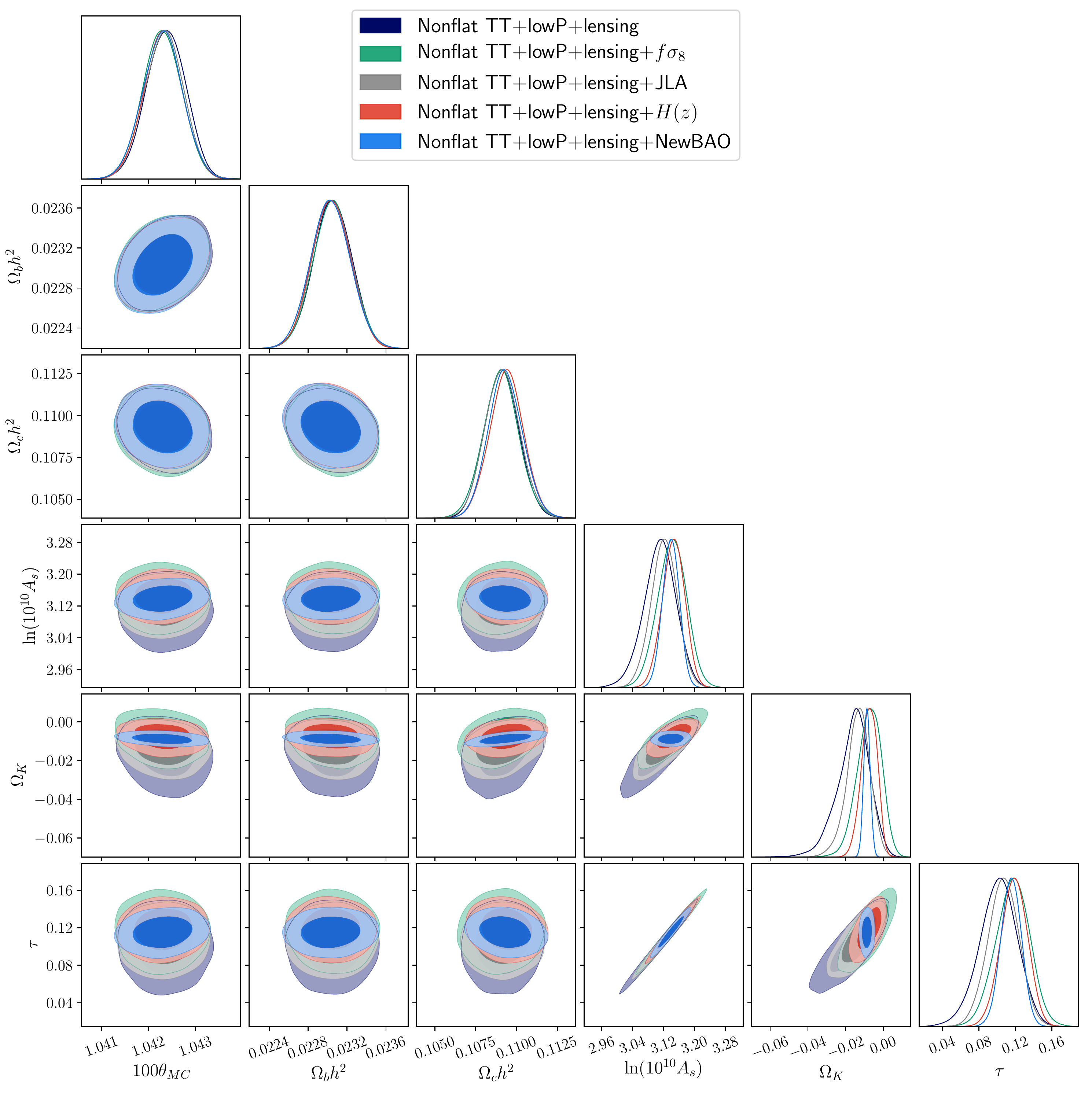}} 
\mbox{\includegraphics[width=87mm]{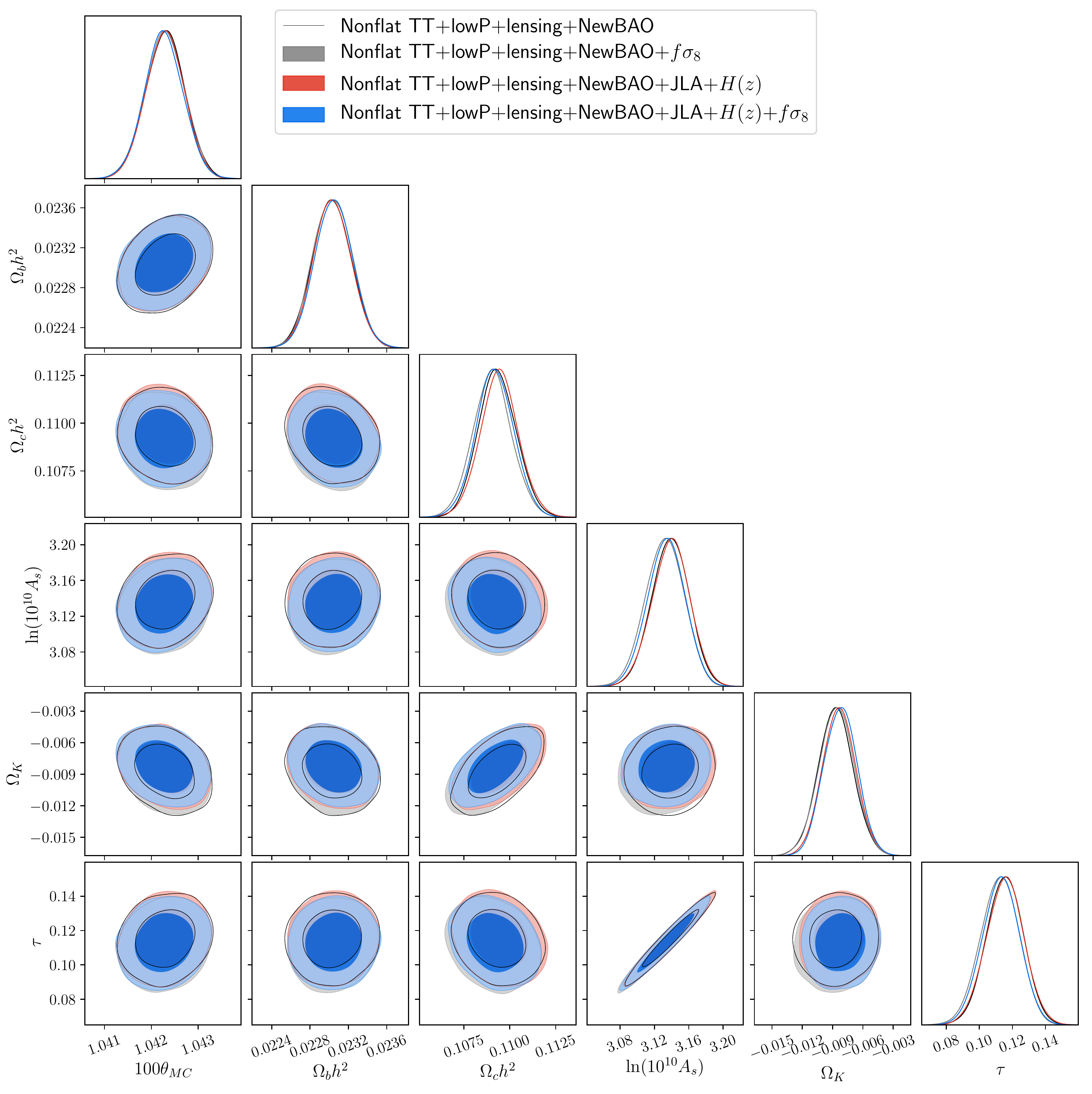}}
\caption{Same as Fig.\ \ref{fig:para_nonflat} but now including
the Planck CMB lensing data.
}
\label{fig:para_nonflat_lensing}
\end{figure*}

We investigate the effect of including non-CMB data sets, with the Planck 
2015 CMB data, on the parameter
constraints of the tilted flat and the untilted non-flat $\Lambda\textrm{CDM}$ models.
The results are presented in Figs.\
\ref{fig:para_flat}--\ref{fig:para_nonflat_lensing}
and Tables \ref{tab:para_flat}--\ref{tab:para_nonflat_lensing}.
In the triangle plots we omit the likelihood contours for
TT + lowP (+ lensing) + JLA + NewBAO data (excluding or including
the Planck lensing data) in both the tilted flat and the untilted non-flat
$\Lambda\textrm{CDM}$ models because they are very similar to
those for TT + lowP (+ lensing) + NewBAO data.

The entries in the CMB-only TT + lowP column of Table \ref{tab:para_flat} 
and those in the TT + lowP + lensing column of Table 
\ref{tab:para_flat_lensing} for the 
tilted flat-$\Lambda$CDM model agree well with the corresponding entries in
Table 4 of \citet{PlanckCollaboration2016}. Similarly, the entries in the 
TT + lowP column of Table \ref{tab:para_nonflat} and those 
in the TT + lowP + lensing column of Table \ref{tab:para_nonflat_lensing} 
for the non-flat $\Lambda$CDM model agree well with the corresponding entries in
Table 1 of \citet{Oobaetal2018a}.

From Tables \ref{tab:para_flat} and \ref{tab:para_flat_lensing} we see that, 
when added to the Planck 2015 CMB anisotropy data, for the tilted 
flat-$\Lambda$CDM model, the NewBAO measurements prove more restrictive 
than either the $H(z)$, $f \sigma_8$, or SNIa observations. We note however 
that our NewBAO compilation includes radial BAO $H(z)$ measurements as well 
as the $f \sigma_8$ measurements of \citet{Alametal2017}. It is likely that 
even if these are moved to the $H(z)$ and $f \sigma_8$ data sets, BAO 
constraints will still be the most restrictive, for the tilted 
flat-$\Lambda$CDM model, but probably closely followed by $H(z)$ and 
$f \sigma_8$ constraints, with SNIa being the least effective.

The situation in the untilted non-flat $\Lambda$CDM case is more interesting.
When CMB lensing data are excluded, Table \ref{tab:para_nonflat}, adding
NewBAO, or JLA SNIa, or $H(z)$, or $f \sigma_8$ data to the CMB data results in 
roughly similarly restrictive constraints on $\Omega_b h^2$, $\Omega_c h^2$, 
$\theta_\textrm{MC}$, $\tau$, $\ln (10^{10} A_s)$, and $\sigma_8$, while CMB + 
NewBAO data provide the tightest constraints on $\Omega_k$, $H_0$, and 
$\Omega_m$.
When the CMB lensing data are included, Table \ref{tab:para_nonflat_lensing},
CMB data with either JLA SNIa, or NewBAO, or $H(z)$, or $f \sigma_8$ data,
provide roughly similarly restrictive constraints on $\Omega_b h^2$,
$\Omega_c h^2$, and $\theta_\textrm{MC}$, while CMB + NewBAO data provide
the tightest constraints on $\tau$, $\ln (10^{10} A_s)$, $\Omega_k$, $H_0$,
$\Omega_m$, and $\sigma_8$.

\begin{table*}
\caption{Tilted flat-$\Lambda\textrm{CDM}$ model parameters constrained
with Planck TT + lowP, JLA SNIa, NewBAO, $H(z)$, and $f\sigma_8$ data (mean and 68.3\% confidence limits).}
\begin{ruledtabular}
\begin{tabular}{lccc}
  Parameter                & TT+lowP                & TT+lowP+JLA              &  TT+lowP+NewBAO        \\[+0mm]
 \hline \\[-2mm]
  $\Omega_b h^2$           & $0.02222 \pm 0.00023$  & $0.02226 \pm 0.00022$    &  $0.02229  \pm 0.00020$    \\[+1mm]
  $\Omega_c h^2$           & $0.1197  \pm 0.0022$   & $0.1193  \pm 0.0020$     &  $0.1187   \pm 0.0012$       \\[+1mm]
  $100\theta_\textrm{MC}$  & $1.04086 \pm 0.00048$  & $1.04092 \pm 0.00047$    &  $1.04099  \pm 0.00041$     \\[+1mm]
  $\tau$                   & $0.078 \pm 0.019$      & $0.080   \pm 0.019$      &  $0.079    \pm 0.017$         \\[+1mm]
  $\ln(10^{10} A_s)$       & $3.089 \pm 0.037$      & $3.092   \pm 0.035$      &  $3.088    \pm 0.034$        \\[+1mm]
  $n_s$                    & $0.9655 \pm 0.0062$    & $0.9666  \pm 0.0057$     &  $0.9678   \pm 0.0044$     \\[+1mm]  
 \hline \\[-2mm]
  $H_0$ [km s$^{-1}$ Mpc$^{-1}$] & $67.32 \pm 0.99$ & $67.52 \pm 0.89$         &  $67.78    \pm 0.55$          \\[+1mm]  
  $\Omega_m$               & $0.315 \pm 0.014$      & $0.312 \pm 0.012$        &  $0.3083   \pm 0.0074$      \\[+1mm]  
  $\sigma_8$               & $0.829 \pm 0.015$      & $0.829 \pm 0.014$        &  $0.826    \pm 0.014$        \\[+1mm]  
  \hline \hline \\[-2mm]
    Parameter              & TT+lowP+$H(z)$         &  TT+lowP+JLA+NewBAO      &  TT+lowP+JLA+NewBAO+$H(z)$  \\[+0mm]
 \hline \\[-2mm]
  $\Omega_b h^2$           & $0.02225 \pm 0.00022$  &  $0.02230  \pm 0.00019$  &  $0.02231 \pm 0.00020$    \\[+1mm]
  $\Omega_c h^2$           & $0.1195  \pm 0.0021$   &  $0.1186   \pm 0.0012$   &  $0.1185  \pm 0.0012$     \\[+1mm]
  $100\theta_\textrm{MC}$  & $1.04091 \pm 0.00047$  &  $1.04101  \pm 0.00041$  &  $1.04103 \pm 0.00041$    \\[+1mm]
  $\tau$                   & $0.079 \pm 0.019$      &  $0.078    \pm 0.017$    &  $0.079   \pm 0.017$      \\[+1mm]
  $\ln(10^{10} A_s)$       & $3.091 \pm 0.036$      &  $3.088    \pm 0.034$    &  $3.089   \pm 0.034$      \\[+1mm]
  $n_s$                    & $0.9661 \pm 0.0060$    &  $0.9679   \pm 0.0044$   &  $0.9682  \pm 0.0042$     \\[+1mm]  
 \hline \\[-2mm]
  $H_0$ [km s$^{-1}$ Mpc$^{-1}$] & $67.44 \pm 0.94$ & $67.82    \pm 0.55$      &  $67.85   \pm 0.52$       \\[+1mm]  
  $\Omega_m$               & $0.313 \pm 0.013$      & $0.3078   \pm 0.0072$    &  $0.3074  \pm 0.0068$     \\[+1mm]  
  $\sigma_8$               & $0.829 \pm 0.014$      & $0.826    \pm 0.014$     &  $0.826   \pm 0.014$      \\[+1mm]  
    \hline \hline \\[-2mm]
    Parameter              & TT+lowP+$f\sigma_8$  & TT+lowP+NewBAO+$f\sigma_8$ &  TT+lowP+JLA+NewBAO+$H(z)$+$f\sigma_8$  \\[+0mm]
 \hline \\[-2mm]
  $\Omega_b h^2$           & $0.02234 \pm 0.00023$  & $0.02231 \pm 0.00020$    &  $0.02232 \pm 0.00020$    \\[+1mm]
  $\Omega_c h^2$           & $0.1174  \pm 0.0020$   & $0.1179  \pm 0.0012$     &  $0.1178  \pm 0.0011$     \\[+1mm]
  $100\theta_\textrm{MC}$  & $1.04110 \pm 0.00046$  & $1.04104 \pm 0.00041$    &  $1.04104 \pm 0.00042$    \\[+1mm]
  $\tau$                   & $0.074 \pm 0.020$      & $0.069   \pm 0.016$      &  $0.069   \pm 0.017$      \\[+1mm]
  $\ln(10^{10} A_s)$       & $3.076 \pm 0.037$      & $3.068   \pm 0.032$      &  $3.068   \pm 0.033$      \\[+1mm]
  $n_s$                    & $0.9702 \pm 0.0061$    & $0.9689  \pm 0.0042$     &  $0.9690  \pm 0.0043$     \\[+1mm]  
 \hline \\[-2mm]
  $H_0$ [km s$^{-1}$ Mpc$^{-1}$] & $68.31 \pm 0.94$ & $68.09 \pm 0.54$         &  $68.12   \pm 0.52$       \\[+1mm]  
  $\Omega_m$               & $0.301 \pm 0.012$      & $0.3040 \pm 0.0069$      &  $0.3035  \pm 0.0067$     \\[+1mm]  
  $\sigma_8$               & $0.817 \pm 0.014$      & $0.815 \pm 0.013$        &  $0.815   \pm 0.013$       \\[+0mm]
\end{tabular}
\end{ruledtabular}
\label{tab:para_flat}
\end{table*}


\begin{table*}
\caption{Tilted flat-$\Lambda\textrm{CDM}$ model parameters constrained with Planck TT + lowP + lensing, JLA, NewBAO, $H(z)$, and $f\sigma_8$ data (mean and 68.3\% confidence limits).}
\begin{ruledtabular}
\begin{tabular}{lccc}
  Parameter                & TT+lowP+lensing        & TT+lowP+lensing+JLA    &  TT+lowP+lensing+NewBAO        \\[+0mm]
 \hline \\[-2mm]
  $\Omega_b h^2$           & $0.02225 \pm 0.00023$  & $0.02227 \pm 0.00022$  &  $0.02227  \pm 0.00020$    \\[+1mm]
  $\Omega_c h^2$           & $0.1186  \pm 0.0020$   & $0.1183  \pm 0.0019$   &  $0.1183   \pm 0.0012$       \\[+1mm]
  $100\theta_\textrm{MC}$  & $1.04102 \pm 0.00046$  & $1.04105 \pm 0.00045$  &  $1.04105  \pm 0.00040$     \\[+1mm]
  $\tau$                   & $0.066 \pm 0.017$      & $0.068   \pm 0.016$    &  $0.066    \pm 0.013$         \\[+1mm]
  $\ln(10^{10} A_s)$       & $3.063 \pm 0.030$      & $3.066   \pm 0.029$    &  $3.064    \pm 0.024$        \\[+1mm]
  $n_s$                    & $0.9677 \pm 0.0060$    & $0.9683  \pm 0.0058$   &  $0.9682   \pm 0.0044$     \\[+1mm]  
 \hline \\[-2mm]
  $H_0$ [km s$^{-1}$ Mpc$^{-1}$] & $67.80 \pm 0.92$ & $67.93 \pm 0.88$       &  $67.92    \pm 0.54$          \\[+1mm]  
  $\Omega_m$               & $0.308 \pm 0.012$      & $0.306 \pm 0.012$      &  $0.3063   \pm 0.0071$      \\[+1mm]  
  $\sigma_8$               & $0.8151 \pm 0.0095$    & $0.8156 \pm 0.0093$    &  $0.8150   \pm 0.0089$        \\[+1mm]  
  \hline \hline \\[-2mm]
    Parameter              & TT+lowP+lensing+$H(z)$ &  TT+lowP+lensing+JLA+NewBAO  &  TT+lowP+lensing+JLA+NewBAO+$H(z)$  \\[+0mm]
 \hline \\[-2mm]
  $\Omega_b h^2$           & $0.02226 \pm 0.00022$  &  $0.02227  \pm 0.00020$ &  $0.02228 \pm 0.00019$    \\[+1mm]
  $\Omega_c h^2$           & $0.1184  \pm 0.0019$   &  $0.1182   \pm 0.0011$   &  $0.1182  \pm 0.0011$     \\[+1mm]
  $100\theta_\textrm{MC}$  & $1.04104 \pm 0.00046$  &  $1.04105  \pm 0.00040$  &  $1.04107 \pm 0.00040$    \\[+1mm]
  $\tau$                   & $0.066 \pm 0.016$      &  $0.067    \pm 0.013$    &  $0.067   \pm 0.012$      \\[+1mm]
  $\ln(10^{10} A_s)$       & $3.063 \pm 0.029$      &  $3.064    \pm 0.024$    &  $3.064   \pm 0.023$      \\[+1mm]
  $n_s$                    & $0.9680 \pm 0.0058$    &  $0.9682   \pm 0.0043$   &  $0.9682  \pm 0.0043$     \\[+1mm]  
 \hline \\[-2mm]
  $H_0$ [km s$^{-1}$ Mpc$^{-1}$]  & $67.86 \pm 0.88$ & $67.95    \pm 0.52$     &  $67.95   \pm 0.51$       \\[+1mm]  
  $\Omega_m$               & $0.307 \pm 0.012$      & $0.3058   \pm 0.0068$    &  $0.3058  \pm 0.0067$     \\[+1mm]  
  $\sigma_8$               & $0.8151 \pm 0.0094$    & $0.8149    \pm 0.0090$   &  $0.8149  \pm 0.0089$      \\[+1mm]  
  \hline  \hline \\[-2mm]
    Parameter              & TT+lowP+lensing+$f\sigma_8$  & TT+lowP+lensing+NewBAO+$f\sigma_8$ &  TT+lowP+lensing+JLA+NewBAO+$H(z)$+$f\sigma_8$  \\[+0mm]
 \hline \\[-2mm]
  $\Omega_b h^2$           & $0.02235 \pm 0.00023$  & $0.02230 \pm 0.00020$    &  $0.02231 \pm 0.00019$    \\[+1mm]
  $\Omega_c h^2$           & $0.1171  \pm 0.0019$   & $0.1178  \pm 0.0011$     &  $0.1177  \pm 0.0011$     \\[+1mm]
  $100\theta_\textrm{MC}$  & $1.04115 \pm 0.00046$  & $1.04104 \pm 0.00041$    &  $1.04106 \pm 0.00040$    \\[+1mm]
  $\tau$                   & $0.070 \pm 0.017$      & $0.065   \pm 0.013$      &  $0.066   \pm 0.012$      \\[+1mm]
  $\ln(10^{10} A_s)$       & $3.068 \pm 0.030$      & $3.059   \pm 0.024$      &  $3.061   \pm 0.023$      \\[+1mm]
  $n_s$                    & $0.9707 \pm 0.0059$    & $0.9690  \pm 0.0043$     &  $0.9692  \pm 0.0042$     \\[+1mm]  
 \hline \\[-2mm]
  $H_0$ [km s$^{-1}$ Mpc$^{-1}$] & $68.44 \pm 0.89$ & $68.13 \pm 0.52$         &  $68.17   \pm 0.50$       \\[+1mm]  
  $\Omega_m$               & $0.299 \pm 0.011$      & $0.3033 \pm 0.0068$      &  $0.3027  \pm 0.0065$     \\[+1mm]  
  $\sigma_8$               & $0.8130 \pm 0.0094$    & $0.8113 \pm 0.0087$      &  $0.8116  \pm 0.0087$      \\[+0mm]
\end{tabular}
\end{ruledtabular}
\label{tab:para_flat_lensing}
\end{table*}


\begin{table*}
\caption{Untilted non-flat $\Lambda\textrm{CDM}$ model parameters constrained with Planck TT + lowP, JLA SNIa, NewBAO, $H(z)$, and $f\sigma_8$ data (mean and 68.3\% confidence limits).}
\begin{ruledtabular}
\begin{tabular}{lccc}
  Parameter               &  TT+lowP                  &  TT+lowP+JLA            &   TT+lowP+NewBAO      \\[+0mm]
 \hline \\[-2mm]
  $\Omega_b h^2$          &  $0.02334 \pm 0.00022$    &  $0.02318 \pm 0.00020$  &   $0.02307 \pm 0.00020$     \\[+1mm]
  $\Omega_c h^2$          &  $0.1093 \pm 0.0011$      &  $0.1094  \pm 0.0011$   &   $0.1096 \pm 0.0011$       \\[+1mm]
  $100\theta_\textrm{MC}$ &  $1.04237 \pm 0.00042$    &  $1.04231 \pm 0.00042$  &   $1.04223 \pm 0.00042$     \\[+1mm]
  $\tau$                  &  $0.089 \pm 0.028$        &  $0.126   \pm 0.018$    &   $0.132 \pm 0.017$       \\[+1mm]
  $\ln(10^{10} A_s)$      &  $3.088 \pm 0.055$        &  $3.162   \pm 0.036$    &   $3.174 \pm 0.034$        \\[+1mm]
  $\Omega_k$              &  $-0.088 \pm 0.040$       &  $-0.0257 \pm 0.0091$   &   $-0.0088 \pm 0.0017$     \\[+1mm]  
 \hline \\[-2mm]
  $H_0$ [km s$^{-1}$ Mpc$^{-1}$] & $49.1 \pm 5.4$     &  $61.5 \pm 2.9$         &   $67.69 \pm 0.66$        \\[+1mm]  
  $\Omega_m$              &  $0.58 \pm 0.14$          &  $0.355 \pm 0.033$      &   $0.2910 \pm 0.0059$     \\[+1mm]  
  $\sigma_8$              &  $0.755 \pm 0.038$        &  $0.815 \pm 0.018$      &   $0.830 \pm 0.015$       \\[+1mm]  
    \hline \hline \\[-2mm]
  Parameter               &  TT+lowP+$H(z)$           &  TT+lowP+JLA+NewBAO     &  TT+lowP+JLA+NewBAO+$H(z)$      \\[+0mm]
 \hline \\[-2mm]
  $\Omega_b h^2$          &  $0.02311 \pm 0.00020$    &  $0.02307 \pm 0.00020$  &  $0.02308  \pm 0.00020$       \\[+1mm]
  $\Omega_c h^2$          &  $0.1097 \pm 0.0011$      &  $0.1096  \pm 0.0011$   &  $0.1097   \pm 0.0011$      \\[+1mm]
  $100\theta_\textrm{MC}$ &  $1.04225 \pm 0.00041$    &  $1.04223 \pm 0.00040$  &  $1.04222  \pm 0.00042$       \\[+1mm]
  $\tau$                  &  $0.134 \pm 0.018$        &  $0.132   \pm 0.016$    &  $0.132    \pm 0.017$         \\[+1mm]
  $\ln(10^{10} A_s)$      &  $3.179 \pm 0.036$        &  $3.173   \pm 0.032$    &  $3.172    \pm 0.034$         \\[+1mm]
  $\Omega_k$              &  $-0.0113 \pm 0.0051$     &  $-0.0087 \pm 0.0016$   &  $-0.0084  \pm 0.0017$        \\[+1mm]  
 \hline \\[-2mm]
  $H_0$ [km s$^{-1}$ Mpc$^{-1}$]  &  $66.7 \pm 2.2$   &  $67.69 \pm 0.63$       &  $67.85 \pm 0.65$             \\[+1mm]  
  $\Omega_m$              &  $0.302 \pm 0.020$        &  $0.2910 \pm 0.0057$    &  $0.2899 \pm 0.0058$            \\[+1mm]  
  $\sigma_8$              &  $0.831 \pm 0.016$        &  $0.830 \pm 0.014$      &  $0.830 \pm 0.015$           \\[+1mm]  
   \hline \hline \\[-2mm]
     Parameter            &  TT+lowP+$f\sigma_8$      & TT+lowP+NewBAO+$f\sigma_8$  &  TT+lowP+JLA+NewBAO+$H(z)$+$f\sigma_8$      \\[+0mm]
 \hline \\[-2mm]
  $\Omega_b h^2$          &  $0.02310 \pm 0.00021$    &  $0.02307 \pm 0.00019$  &  $0.02307  \pm 0.00020$       \\[+1mm]
  $\Omega_c h^2$          &  $0.1090  \pm 0.0011$     &  $0.1092  \pm 0.0010$   &  $0.1093   \pm 0.0010$      \\[+1mm]
  $100\theta_\textrm{MC}$ &  $1.04225 \pm 0.00042$    &  $1.04224 \pm 0.00040$  &  $1.04222  \pm 0.00042$       \\[+1mm]
  $\tau$                  &  $0.121 \pm 0.019$        &  $0.121   \pm 0.016$    &  $0.121    \pm 0.017$         \\[+1mm]
  $\ln(10^{10} A_s)$      &  $3.150 \pm 0.038$        &  $3.151   \pm 0.032$    &  $3.151    \pm 0.033$         \\[+1mm]
  $\Omega_k$              &  $-0.0120 \pm 0.0085$     &  $-0.0087 \pm 0.0017$   &  $-0.0082  \pm 0.0016$        \\[+1mm]  
 \hline \\[-2mm]
  $H_0$ [km s$^{-1}$ Mpc$^{-1}$]  &  $66.8 \pm 3.6$   &  $67.87 \pm 0.64$       &  $68.04 \pm 0.62$             \\[+1mm]  
  $\Omega_m$              &  $0.301 \pm 0.032$        &  $0.2886 \pm 0.0057$    &  $0.2874 \pm 0.0055$            \\[+1mm]  
  $\sigma_8$              &  $0.816 \pm 0.019$        &  $0.819 \pm 0.014$      &  $0.820 \pm 0.014$                \\[+0mm]
\end{tabular}
\end{ruledtabular}
\label{tab:para_nonflat}
\end{table*}


\begin{table*}
\caption{Untilted non-flat $\Lambda\textrm{CDM}$ model parameters constrained with Planck TT + lowP + lensing, JLA, NewBAO, $H(z)$, and $f\sigma_8$ data (mean and 68.3\% confidence limits).}
\begin{ruledtabular}
\begin{tabular}{lccc}
  Parameter               &  TT+lowP+lensing          & TT+lowP+lensing+JLA     &   TT+lowP+lensing+NewBAO      \\[+0mm]
 \hline \\[-2mm]
  $\Omega_b h^2$          &  $0.02305 \pm 0.00020$    &  $0.02304 \pm 0.00020$  &   $0.02303 \pm 0.00020$       \\[+1mm]
  $\Omega_c h^2$          &  $0.1091 \pm 0.0010$      &  $0.1091 \pm 0.0011$    &   $0.1093 \pm 0.0010$         \\[+1mm]
  $100\theta_\textrm{MC}$ &  $1.04237 \pm 0.00042$    &  $1.04233 \pm 0.00041$  &   $1.04230 \pm 0.00042$       \\[+1mm]
  $\tau$                  &  $0.101 \pm 0.021$        &  $0.107 \pm 0.017$      &   $0.115 \pm 0.011$        \\[+1mm]
  $\ln(10^{10} A_s)$      &  $3.110 \pm 0.041$        &  $3.120 \pm 0.034$      &   $3.139 \pm 0.022$           \\[+1mm]
  $\Omega_k$              &  $-0.0160 \pm 0.0087$     &  $-0.0133 \pm 0.0063$   &   $-0.0087 \pm 0.0017$        \\[+1mm]  
 \hline \\[-2mm]
  $H_0$ [km s$^{-1}$ Mpc$^{-1}$] &  $65.1 \pm 3.3$    &  $66.0 \pm 2.5$         &   $67.81 \pm 0.66$            \\[+1mm]  
  $\Omega_m$              &  $0.316\pm 0.033$         &  $0.306 \pm 0.023$      &   $0.2893 \pm 0.0057$          \\[+1mm]  
  $\sigma_8$              &  $0.799 \pm 0.021$        &  $0.805 \pm 0.017$      &   $0.8148 \pm 0.0097$         \\[+1mm]  
  \hline\hline \\[-2mm]
  Parameter               &  TT+lowP+lensing+$H(z)$   &  TT+lowP+lensing+JLA+NewBAO  &  TT+lowP+lensing+JLA+NewBAO+$H(z)$      \\[+0mm]
 \hline \\[-2mm]
  $\Omega_b h^2$          &  $0.02304 \pm 0.00020$    &  $0.02302 \pm 0.00020$  &  $0.02303  \pm 0.00019$     \\[+1mm]
  $\Omega_c h^2$          &  $0.1094 \pm 0.0010$      &  $0.1094  \pm 0.0010$   &  $0.1094   \pm 0.0010$    \\[+1mm]
  $100\theta_\textrm{MC}$ &  $1.04231 \pm 0.00041$    &  $1.04228 \pm 0.00042$  &  $1.04229  \pm 0.00041$     \\[+1mm]
  $\tau$                  &  $0.119 \pm 0.015$        &  $0.115   \pm 0.011$    &  $0.115    \pm 0.011$       \\[+1mm]
  $\ln(10^{10} A_s)$      &  $3.145 \pm 0.029$        &  $3.138   \pm 0.022$    &  $3.139    \pm 0.022$       \\[+1mm]
  $\Omega_k$              &  $-0.0075 \pm 0.0042$     &  $-0.0086 \pm 0.0017$   &  $-0.0083  \pm 0.0016$      \\[+1mm]  
 \hline \\[-2mm]
  $H_0$ [km s$^{-1}$ Mpc$^{-1}$]  &  $68.4 \pm 1.9$   &  $67.82 \pm 0.66$       &  $67.93 \pm 0.63$           \\[+1mm]  
  $\Omega_m$              &  $0.285 \pm 0.016$        &  $0.2893 \pm 0.0058$    &  $0.2885 \pm 0.0055$         \\[+1mm]  
  $\sigma_8$              &  $0.818  \pm 0.014$       &  $0.8148 \pm 0.0098$    &  $0.8156 \pm 0.0098$        \\[+1mm]  
    \hline\hline \\[-2mm]
  Parameter               &  TT+lowP+lensing+$f\sigma_8$ &  TT+lowP+lensing+NewBAO+$f\sigma_8$  &  TT+lowP+lensing+JLA+NewBAO+$H(z)$+$f\sigma_8$  \\[+0mm]
 \hline \\[-2mm]
  $\Omega_b h^2$          &  $0.02305 \pm 0.00020$    &  $0.02303 \pm 0.00020$  &  $0.02305  \pm 0.00020$     \\[+1mm]
  $\Omega_c h^2$          &  $0.1090 \pm 0.0011$      &  $0.1091  \pm 0.0011$   &  $0.1092   \pm 0.0010$      \\[+1mm]
  $100\theta_\textrm{MC}$ &  $1.04229 \pm 0.00041$    &  $1.04229 \pm 0.00041$  &  $1.04226  \pm 0.00041$     \\[+1mm]
  $\tau$                  &  $0.117 \pm 0.019$        &  $0.112   \pm 0.011$    &  $0.113    \pm 0.011$       \\[+1mm]
  $\ln(10^{10} A_s)$      &  $3.141 \pm 0.037$        &  $3.132   \pm 0.022$    &  $3.134    \pm 0.022$       \\[+1mm]
  $\Omega_k$              &  $-0.0076 \pm 0.0064$     &  $-0.0086 \pm 0.0017$   &  $-0.0082  \pm 0.0016$      \\[+1mm]  
 \hline \\[-2mm]
  $H_0$ [km s$^{-1}$ Mpc$^{-1}$]  &  $68.7 \pm 3.0$   &  $67.93 \pm 0.67$       &  $68.07 \pm 0.63$           \\[+1mm]  
  $\Omega_m$              &  $0.283 \pm 0.024$        &  $0.2877 \pm 0.0058$    &  $0.2868 \pm 0.0055$         \\[+1mm]  
  $\sigma_8$              &  $0.815 \pm 0.019$        &  $0.8111 \pm 0.0098$    &  $0.8124 \pm 0.0095$         \\[+0mm]
\end{tabular}
\end{ruledtabular}
\label{tab:para_nonflat_lensing}
\end{table*}

If we focus on CMB TT + lowP + lensing data, Figs.\
\ref{fig:para_flat_lensing} and
\ref{fig:para_nonflat_lensing} and Tables \ref{tab:para_flat_lensing}
and \ref{tab:para_nonflat_lensing}, we see that adding only one 
of the four non-CMB data sets at a time to the CMB measurements
(left triangle plots in the two figures) results in four sets of
contours that are quite consistent with each other, as well as with
the original CMB alone contours, for both the tilted flat-$\Lambda$CDM
case and for the untilted non-flat $\Lambda$CDM model. The same holds true for
the tilted flat-$\Lambda$CDM model when the CMB lensing data are excluded
(left triangle plot of Fig.\ \ref{fig:para_flat}).
However, in the untilted non-flat $\Lambda$CDM case without the lensing data
when any of the four non-CMB data sets are added to the CMB data
(left triangle plot of Fig.\ \ref{fig:para_nonflat}),
they each pull the results towards a smaller $|\Omega_k|$ 
(closer to the flat model) and slightly larger $\tau$ and $\ln(10^{10} A_s)$
and smaller $\Omega_b h^2$ than is favored by the CMB data alone,
although all five sets of constraint contours are largely mutually consistent.
It is reassuring that the four non-CMB data sets do not pull the CMB
constraints in significantly different directions.   

As noted above, adding the NewBAO data to the CMB data typically makes 
the biggest
difference, but the other three non-CMB data sets also contribute. 
Focusing on the TT + lowP + lensing data, we see from Table 
\ref{tab:para_flat_lensing} for the tilted flat-$\Lambda$CDM case that
the NewBAO data tightly constrains model parameters, particularly
$\Omega_c h^2$, while the growth rate ($f\sigma_8$) data shifts
$\Omega_b h^2$ and $n_s$ to larger values and $\Omega_c h^2$ to a 
smaller value. In this case $\Omega_m$ is the quantity whose error bar is
reduced the most by the full combination of data relative to the CMB
and NewBAO compilation, followed by the $H_0$ error bar reduction. For the 
untilted non-flat $\Lambda$CDM model, from Table \ref{tab:para_nonflat_lensing},
$\Omega_c h^2$ and $\tau$ error bars from the 
CMB and NewBAO data are not reduced by including the $H(z)$, $f \sigma_8$, 
and JLA SNIa measurements in the mix.
In all cases, adding JLA SNIa or growth rate ($f\sigma_8$) data to
the combination of CMB + NewBAO data does not much improve the observational
constraints.\footnote{We did not check what happens when just $H(z)$ data is
added to the CMB and NewBAO combination but suspect a similar conclusion 
holds for this case also.} 

Again concentrating on the TT + lowP + lensing data, Tables   
\ref{tab:para_flat_lensing} and \ref{tab:para_nonflat_lensing}, we see that 
for the tilted flat-$\Lambda$CDM model, adding the four non-CMB data sets 
to the mix most affects $\Omega_c h^2$ and $\Omega_m$, with both central 
values moving down about 0.5$\sigma$ of the CMB data alone error bars. The 
situation in the untilted non-flat $\Lambda$CDM case is a little more dramatic, with 
$\Omega_k$ moving closer to flatness by about 1$\sigma$, $H_0$ and $\Omega_m$ 
also moving by about 1$\sigma$, and the $\sigma_8$, 
$\ln (10^{10} A_s)$, and $\tau$ central values moving by about 0.5$\sigma$.

Perhaps the biggest consequence of including the four non-CMB data sets in the 
analyses is the significant strengthening of the evidence for non-flatness 
in the untilted non-flat $\Lambda$CDM case, with it increasing from 1.8$\sigma$ away 
from flatness for the CMB alone case, to 5.1$\sigma$ away from flatness for
the full data combination in Table 
\ref{tab:para_nonflat_lensing},\footnote{It is possible to assume that all 
one-dimensional likelihoods are close to Gaussian, except for $\Omega_k$ 
estimated using the TT + lowP, TT + lowP + $H(z)$, and TT + lowP + 
$f\sigma_8$ data.} 
where the NewBAO data plays the most important role among the four non-CMB
data sets. This is consistent with, but stronger than, the 
\citet{Oobaetal2018a} results. The same situation is also seen when the 
lensing data are excluded, as shown in Table \ref{tab:para_nonflat}. We also 
note that combining CMB data with either JLA SNIa, $H(z)$, or growth rate 
data do not strongly support non-flatness. When combined with CMB data with
lensing, SNIa, $H(z)$, and $f \sigma_8$ data result in $\Omega_k$ being 
2.1$\sigma$, 1.8$\sigma$, and 1.2$\sigma$ away from flatness, while CMB and 
NewBAO data favor $\Omega_k$ being 5.1$\sigma$ away from flatness 
(Table \ref{tab:para_nonflat_lensing}). In the untilted non-flat $\Lambda\textrm{CDM}$ 
case, the effect of growth rate data on the model constraints differs 
from that of the NewBAO data.
The results for the untilted non-flat $\Lambda\textrm{CDM}$ model from
TT + lowP + $f\sigma_8$ observations excluding (including) the lensing data
shows that the growth rate measurements favor $\Omega_k$ moving closer
to spatial flatness with a deviation of only 1.4$\sigma$ (1.2$\sigma$) from 
zero spatial curvature. Adding $f\sigma_8$ data to TT + lowP (+ lensing) + 
NewBAO measurements --- that favor the closed model by $5.2\sigma$ 
($5.1\sigma$) --- gives a negative $\Omega_k$ deviating from flatness by
$5.1\sigma$ ($5.1\sigma$). Thus the negativeness of the curvature parameter
persists for the combination of BAO and growth rate data,
which also implies that the BAO data most tightly constrains the curvature 
parameter compared to the other non-CMB data. 

For the full data combination, $H_0$ measured in the two models (with
lensing data) in Tables   
\ref{tab:para_flat_lensing} and \ref{tab:para_nonflat_lensing}, 
$68.17 \pm 0.50$ and $68.07 \pm 0.63$ km s$^{-1}$ Mpc$^{-1}$,
are very consistent with each other, agreeing to within
0.12$\sigma$ (of the quadrature sum of the two error bars).\footnote{Potential
systematic errors, ignored here, have been discussed by \citet{Addisonetal2016}
and \citet{PlanckCollaboration2017}.}
These values are consistent with the most recent median statistics estimate
$H_0=68\pm2.8$ km s$^{-1}$ Mpc$^{-1}$ 
\citep{ChenRatra2011a}, which is consistent with earlier median statistics 
estimates \citep{Gottetal2001, Chenetal2003}. Many recent
estimates of $H_0$ are also quite consistent with these
measurements \citep{Calabreseetal2012,Hinshawetal2013,Sieversetal2013,Aubourgetal2015,PlanckCollaboration2016,LHuillierShafieloo2017,Chenetal2017,Lukovicetal2016,Wangetal2017,LinIshak2017,DESCollaboration2017b,Yuetal2018,Haridasuetal2018}, but, as is well known, they are 
lower than the local measurement of $H_0 = 73.06\pm1.74$ km s$^{-1}$ Mpc$^{-1}$
\citep{AndersonRiess2017}.\footnote{This local measurement is 2.7$\sigma$ (of 
the quadrature sum of the two error bars) higher than $H_0$ measured in both
models. We note that some other local expansion rate
measurements find a slightly lower $H_0$ with larger error bars
\citep{Rigaultetal2015, Zhangetal2017, Dhawanetal2018, FernandezArenasetal2018}.}

In our analyses here, $H_0$ and $\sigma_8$ (discussed below) are the only 
cosmological parameters that are determined in a cosmological model (spatial
curvature and tilt) independent 
manner. For instance, $\Omega_m$ determined using the tilted flat-$\Lambda$CDM
model differs from that measured in the untilted non-flat $\Lambda$CDM model by about
1.9$\sigma$ (of the quadrature sum of the error bars), however both estimates
are consistent with many other determinations 
\citep[see e.g.,][]{ChenRatra2003}. 
 
Like $\Omega_m$, measurements of $\theta_\textrm{MC}$, $\Omega_b h^2$, 
$\ln(10^{10} A_s)$, and $\tau$ are more model dependent, differing by 
2.1$\sigma$, 2.3$\sigma$, 2.7$\sigma$, and 2.9$\sigma$ between the two 
models. The measurements of $\Omega_c h^2$ differ by 5.7$\sigma$, so the 
cosmological model 
dependence of this measurement is much more important than the statistical 
errors determined by using a given cosmological model. It is important to 
account for such model dependence when comparing a cosmologically estimated 
value to that estimated using a different technique. This model dependence 
can have very striking consequences. For instance, as discussed in 
\citet{Mitraetal2018,Mitraetal2019}, the much larger value of $\tau$ in the 
untilted non-flat case 
significantly alters the cosmological reionization scenario, although we note 
that using the more extensive non-CMB data compilation here we find a 
0.6$\sigma$ reduction in $\tau$ compared to the larger value found in 
\citet{Oobaetal2018a} thus somewhat alleviating
the potential tension discovered in \citet{Mitraetal2018,Mitraetal2019} for the 
higher $\tau$ value.  From Tables \ref{tab:para_flat_lensing} and 
\ref{tab:para_nonflat_lensing}, for the full data compilation including 
CMB lensing observations, we find in the tilted flat-$\Lambda$CDM (non-flat
$\Lambda$CDM) model $0.02193 \le \Omega_b h^2 \le 0.02269$ 
($0.02265 \le \Omega_b h^2 \le 0.02345$) at 2$\sigma$, which are almost 
disjoint.\footnote{A compilation of measured primordial deuterium abundances mildly favors the flat case \citep{Pentonetal2018}.} Clearly it is not possible to robustly measure $\Omega_b h^2$
(and some other cosmological parameters) in a model independent way from 
cosmological data and care must be taken when comparing a value measured
in a cosmological model to a value determined using some other technique
\citep[see, e.g.][]{Cookeetal2018}.

\begin{figure*}
\centering
\mbox{\includegraphics[width=65mm]{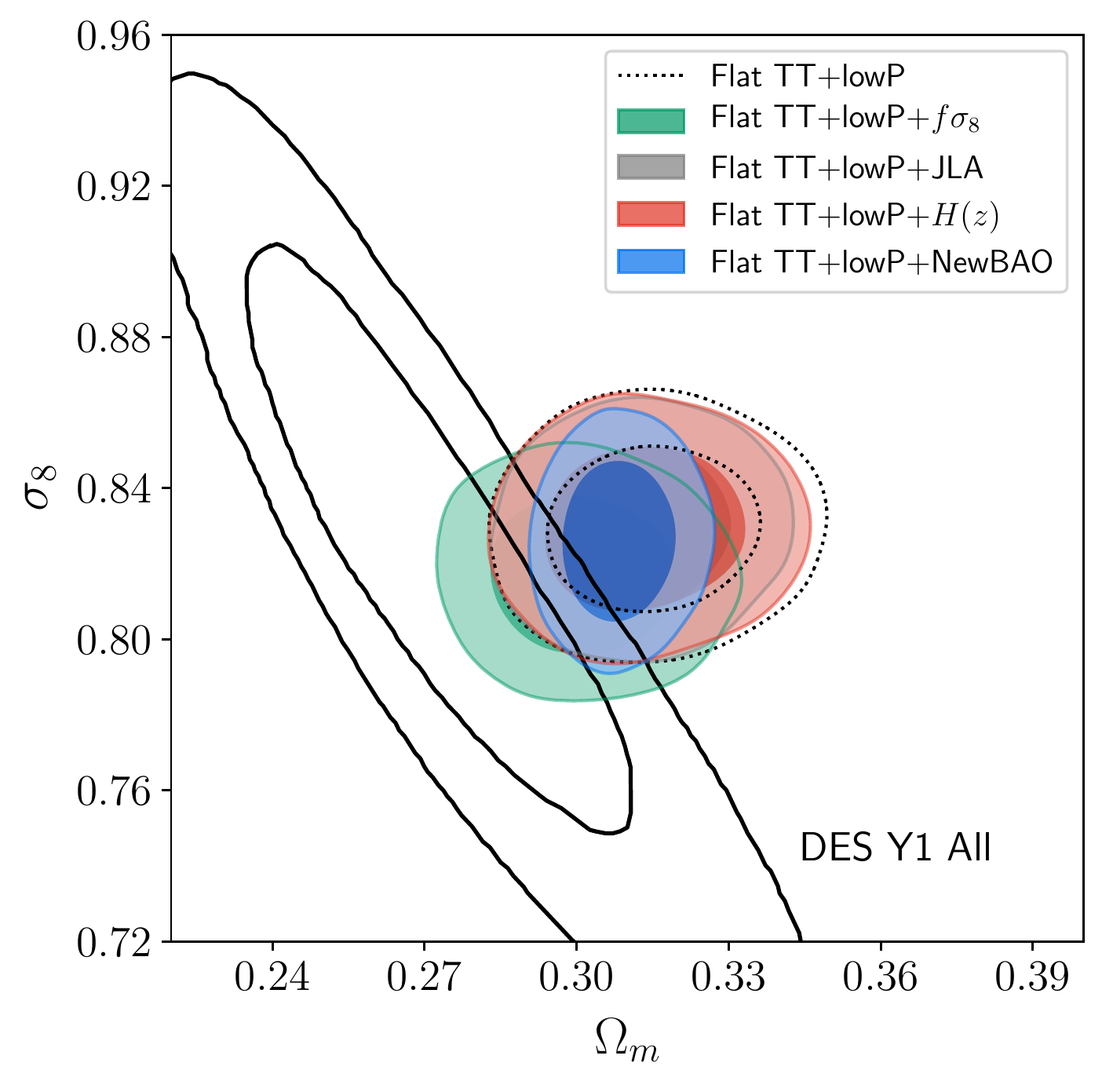}}
\mbox{\includegraphics[width=65mm]{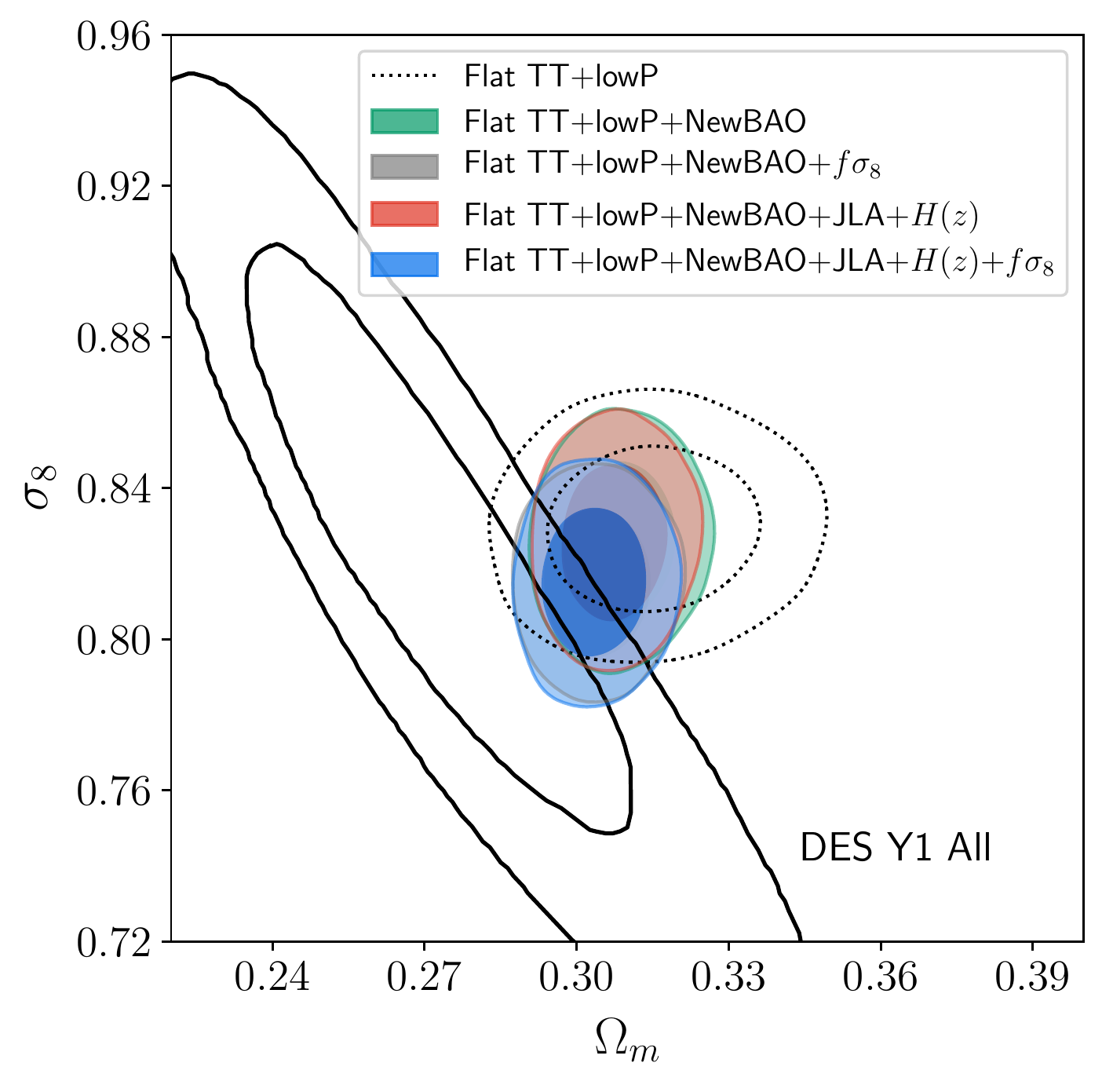}} \\
\mbox{\includegraphics[width=65mm]{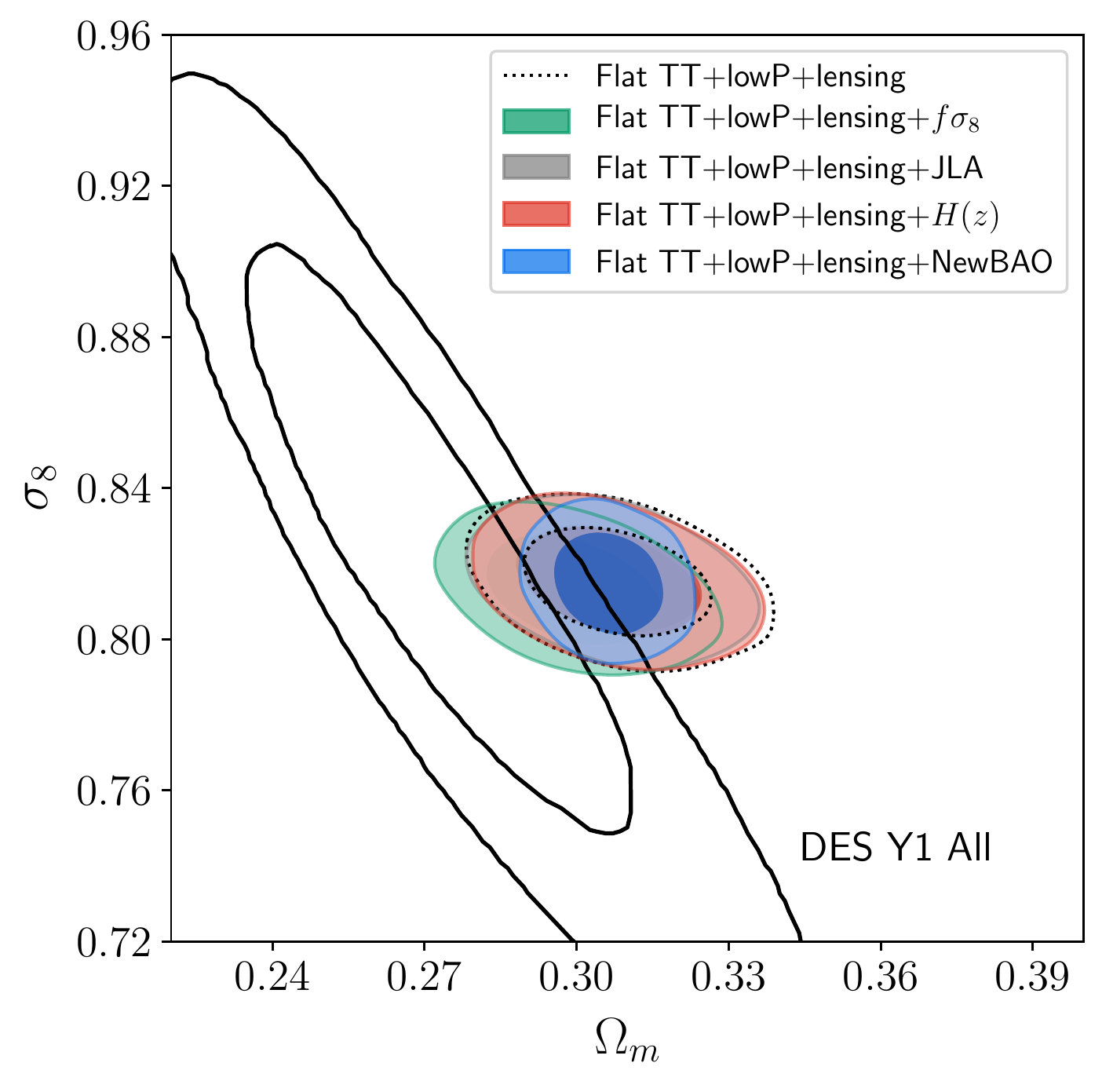}}
\mbox{\includegraphics[width=65mm]{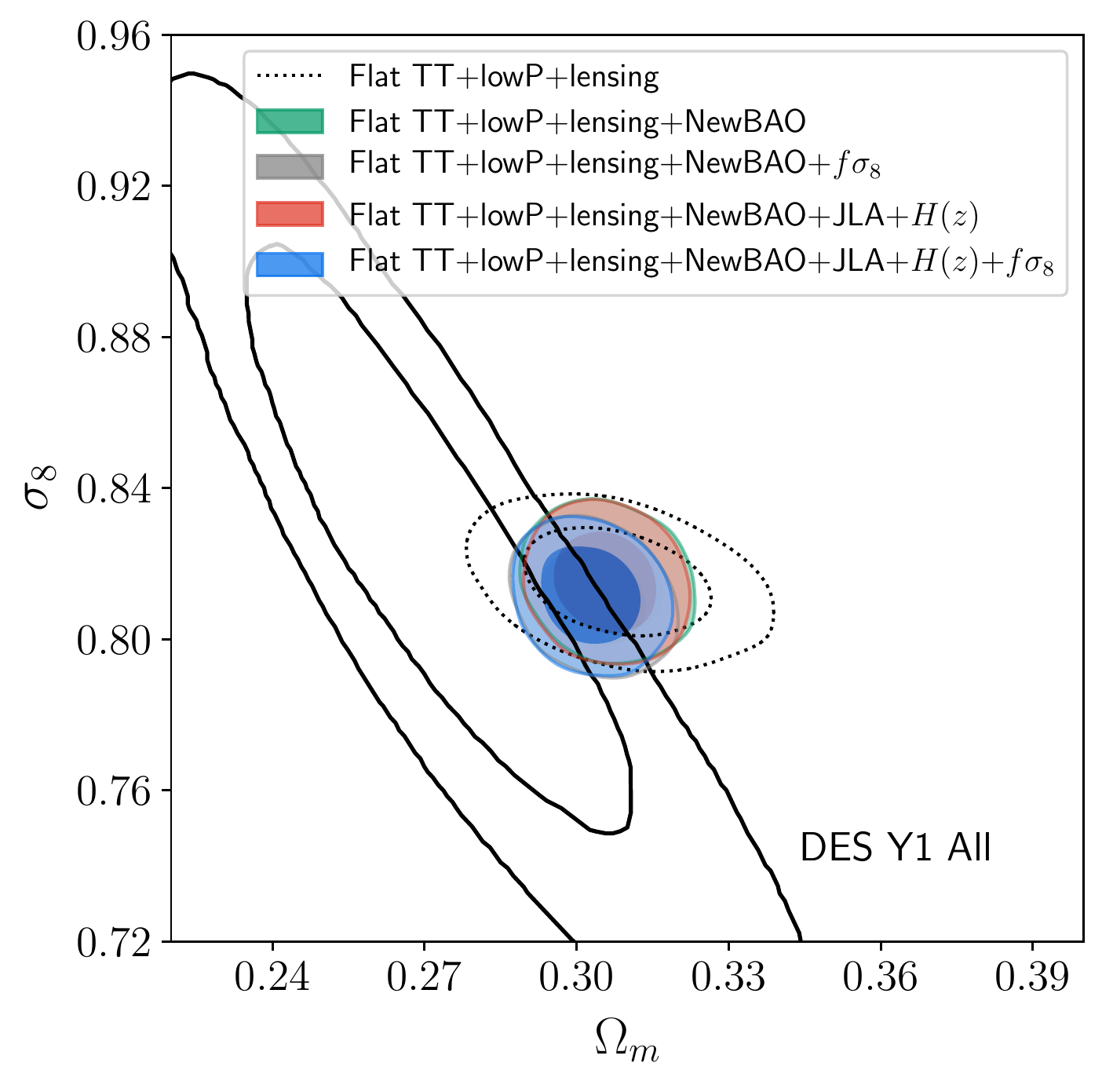}}
\caption{Likelihood distributions in the $\Omega_m$--$\sigma_8$ plane for
the tilted flat-$\Lambda\textrm{CDM}$ model constrained by Planck CMB
TT + lowP (+lensing), JLA SNIa, NewBAO, $H(z)$, and $f\sigma_8$ data.
In each panel the $\Lambda\textrm{CDM}$ constraints (68.3\% and 95.4\% confidence limits)
obtained from the first-year Dark Energy Survey (DES Y1 All) \citep{DESCollaboration2017a}
are shown as thick solid curves for comparison.
}
\label{fig:omm_sig8_flat}
\end{figure*}

\begin{figure*}
\centering
\mbox{\includegraphics[width=65mm]{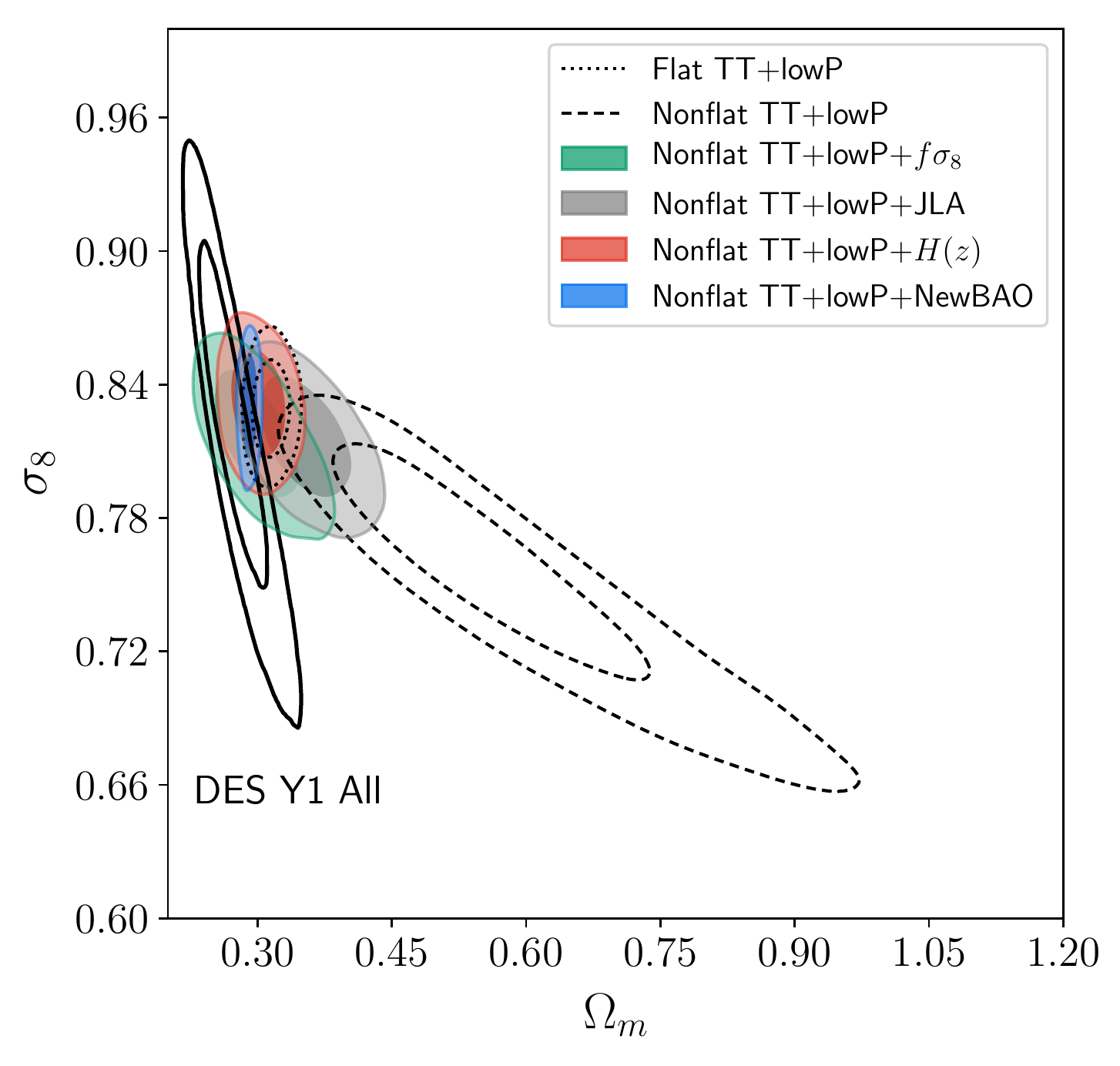}}
\mbox{\includegraphics[width=65mm]{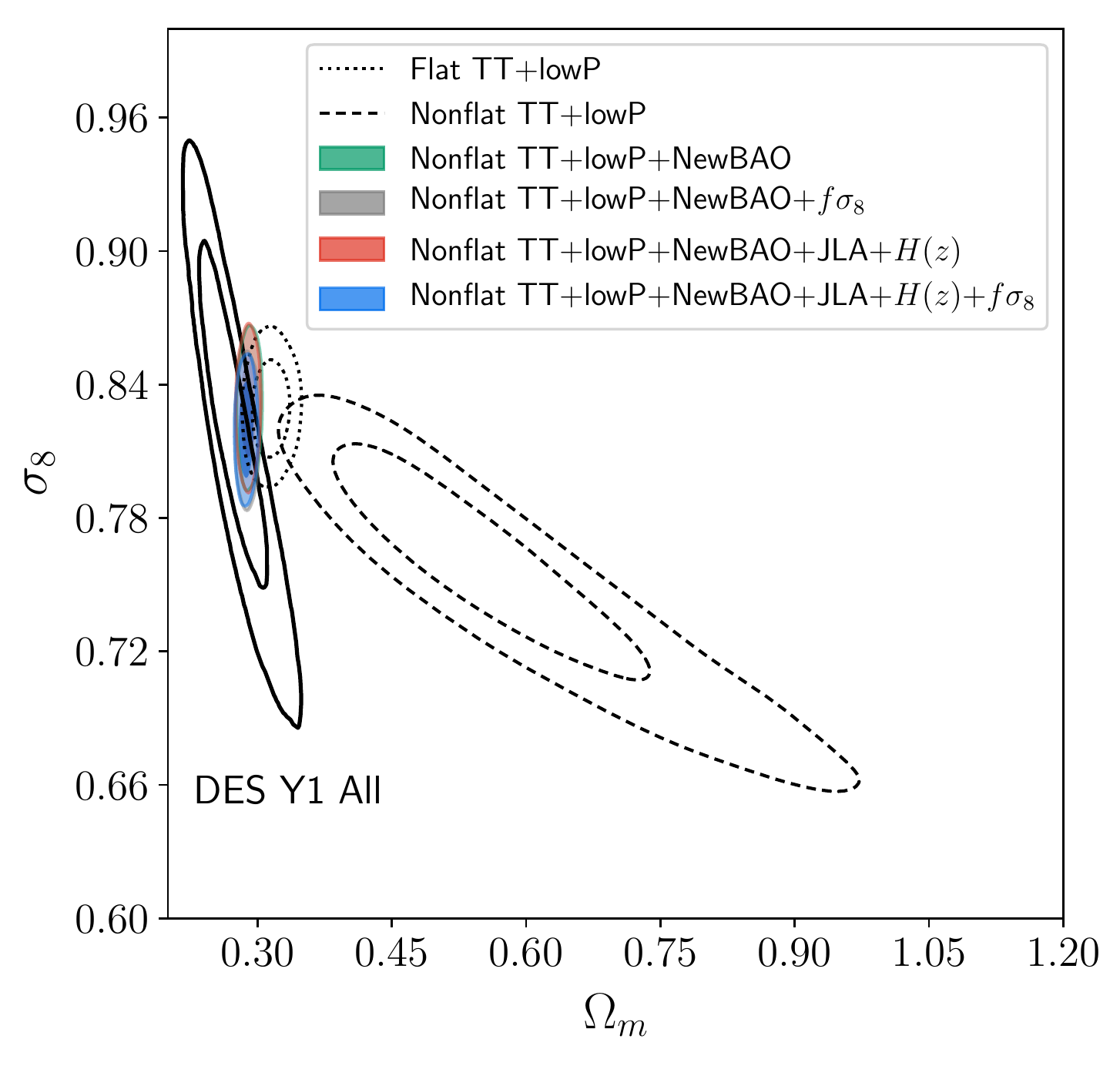}} \\
\mbox{\includegraphics[width=65mm]{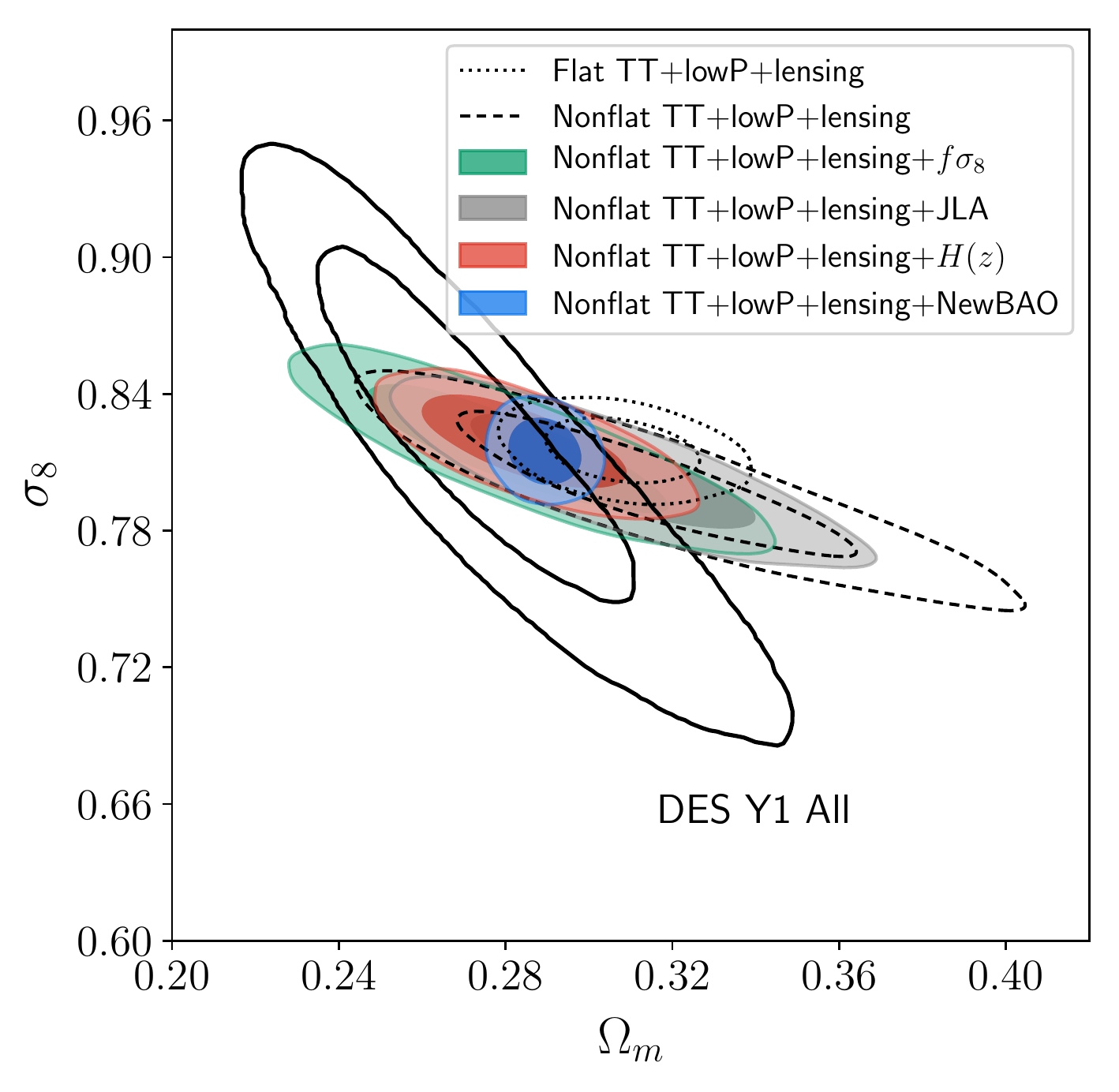}}
\mbox{\includegraphics[width=65mm]{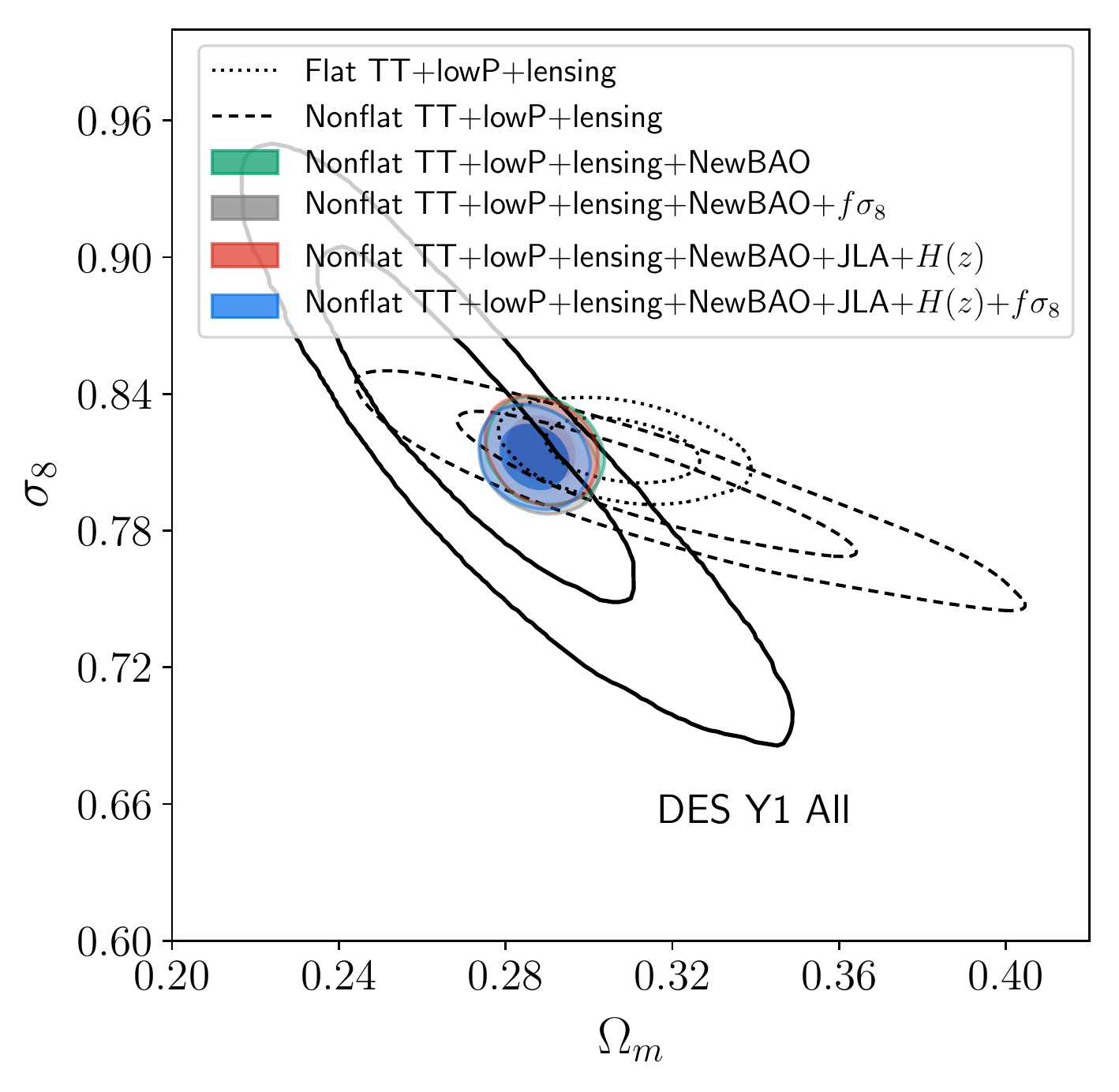}}
\caption{Same as Fig.\ \ref{fig:omm_sig8_flat} but for the untilted non-flat
$\Lambda\textrm{CDM}$ model.
}
\label{fig:omm_sig8_nonflat}
\end{figure*}

For the full data combination, $\sigma_8$'s measured in the two models (with 
CMB lensing data), Tables \ref{tab:para_flat_lensing} and 
\ref{tab:para_nonflat_lensing}, agree to 0.062$\sigma$ (of the quadrature 
sum of the two error bars). Figures \ref{fig:omm_sig8_flat} and 
\ref{fig:omm_sig8_nonflat} show the marginalized two-dimensional likelihood 
distribution of $\Omega_m$ and $\sigma_8$ for the tilted flat and untilted non-flat 
$\Lambda\textrm{CDM}$
models constrained by the Planck CMB and the non-CMB data sets.
For comparison, in each panel we present the $\Lambda\textrm{CDM}$ constraints
obtained from a combined analysis of galaxy clustering and weak gravitational
lensing based on the first year result of the Dark Energy Survey (DES Y1 All)
\citep{DESCollaboration2017a}, whose 68.3\% confidence limits are 
$\Omega_m=0.264_{-0.019}^{+0.032}$ and $\sigma_8=0.807_{-0.041}^{+0.062}$.
The likelihood distribution in the $\Omega_m$--$\sigma_8$ plane obtained by
adding each non-CMB data set to the Planck CMB data are consistent with each 
other. As expected, the NewBAO data or the NewBAO data combined with other 
non-CMB data sets give tighter constraints in all cases.
As shown in Figs.\ \ref{fig:omm_sig8_flat} and \ref{fig:omm_sig8_nonflat}, 
there is tension between both $\Lambda\textrm{CDM}$ models constrained by 
Planck TT + lowP data (dotted and dashed curves in the top panels) and the 
DES constraints. This tension disappears when the CMB lensing data are included 
(bottom panels).

Although our $\sigma_8$ constraints from the flat and non-flat models (excluding
and including CMB lensing data) are similar to the DES Y1 All result, 
our $\Omega_m$ constraints favor a larger value by over 1$\sigma$ for 
the flat-$\Lambda\textrm{CDM}$ model. Including the CMB lensing data reduces 
the tension to 1.2$\sigma$. We note that the best-fit point for the non-flat 
$\Lambda\textrm{CDM}$ model constrained by the Planck CMB data (including 
lensing) combined with all non-CMB data enters well into the 1$\sigma$ 
region of the DES Y1 All constraint contour (Fig.\ \ref{fig:omm_sig8_nonflat} 
lower right panel), unlike the case for the tilted flat-$\Lambda$CDM model 
(Fig.\ \ref{fig:omm_sig8_flat} lower right panel).

\begin{table*}
\caption{Individual and total $\chi^2$ values for the best-fit tilted flat and untilted non-flat $\Lambda\textrm{CDM}$ inflation models.}
\begin{ruledtabular}
\begin{tabular}{lccccccccccc}
    Data sets   & $\chi_{\textrm{Plik}}^2$  & $\chi_{\textrm{lowTEB}}^2$  &  $\chi_{\textrm{lensing}}^2$  &  $\chi_{\textrm{JLA}}^2$  & $\chi_{\textrm{NewBAO}}^2$  &  $\chi_{H(z)}^2$   &  $\chi_{f\sigma_8}^2$ & $\chi_{\textrm{prior}}^2$  &  Total $\chi^2$      & $\Delta\chi^2$    &  $\ln B$  \\[+0mm]
 \hline \\[-2mm]
 \multicolumn{12}{c}{Tilted flat-$\Lambda\textrm{CDM}$ model} \\
  \hline \\[-2mm]
   TT+lowP                                    & 763.57 & 10496.41 &       &        &       &       &       &  1.96 & 11261.93 &      & \\[+1mm]
   \quad\quad +JLA                            & 763.60 & 10496.48 &       & 695.32 &       &       &       &  1.92 & 11957.32 &      & \\[+1mm]
   \quad\quad +NewBAO                         & 762.50 & 10497.73 &       &        & 13.46 &       &       &  2.36 & 11276.05 &      & \\[+1mm]
   \quad\quad +$H(z)$                         & 763.98 & 10496.36 &       &        &       & 14.89 &       &  1.70 & 11276.93 &      & \\[+1mm]
   \quad\quad +$f\sigma_8$                    & 766.83 & 10494.95 &       &        &       &       & 12.15 &  1.87 & 11275.80 &      & \\[+1mm]
   \quad\quad +NewBAO+$f\sigma_8$             & 766.47 & 10494.93 &       &        & 12.63 &       & 12.45 &  2.02 & 11288.50 &      & \\[+1mm]
   \quad\quad +JLA+NewBAO                     & 764.11 & 10496.10 &       & 695.17 & 12.91 &       &       &  2.21 & 11970.51 &      & \\[+1mm]
   \quad\quad +JLA+NewBAO+$H(z)$              & 764.30 & 10496.06 &       & 695.19 & 12.94 & 14.81 &       &  1.95 & 11985.25 &      & \\[+1mm]
   \quad\quad +JLA+NewBAO+$H(z)$+$f\sigma_8$  & 766.81 & 10494.80 &       & 695.12 & 12.73 & 14.79 & 12.15 &  2.05 & 11998.43 &      & \\[+1mm]
 \hline\\[-2mm]
   TT+lowP+lensing                            & 766.20 & 10494.93 &  9.30 &        &       &       &       &  2.00 & 11272.44 &      & \\[+1mm]
   \quad\quad +JLA                            & 767.15 & 10494.77 &  8.98 & 695.07 &       &       &       &  2.18 & 11968.15 &      & \\[+1mm]
   \quad\quad +NewBAO                         & 766.37 & 10494.86 &  9.12 &        & 12.59 &       &       &  2.11 & 11285.06 &      & \\[+1mm]
   \quad\quad +$H(z)$                         & 766.20 & 10494.92 &  9.27 &        &       & 14.83 &       &  2.04 & 11287.27 &      & \\[+1mm]
   \quad\quad +$f\sigma_8$                    & 768.26 & 10494.43 &  8.67 &        &       &       & 11.31 &  1.94 & 11284.62 &      & \\[+1mm]
   \quad\quad +NewBAO+$f\sigma_8$             & 767.47 & 10494.57 &  8.73 &        & 12.66 &       & 11.80 &  2.11 & 11297.33 &      & \\[+1mm]
   \quad\quad +JLA+NewBAO                     & 766.42 & 10494.85 &  9.16 & 695.19 & 12.61 &       &       &  2.01 & 11980.24 &      & \\[+1mm]
   \quad\quad +JLA+NewBAO+$H(z)$              & 766.57 & 10494.76 &  9.04 & 695.16 & 12.59 & 14.81 &       &  2.15 & 11995.08 &      & \\[+1mm]
   \quad\quad +JLA+NewBAO+$H(z)$+$f\sigma_8$  & 767.50 & 10494.56 &  8.74 & 695.12 & 12.65 & 14.79 & 11.79 &  2.07 & 12007.21 &      & \\[+1mm]
   \hline \\[-2mm]
    \multicolumn{12}{c}{Untilted non-flat $\Lambda\textrm{CDM}$ model}  \\
   \hline \\[-2mm]
   TT+lowP                                    & 774.34 & 10495.42 &       &        &       &       &       &  2.33 & 11272.10 & 10.17 & $-5.63$ \\[+1mm]
   \quad\quad +JLA                            & 771.82 & 10503.02 &       & 696.54 &       &       &       &  2.40 & 11973.80 & 16.48 & $-7.09$ \\[+1mm]
   \quad\quad +NewBAO                         & 779.37 & 10499.98 &       &        & 14.25 &       &       &  1.95 & 11295.55 & 19.50 & $-11.2$ \\[+1mm]
   \quad\quad +$H(z)$                         & 777.14 & 10500.93 &       &        &       & 17.11 &       &  1.96 & 11297.15 & 20.22 & $-11.1$ \\[+1mm]
   \quad\quad +$f\sigma_8$                    & 783.38 & 10497.49 &       &        &       &       & 11.51 &  2.41 & 11294.79 & 18.99 & $-9.72$ \\[+1mm]
   \quad\quad +NewBAO+$f\sigma_8$             & 781.01 & 10500.07 &       &        & 14.32 &       & 11.25 &  1.97 & 11308.62 & 20.12 & $-10.1$ \\[+1mm]
   \quad\quad +JLA+NewBAO                     & 784.80 & 10496.68 &       & 695.18 & 13.95 &       &       &  2.29 & 11992.90 & 22.39 & $-9.42$ \\[+1mm]
   \quad\quad +JLA+NewBAO+$H(z)$              & 779.51 & 10500.26 &       & 695.17 & 14.18 & 16.08 &       &  1.98 & 12007.20 & 21.95 & $-11.9$ \\[+1mm]
   \quad\quad +JLA+NewBAO+$H(z)$+$f\sigma_8$  & 784.11 & 10497.49 &       & 695.20 & 13.95 & 16.36 & 10.61 &  1.82 & 12019.53 & 21.10 & $-10.8$ \\[+1mm]
 \hline\\[-2mm]
   TT+lowP+lensing                            & 786.87 & 10493.86 &  9.77 &        &       &       &       &  1.79 & 11292.29 & 19.85 & $-10.5$ \\[+1mm]
   \quad\quad +JLA                            & 786.10 & 10494.47 &  9.87 & 695.22 &       &       &       &  2.07 & 11987.73 & 19.58 & $-8.99$ \\[+1mm]
   \quad\quad +NewBAO                         & 784.85 & 10496.56 &  9.42 &        & 13.81 &       &       &  1.87 & 11306.51 & 21.45 & $-9.89$ \\[+1mm]
   \quad\quad +$H(z)$                         & 786.87 & 10496.02 &  8.66 &        &       & 16.36 &       &  2.19 & 11310.10 & 22.83 & $-11.4$ \\[+1mm]
   \quad\quad +$f\sigma_8$                    & 786.41 & 10496.00 &  8.75 &        &       &       &  9.79 &  1.99 & 11302.93 & 18.31 & $-9.60$ \\[+1mm]
   \quad\quad +NewBAO+$f\sigma_8$             & 786.11 & 10495.99 &  8.86 &        & 13.80 &       &  9.96 &  1.84 & 11316.56 & 19.23 & $-10.3$ \\[+1mm]
   \quad\quad +JLA+NewBAO                     & 786.53 & 10495.26 &  8.83 & 695.17 & 13.67 &       &       &  1.93 & 12001.40 & 21.16 & $-11.3$ \\[+1mm]
   \quad\quad +JLA+NewBAO+$H(z)$              & 787.20 & 10495.41 &  8.56 & 695.19 & 13.78 & 16.36 &       &  1.85 & 12018.35 & 23.27 & $-12.0$ \\[+1mm]
   \quad\quad +JLA+NewBAO+$H(z)$+$f\sigma_8$  & 786.97 & 10495.48 &  8.71 & 695.20 & 13.74 & 16.29 &  9.87 &  1.92 & 12028.17 & 20.96 & $-11.0$ \\[+1mm]
\end{tabular}
\\[+1mm]
Note: $\Delta\chi^2$ and $\ln B$ of a non-flat $\Lambda\textrm{CDM}$ model estimated for a combination of data sets represent the excess $\chi^2$ value and ratio of Bayesian evidence, respectively, relative to the tilted flat model for the same combination of data sets.
\end{ruledtabular}
\label{tab:chi2}
\end{table*}

Table \ref{tab:chi2} lists the individual and total $\chi^2$
values for the best-fit tilted flat and untilted non-flat $\Lambda\textrm{CDM}$ models.
The best-fit position in the parameter space is found with the COSMOMC built-in
routine that obtains the minimum $\chi^2$ by using Powell's minimization method.
This method searches for the local minimum by differentiating the likelihood distribution
and is efficient at finding the accurate location of the minimum $\chi^2$.\footnote{Our
minimum $\chi^2$ values are very similar to those supplied by the Planck team.
For the tilted flat-$\Lambda\textrm{CDM}$ model constrained with TT + lowP data,
the Planck team provides $\chi^2$ estimated from Powell's minimization method:
$\chi_{\textrm{Plik}}^2 =763.37$, $\chi_{\textrm{lowTEB}}^2 =10496.47$,
$\chi_{\textrm{prior}}^2 =2.08$, with total $\chi^2=11261.9$.}
We present the individual contribution of each data set used to constrain
the model parameters. The total $\chi^2$ is the sum of those from the 
high-$\ell$ CMB TT likelihood ($\chi_{\textrm{PlikTT}}^2$), the low-$\ell$ CMB 
power spectra ($\chi_{\textrm{lowTEB}}^2$), lensing ($\chi_{\textrm{lensing}}^2$),
JLA SNIa ($\chi_{\textrm{JLA}}^2$), NewBAO ($\chi_{\textrm{NewBAO}}^2$),
$H(z)$ ($\chi_{H(z)}^2$), $f\sigma_8$ data ($\chi_{f\sigma_8}^2$),
and the contribution from the foreground nuisance parameters
($\chi_{\textrm{prior}}^2$).
The nonstandard normalization of the Planck 2015 CMB data likelihoods 
means that only the 
difference of $\chi^2$ of one model relative to the other is meaningful 
for the Planck CMB data. In Table \ref{tab:chi2}, for the untilted non-flat 
$\Lambda\textrm{CDM}$ model, we list $\Delta\chi^2$, the excess $\chi^2$ 
over the value of the tilted flat-$\Lambda\textrm{CDM}$ model constrained 
with the same combination of data sets. For the non-CMB data sets, the numbers 
of degrees of freedom are 735, 15, 31, 10 for JLA SNIa, NewBAO, $H(z)$, 
$f\sigma_8$ data sets, respectively, for a total of 791 degrees of
freedom. The reduced $\chi^2$'s for the individual non-CMB data sets are 
$\chi^2 / \nu \lesssim 1$. There are 189 points in the TT + lowP Planck 
2015 data (binned angular power spectrum) and 197 when the CMB lensing
observations are included. 

Let us first focus on how the model fits the individual data sets. Compared to 
the tilted flat-$\Lambda\textrm{CDM}$ model, the untilted non-flat $\Lambda\textrm{CDM}$ 
model constrained with the Planck CMB data alone (excluding and including 
CMB lensing data) does worse at fitting the Planck high-$\ell$ $C_\ell$'s 
while it fits the low-$\ell$ ones a bit better. Inclusion of the non-CMB
data with the CMB data also results in the best-fit untilted non-flat model providing
a poorer fit to the high-$\ell$ TT measurements, both with and without the 
lensing data, compared to the tilted flat-$\Lambda$CDM case.
Adding JLA SNIa or NewBAO data to the Planck TT + lowP + lensing data 
improves the untilted non-flat model fit to the Planck low-$\ell$ TEB data. There 
is a tendency for the  non-flat models to more poorly fit the NewBAO and $H(z)$ 
data (with larger values of $\chi_{\textrm{NewBAO}}^2$ and $\chi_{H(z)}^2$)
than the flat models do, while the opposite is true for the case of the growth
rate ($f\sigma_8$) measurements.  
 
Comparing results for the TT + lowP + lensing analyses, $\Delta\chi^2 = 21$ 
for the full data compilation, for the non-flat $\Lambda$CDM case relative 
to the flat-$\Lambda$CDM model (last column in the last row of Table 
\ref{tab:chi2}). Unfortunately it is unclear how to turn this into a 
quantitative relative probability as the two six parameter models are not 
nested (and the number of degrees of freedom of the Planck CMB anisotropy data
is not available). Rather the best-fit versions of each six parameter model 
provide distinct local likelihood maxima in a larger seven parameter model 
space.\footnote{The energy density inhomogeneity power spectrum for this 
seven parameter tilted non-flat $\Lambda$CDM model is not known.} 
However, it is clear that the untilted non-flat $\Lambda$CDM model does not do as good a job in 
fitting the higher-$\ell$ $C_\ell$'s as it does in fitting the lower-$\ell$ 
ones. In this context it might be relevant to note that there has been 
some discussion about systematic differences between constraints derived 
using the higher-$\ell$ and the 
lower-$\ell$ Planck 2015 CMB data \citep{Addisonetal2016,PlanckCollaboration2017}. Additionally, in the 
flat-$\Lambda$CDM model, there appear to be inconsistencies between the 
higher-$\ell$ Planck 2015 CMB anisotropy data and the South Pole Telescope 
CMB anisotropy data \citep{Ayloretal2017}.

To compare untilted non-flat $\Lambda$CDM model with
the tilted flat one, we may also use the Bayes factor
$B=E[\textrm{nonflat}]/E[\textrm{flat}]$ that is defined as a ratio of Bayesian evidence
of non-flat model relative to the flat one for the same combination of data sets.
The Bayesian evidence of a model $M$ is given by
\begin{equation}
   E = p(\mbox{\boldmath $x$} | M) = \int d\mbox{\boldmath $\theta$} p(\mbox{\boldmath $x$}| \mbox{\boldmath $\theta$}, M) \pi(\mbox{\boldmath $\theta$}|M),
\end{equation}
where $\mbox{\boldmath $x$}$ indicates a data set, $\mbox{\boldmath $\theta$}$ is a vector containing
parameters of the model $M$, and $\pi(\mbox{\boldmath $\theta$}|M)$ is the prior on the parameters.
We calculate the Bayesian evidence using the algorithm developed by \citet{Heavensetal2017}
in which the posterior for the Bayesian evidence is obtained with the nearest-neighbor distances 
in parameter space. In Table \ref{tab:chi2} we list the logarithm of Bayes factor $\ln B$ for each untilted non-flat $\Lambda$CDM model
relative to the tilted flat one. In all cases, we find that $\ln B < -5$, which indicates
very strong evidence that the untilted non-flat $\Lambda$CDM model is less favored than the tilted flat one
(\citealt{Trotta2008}). However, we again take note of possible systematic differences in the CMB data mentioned at the end of the previous paragraph \citep{Addisonetal2016,PlanckCollaboration2017,Ayloretal2017} which, if real, could alter the Bayesian evidence in either direction. In addition, the Bayesian evidence we have computed here does not account for the fact that the best-fit untilted non-flat model has a lower $\Omega_m$ than does the best-fit tilted flat model (when both are fit to the cosmological data compilation we have used in our analyses here), and so is in better agreement with the lower $\Omega_m$ determined from weak-lensing measurements.

\begin{figure*}[htbp]
\centering
\mbox{\includegraphics[width=100mm,bb=15 230 570 750]{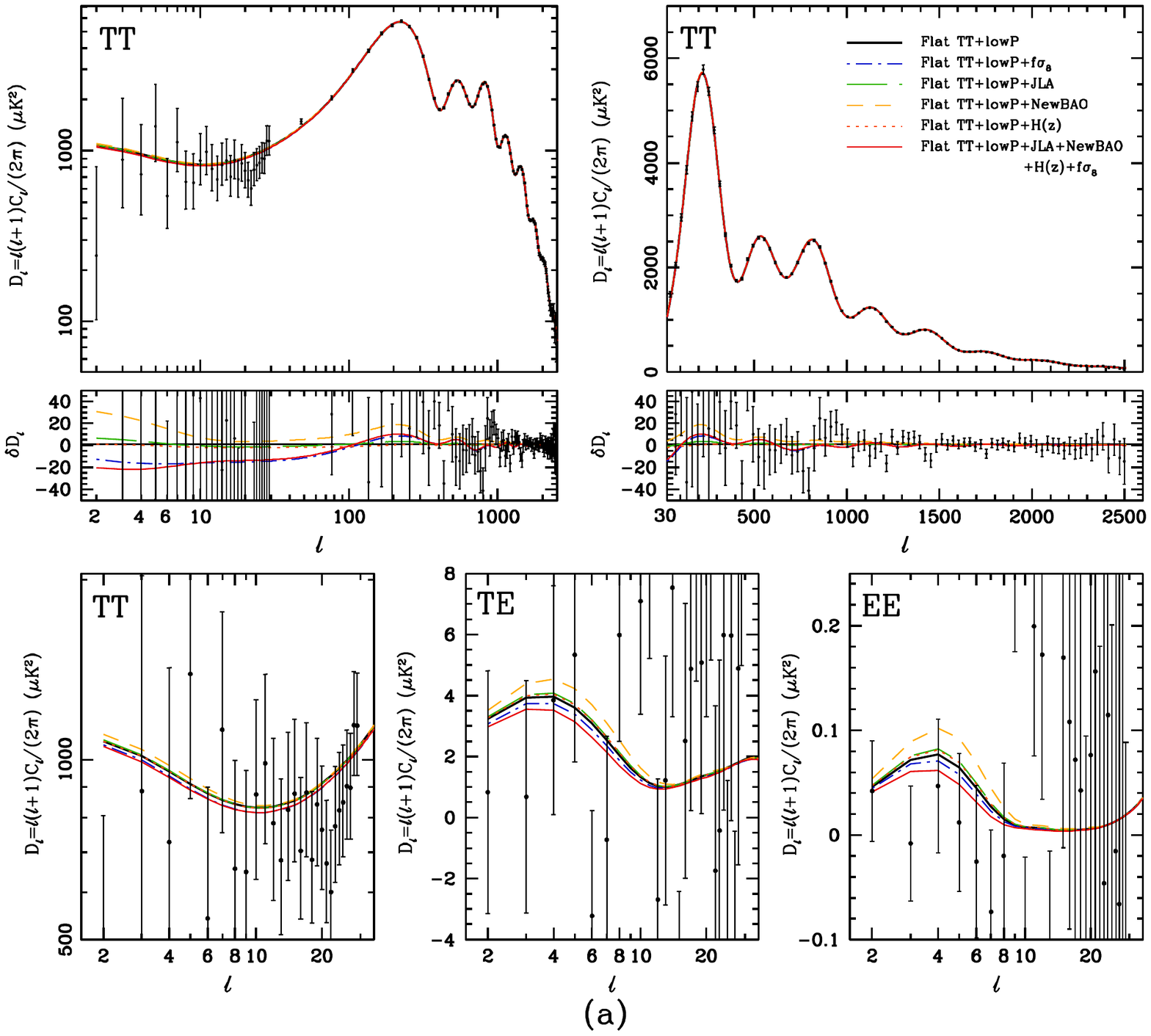}} \\
\mbox{\includegraphics[width=100mm,bb=15 210 570 750]{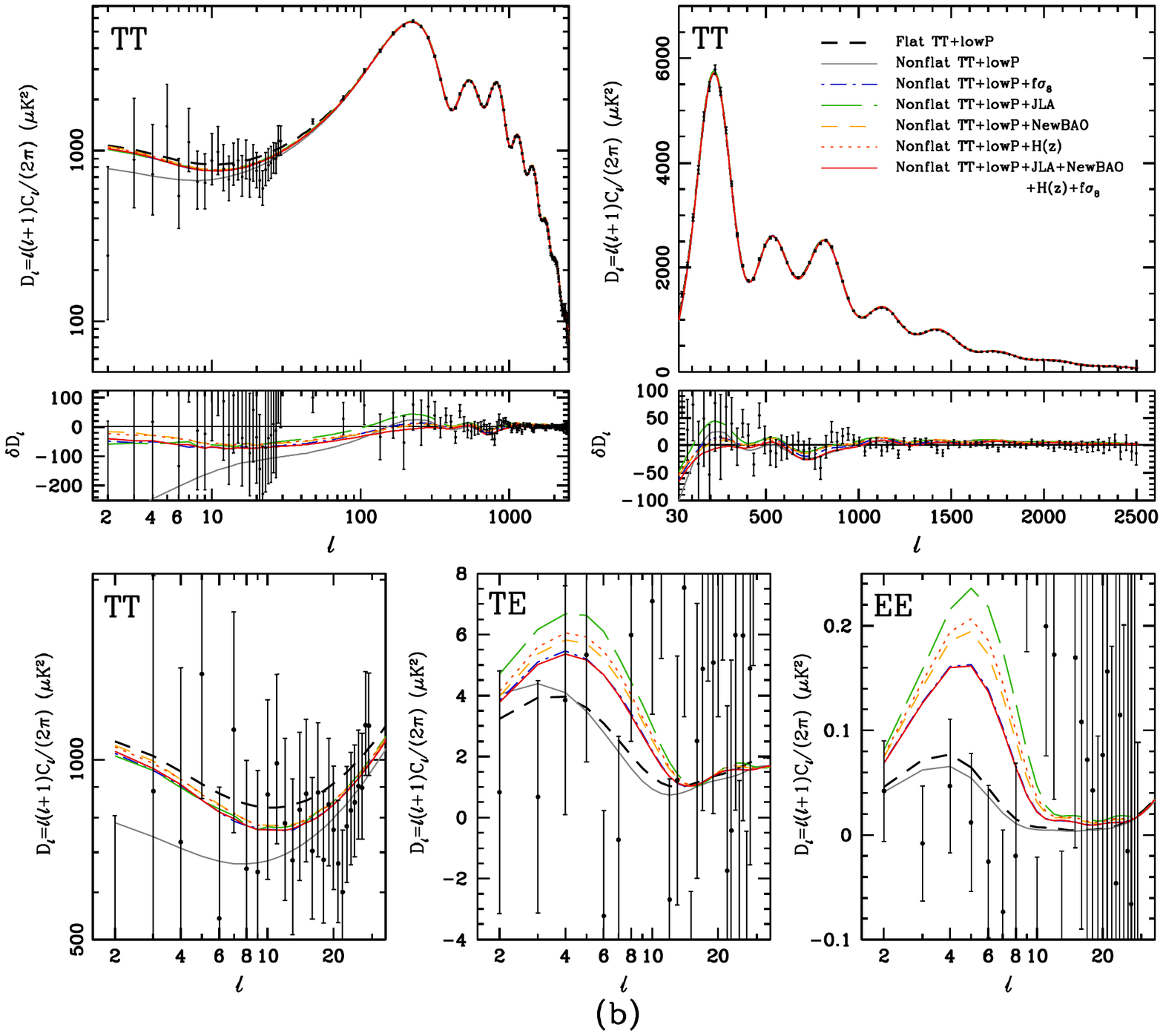}} 
\caption{Best-fit power spectra of (a) tilted flat (top five panels) and (b)
untilted non-flat $\Lambda\textrm{CDM}$ models (bottom five panels) constrained
by the Planck CMB TT + lowP data (excluding the lensing data) together with JLA
SNIa, NewBAO, $H(z)$, and $f\sigma_8$ data.
}
\label{fig:ps_cmb}
\end{figure*}

\begin{figure*}[htbp]
\centering
\mbox{\includegraphics[width=100mm,bb=15 230 570 750]{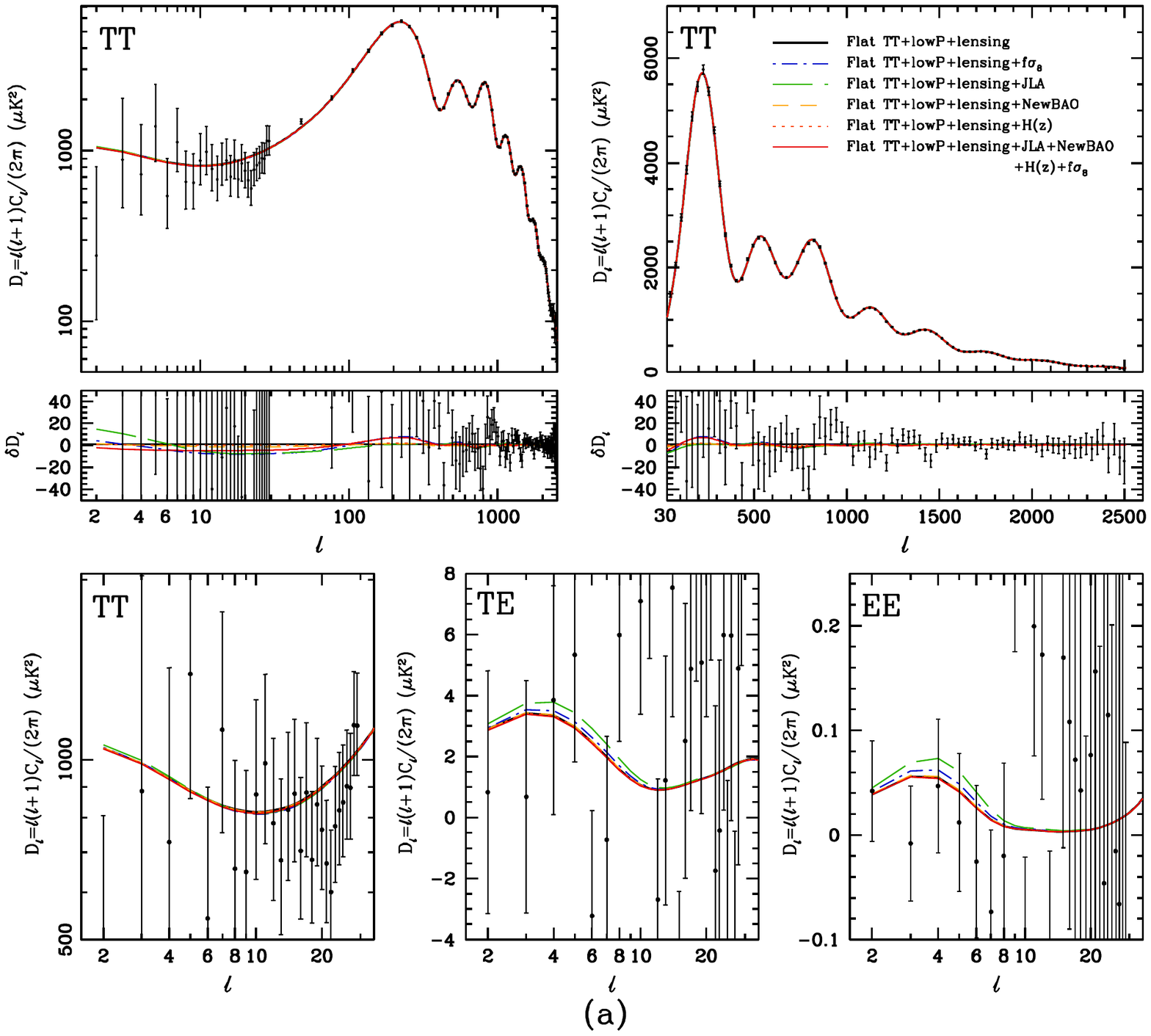}} \\
\mbox{\includegraphics[width=100mm,bb=15 210 570 750]{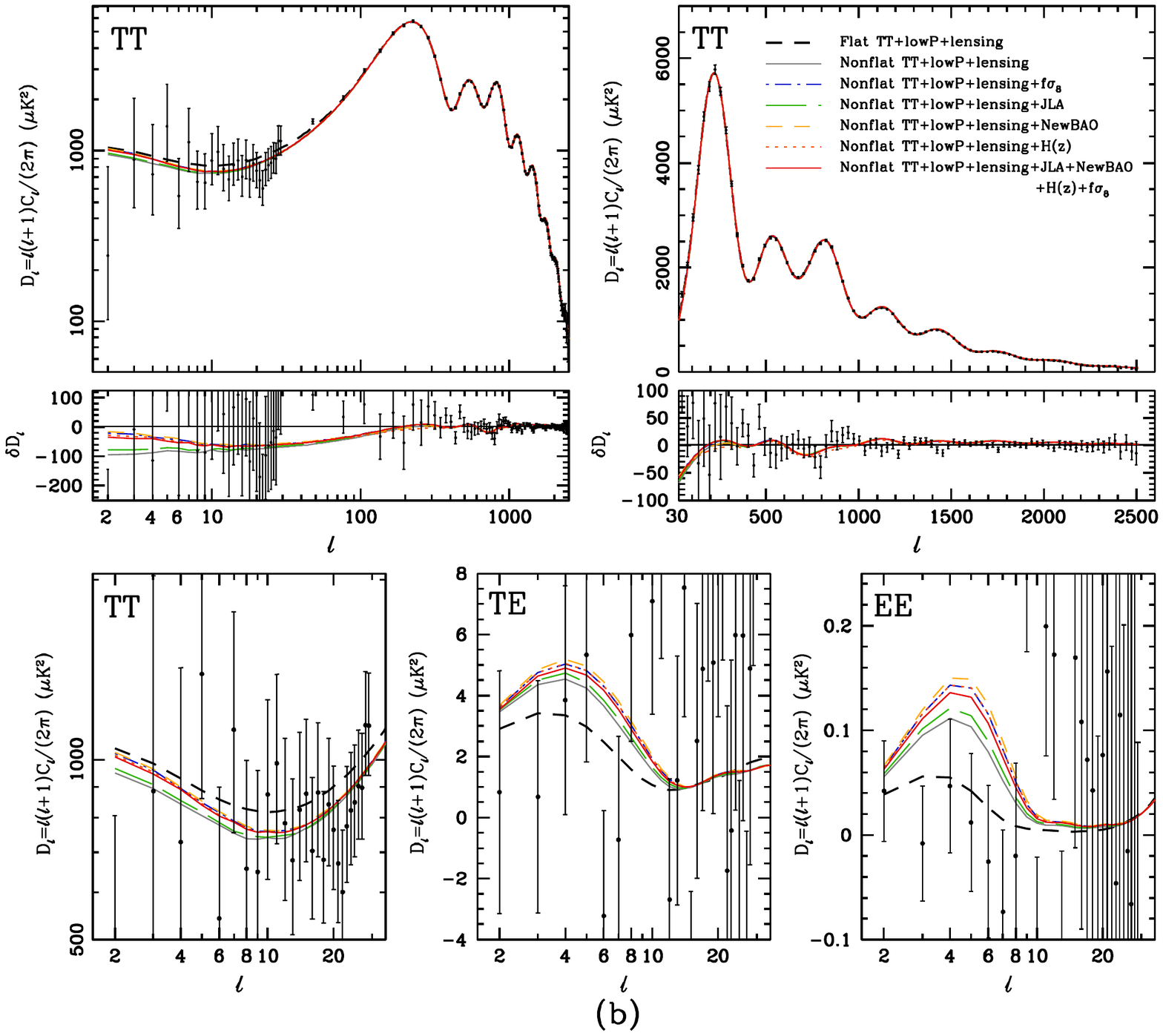}}
\caption{Same as Fig. \ref{fig:ps_cmb} but now including the lensing data.
}
\label{fig:ps_cmb_lensing}
\end{figure*}

Figures \ref{fig:ps_cmb} and \ref{fig:ps_cmb_lensing} show the CMB 
high-$\ell$ TT, and the low-$\ell$ TT, TE, EE power spectra of the best-fit 
tilted flat and untilted non-flat $\Lambda\textrm{CDM}$ models, excluding and including
the lensing data, respectively. The non-flat $\Lambda\textrm{CDM}$ model
constrained by adding each non-CMB data set to the Planck 2015 CMB anisotropy
observations generally gives a poorer fit to the low-$l$ EE power spectrum 
while it better fits the low-$\ell$ TT power spectrum (see the bottom left 
panel of Figs.\ \ref{fig:ps_cmb} and \ref{fig:ps_cmb_lensing}).
The shape of the best-fit $C_\ell$ power spectra of various models relative to 
the Planck CMB data points are consistent with the $\chi^2$ values listed in Table \ref{tab:chi2}.

\begin{figure*}[htbp]
\mbox{\includegraphics[width=85mm,bb=30 170 500 620]{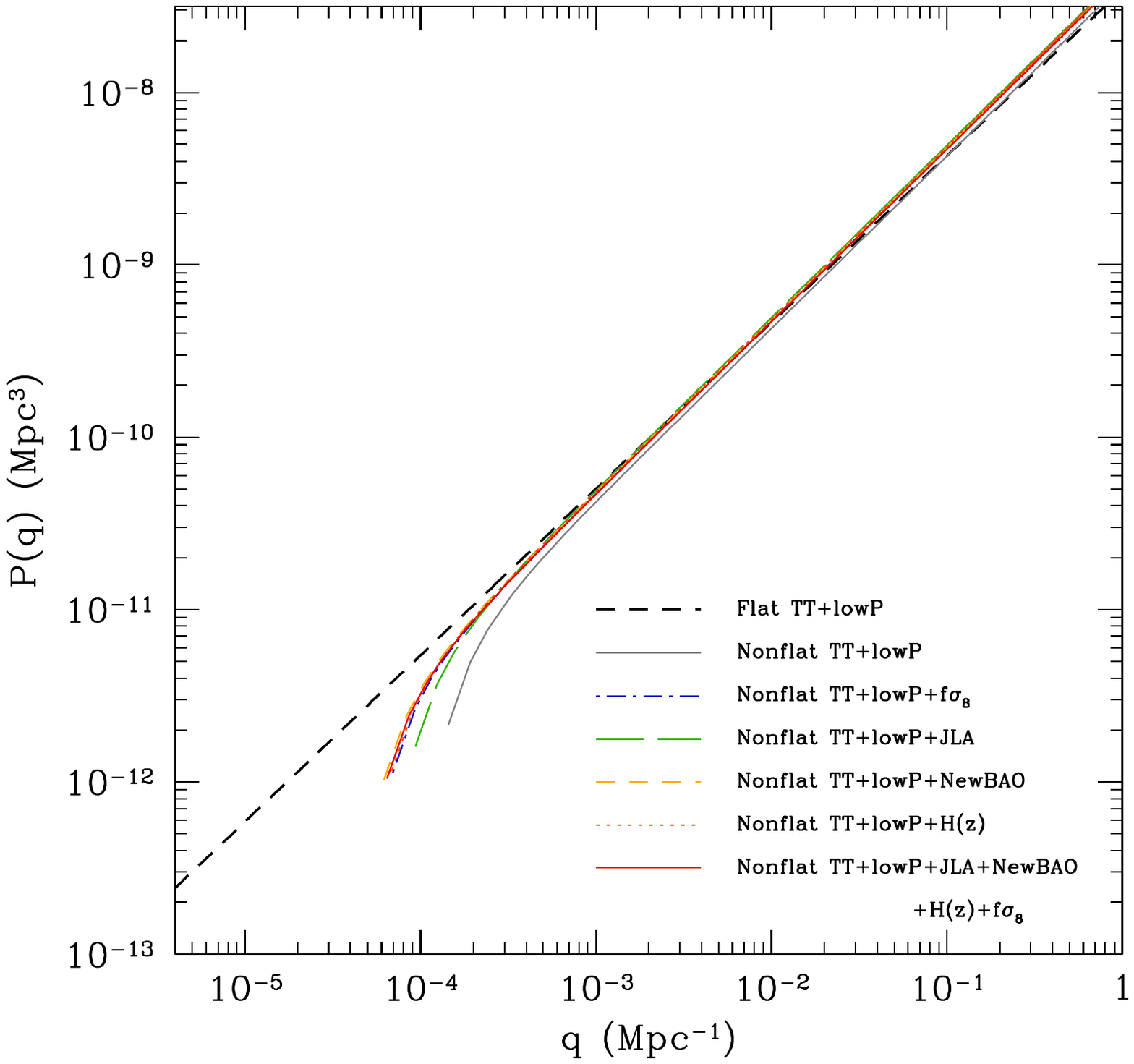}} 
\mbox{\includegraphics[width=85mm,bb=30 170 500 620]{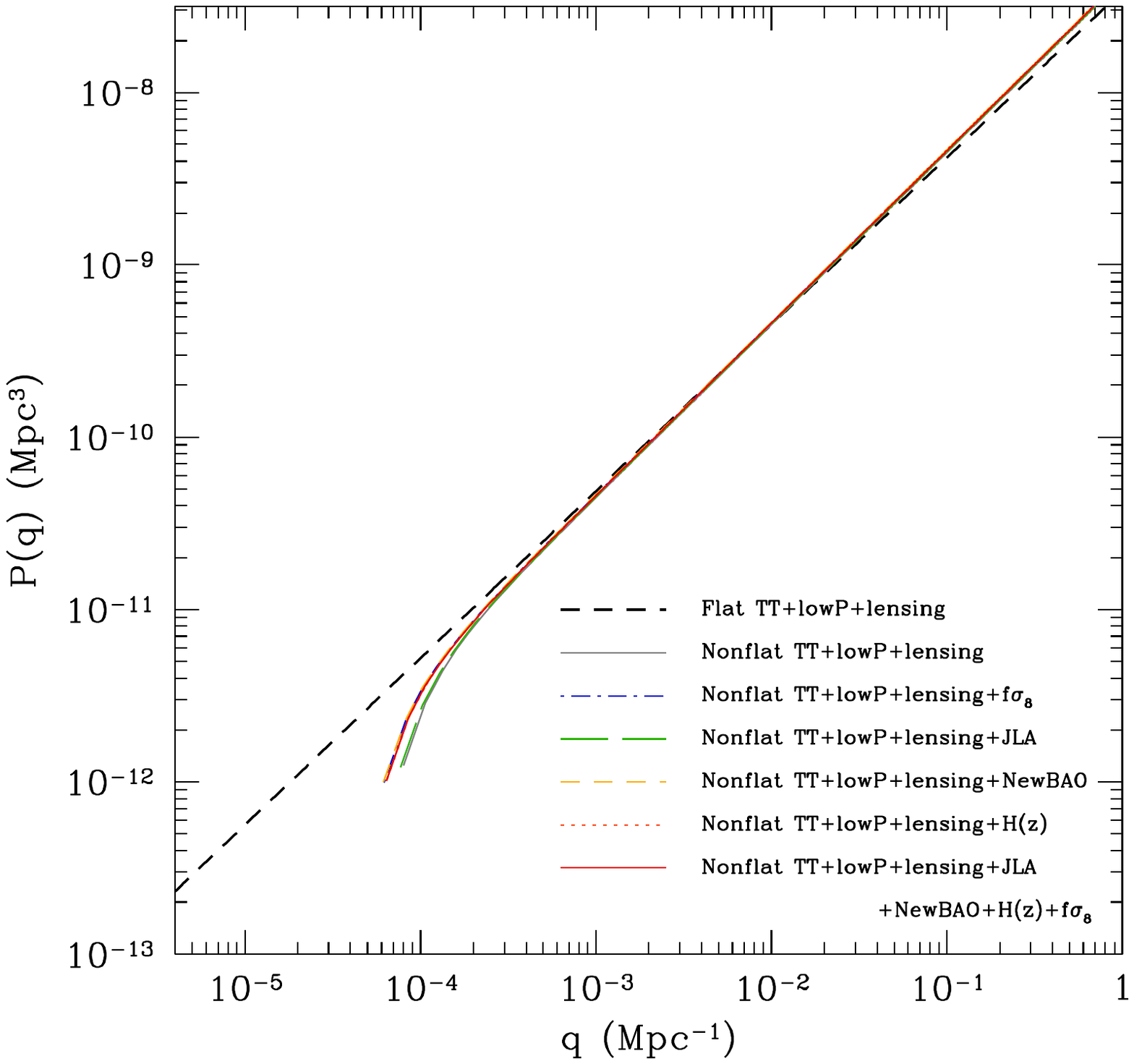}} 
\caption{Power spectra of primordial scalar-type perturbations constrained by
Planck TT + lowP data (left panel) and TT + lowP + lensing data (right panel).
In both panels the primordial power spectrum of the best-fit tilted flat-$\Lambda$CDM 
model is shown as dashed curves. 
Note that the power spectrum of each untilted non-flat model has discrete points of 
normal modes with positive integers $\nu=qK^{-1/2}=3,4,5,\cdots$, where $q=\sqrt{k^2 + K}$
and $K$ is the spatial curvature. The power spectrum is normalized
to $P(q)=A_s$ at the pivot scale $k_0=0.05~\textrm{Mpc}^{-1}$. 
}
\label{fig:pq}
\end{figure*}

Figure \ref{fig:pq} shows the best-fit initial power spectra of scalar-type 
fractional energy density perturbations for the non-flat $\Lambda$CDM model 
constrained by the Planck TT + lowP (left) and TT + lowP + lensing 
(right panel) data together with other non-CMB data sets. The reduction in 
power at low $q$ in the best-fit closed-$\Lambda$CDM inflation model power 
spectra shown in Fig.\ \ref{fig:pq} is partially responsible for the low-$\ell$ 
TT power reduction of the best-fit untilted closed model $C_\ell$'s (shown in the 
lower panels of Figs.\ \ref{fig:ps_cmb} and \ref{fig:ps_cmb_lensing}) relative
to the best-fit tilted flat model $C_\ell$'s. Other effects, including the 
usual and integrated Sachs-Wolfe effects, also play a role in affecting the
shape of the low-$\ell$ $C_\ell$'s. For a detailed discussion of how the 
interplay among these effects influences the low-$\ell$ shape of the $C_\ell$'s
in the open inflation case see \citet{Gorskietal1998}.   

\section{Conclusion}

We use the tilted flat-$\Lambda$CDM and the untilted non-flat $\Lambda$CDM inflation 
models to measure cosmological parameters from a carefully gathered compilation 
of observational data, the largest such collection utilized to date.

Our main results, in summary, are:
\begin{itemize}
\item Using a consistent power spectrum for energy density inhomogeneities 
in the untilted non-flat model, we confirm, with greater significance, the 
\citet{Oobaetal2018a} result that cosmological data does not demand 
spatially-flat hypersurfaces. These data (including CMB lensing measurements)
favor a closed Universe at more than 5$\sigma$ significance, with spatial 
curvature contributing about a percent to the current cosmological energy 
budget.
\item The best-fit untilted non-flat $\Lambda$CDM model provides a better fit to the 
low-$\ell$ temperature anisotropy $C_\ell$'s and better agrees with the 
$\sigma_8$--$\Omega_m$ DES constraints, but does worse than the best-fit 
tilted flat-$\Lambda$CDM model in fitting the higher-$\ell$ temperature 
anisotropy $C_\ell$'s.\footnote{We note that the tilted flat XCDM and $\phi$CDM models, with dynamical dark energy, provide slightly better fits to the data than does the tilted flat $\Lambda$CDM model \citep[][and references therin]{Oobaetal2018d,ParkRatra2019,ParkRatra2018a,Solaetal2018}.}
\item $H_0$ measured in both models are almost identical, and consistent
with most other measurements of $H_0$. However, as is well known, an 
estimate of the local expansion rate \citep{AndersonRiess2017} is 
2.7$\sigma$ larger.   
\item $\sigma_8$ measured in both models are identical and consistent 
with the recent DES measurement \citep{DESCollaboration2017a}.
\item The measured $\Omega_m$ is more model dependent than the measured 
$\sigma_8$ and the $\Omega_m$ value
measured using the non-flat $\Lambda$CDM model is more consistent with 
the recent DES measurement \citep{DESCollaboration2017a}. 
\item $\Omega_b h^2$, $\tau$, $\Omega_c h^2$, and some of the other 
measured cosmological parameter values are quite model dependent.
For such parameters, caution is called for when comparing a value measured 
in a cosmological model to a value determined using another technique. 
\end{itemize}

Overall, the tilted flat-$\Lambda$CDM model has a lower $\chi^2$
than the untilted non-flat $\Lambda$CDM case and so is more favored.  
On the other hand, the untilted non-flat $\Lambda$CDM model has other advantages, 
including having a lower $\sigma_8$.
It is possible that a more complete understanding of systematic 
differences between constraints derived using the lower-$\ell$ and 
higher-$\ell$ Planck CMB anisotropy data, as well as a more complete understanding of the differences between the Planck and South Pole Telescope CMB anisotropy data, might have some bearing on these issues.

%
%
\acknowledgements{
We acknowledge valuable discussions with J.\ Ooba.
C.-G.P.\ was supported by the Basic Science Research Program through the National Research Foundation of Korea (NRF) funded by the Ministry of Education (No. 2017R1D1A1B03028384).
B.R.\ was supported in part by DOE grant DE-SC0019038.
}


%
%

\def\and{{and }}
\bibliographystyle{yahapj}

\begin{thebibliography}{99}

\bibitem[{{Addison} {et~al.}(2016)}]{Addisonetal2016}
Addison, G.~E., Huang, Y., Watts, D.~J., {et~al.} 2016, \apj, 818, 132 
[arXiv:1511.00055]

\bibitem[{{Alam} {et~al.}(2017)}]{Alametal2017}
{Alam}, S., {Ata}, M., {Bailey}, S., {et~al.} 2017, \mnras, 470, 2617
[arXiv:1607.03155]

\bibitem[{{Anagnostopoulos} \& {Basilakos}(2017)}]{AnagnostopoulosBasilakos2017}
{Anagnostopoulos}, F., \& {Basilakos}, S.\ 2017, arXiv:1709.02356

\bibitem[{{Anderson} {et~al.}(2014)}]{Andersonetal2014}
{Anderson}, L., {Aubourg}, {\'E}., {Bailey}, S., {et~al.} 2014, \mnras, 441, 24
[arXiv:1312.4877]

\bibitem[{{Anderson} \& {Riess}(2017)}]{AndersonRiess2017}
{Anderson}, R.~I., \& {Riess}, A.~G.\ 2017, arXiv:1712.01065

\bibitem[{{Ata} {et~al.}(2018)}]{Ataetal2018}
{Ata}, M., {Baumgarten}, F., {Bautista}, J., {et~al.} 2018, \mnras, 473, 4773 
[arXiv:1705.06373]

\bibitem[{{Aubourg} {et~al.}(2015)}]{Aubourgetal2015}
{Aubourg}, {\'E}., {Bailey}, S., {Bautista}, J.~E., {et~al.} 2015, 
\prd, 92, 123516 [arXiv:1411.1074]

\bibitem[{{Audren} {et~al.}(2013)}]{Audrenetal2013}
Audren, B,, Lesgourgues, J., Benabed, K., \& Prunet, S.\ 2013, 
JCAP, 1302, 001 [arXiv:1210.7183]

\bibitem[{{Aylor} {et~al.}(2017)}]{Ayloretal2017}
Aylor, K., Hou, Z., Knox, L., {et~al.} 2017, \apj, 850, 101 
[arXiv:1706.10286]

\bibitem[{{Bautista} {et~al.}(2017)}]{Bautistaetal2017}
{Bautista}, J.~E., {Busca}, N.~G., {Guy}, J., {et~al.} 2017, \aap, 603, A12
[arXiv:1702.00176]

\bibitem[{{Betoule} {et~al.}(2014)}]{Betouleetal2014}
{Betoule}, M., {Kessler}, R., {Guy}, J., {et~ al.} 2014, \aap, 568, A22
[arXiv:1401.4064]

\bibitem[{{Beutler} {et~al.}(2011)}]{Beutleretal2011}
{Beutler}, F., {Blake}, C., {Colless}, M., {et~al.} 2011, \mnras, 416, 3017
[arXiv:1106.3366]

\bibitem[{{Beutler} {et~al.}(2012)}]{Beutleretal2012}
{Beutler}, F., {Blake}, C., {Colless}, M., {et~al.} 2012, \mnras, 423, 3430
[arXiv:1204.4725]

\bibitem[{{Blake} {et~al.}(2013)}]{Blakeetal2013}
{Blake}, C., {Baldry}, I.~K., {Bland-Hawthorn}, J., {et~al.} 2013, 
\mnras, 436, 3089 [arXiv:1309.5556]

\bibitem[{{Blas} {et~al.}(2011)}]{Blasetal2011}
Blas, D,, Lesgourgues, J., \& Tram, T.\ 2011, 
JCAP, 1107, 034 [arXiv:1104.2933]

\bibitem[{{Cai} {et~al.}(2016)}]{Caietal2016}
{Cai}, R.-G., {Guo}, Z.-K., \& {Yang}, T.\ 2016, \prd, 93, 043517 
[arXiv:1509.06283]

\bibitem[{{Calabrese} {et~al.}(2012)}]{Calabreseetal2012}
{Calabrese}, E., {Archidiacono}, M., {Melchiorri}, A., \& {Ratra}, B.\ 2012,
\prd,  86, 043520 [arXiv:1205.6753]

\bibitem[{{Cao} {et~al.}(2018)}]{Caoetal2018}
{Cao}, S.-L., {Duan}, X.-W., {Meng}, X.-L., \& {Zhang}, T.-J.\ 
2018, Eur.\ Phys.\ J., C78, 313 [arXiv:1712.01703]

\bibitem[{{Challinor} \& {Lasenby}(1999)}]{ChallinorLasenby1999}
{Challinor}, A., \& {Lasenby}, A.\ 1999, \apj, 513, 1 [arXiv:astro-ph/9804301]

\bibitem[{{Chen} {et~al.}(2003)}]{Chenetal2003}
{Chen}, G., {Gott}, J.~R., \& {Ratra}, B.\ 2003, \pasp, 115, 1269 [arXiv:astro-ph/0308099]

\bibitem[{{Chen} \& {Ratra}(2003)}]{ChenRatra2003}
{Chen}, G., \& {Ratra}, B.\ 2003, \pasp, 115, 1143 [arXiv:astro-ph/0302002]

\bibitem[{{Chen} \& {Ratra}(2011{\natexlab{a}})}]{ChenRatra2011a}
{Chen}, G., \& {Ratra}, B.\ 2011{\natexlab{a}}, \pasp, 123, 1127
[arXiv:1105.5206]

\bibitem[{{Chen} {et~al.}(2017)}]{Chenetal2017}
{Chen}, Y., {Kumar}, S., \& {Ratra}, B.\ 2017, \apj, 835, 86
[arXiv:1606.07316]

\bibitem[{{Chen} \& {Ratra}(2011{\natexlab{b}})}]{ChenRatra2011b}
{Chen}, Y., \& {Ratra}, B.\ 2011{\natexlab{b}}, Phys.\ Lett.\ B, 703, 406
[arXiv:1106.4294]

\bibitem[{{Chen} {et~al.}(2016)}]{Chenetal2016}
{Chen}, Y., {Ratra}, B., {Biesiada}, M., {Li}, S., \& {Zhu}, Z.-H.\ 2016, 
\apj, 829, 61 [arXiv:1603.07115]

\bibitem[{{Cooke} {et~al.}(2018)}]{Cookeetal2018}
{Cooke}, R.~J., {Pettini}, M., \& {Steidel}, C.~C.\ 2018, \apj, 855, 102 
[arXiv:1710.11129]

\bibitem[{{DES Collaboration}(2017a)}]{DESCollaboration2017a}
{DES Collaboration}, {Abbott}, T.~M.~C., {Abdalla}, F.~B., {Alarcon}, A., 
{et~al.} 2017a, arXiv:1708.01530

\bibitem[{{DES Collaboration}(2017b)}]{DESCollaboration2017b}
{DES Collaboration}, {Abbott}, T.~M.~C., {Abdalla}, F.~B., {Annis}, J., 
{et~al.} 2017b, arXiv:1711.00403

\bibitem[{{Dhawan} {et~al.}(2018)}]{Dhawanetal2018}
{Dhawan}, S., {Jha}, S.~W., \& {Leibundgut}, B.\ 2018, \aap, 609, A72 
[arXiv:1707.00715]

\bibitem[{{Farooq} {et~al.}(2017)}]{Farooqetal2017}
{Farooq}, O., {Madiyar}, F.~R., {Crandall}, S., \& {Ratra}, B.\ 2017, \apj,
  835, 26 [arXiv:1607.03537]

\bibitem[{{Farooq} {et~al.}(2015)}]{Farooqetal2015}
Farooq, O., Mania, D., \& Ratra, B.\ 2015, ApSS, 357, 11 [arXiv:1308.0834]

\bibitem[{{Farooq} \& {Ratra}(2013)}]{FarooqRatra2013}
{Farooq}, O., \& {Ratra}, B.\ 2013, \apj, 766, L7 [arXiv:1301.5243]

\bibitem[{{Feix} {et~ al.}(2015)}]{Feixetal2015} 
{Feix}, M., {Nusser}, A., \& {Branchini}, E.\ 2015, \prl, 115, 011301 
[arXiv:1503.05945]

\bibitem[{{Fern{\'a}ndez Arenas} {et~al.}(2018)}]{FernandezArenasetal2018}
{Fern{\'a}ndez Arenas}, D., {Terlevich}, E., {Terlevich}, R., {et~al.} 2018,
\mnras, 474, 1250 [arXiv:1710.05951]

\bibitem[{{Fixsen}(2009)}]{Fixsen2009} 
{Fixsen}, D.~J.\ 2009, \apj, 707, 916 [arXiv:0911.1955]

\bibitem[{{Font-Ribera} {et~al.}(2014)}]{Font-Riberaetal2014}
{Font-Ribera}, A., {Kirkby}, D., {Busca}, N., {et~al.} 2014, JCAP, 1405, 027
[arXiv:1311.1767]

\bibitem[{{G{\'o}rski} {et~al.}(1998)}]{Gorskietal1998}
{G{\'o}rski}, K.~M., {Ratra}, B., {Stompor}, R., {Sugiyama}, N., \& 
{Banday}, A.~J.\ 1998, \apjs, 114, 1 [arXiv:astro-ph/9608054]

\bibitem[{{G{\'o}rski} {et~al.}(1995)}]{Gorskietal1995}
{G{\'o}rski}, K.~M., {Ratra}, B., {Sugiyama}, N., \& {Banday}, A.~J.\ 1995,
\apj, 444, L65 [arXiv:astro-ph/9502034]

\bibitem[{{Gott}(1982)}]{Gott1982}
{Gott}, J.~R.\ 1982, \nat, 295, 304

\bibitem[{{Gott} {et~al.}(2001)}]{Gottetal2001}
{Gott}, J.~R., {Vogeley}, M.~S., {Podariu}, S., \& {Ratra}, B.\ 2001, \apj, 549, 1 [arXiv:astro-ph/0006103]

\bibitem[{{Haridasu} {et~al.}(2018)}]{Haridasuetal2018}
{Haridasu}, B.~S., {Lukovi{\'c}}, V.~V., \& {Vittorio}, N.\ 2018, JCAP, 1805, 
033 [arXiv:1711.03929]

\bibitem[{{Hawking}(1984)}]{Hawking1984}
{Hawking}, S.~W.\ 1984, Nucl.\ Phys.\ B, 239, 257

\bibitem[{{Heavens} {et~al.}(2017)}]{Heavensetal2017}
{Heavens}, A., {Fantaye}, Y., {Mootoovaloo}, A., et al.\ 2017, arXiv:1704.03472

\bibitem[{{Hinshaw} {et~al.}(2013)}]{Hinshawetal2013}
{Hinshaw}, G., {Larson}, D., {Komatsu}, E., {et~al.} 2013, \apjs, 208, 19 
[arXiv:1212.5226]

\bibitem[{{Howlett} {et~al.}(2015)}]{Howlettetal2015} 
{Howlett}, C., {Ross}, A.~J., {Samushia}, L., {Percival}, W.~J., 
\& {Manera}, M.\ 2015, \mnras, 449, 848 
[arXiv:1409.3238]

\bibitem[{{Hudson} \& {Turnbull}(2012)}]{HudsonTurnbull2012} 
{Hudson}, M.~J., \& {Turnbull}, S.~J.\ 2012, \apj, 751, L30 
[arXiv:1203:4814]

\bibitem[{{Huterer} \& {Shafer} (2017)}]{HutererShafer2017}
{Huterer}, D., \& {Shafer}, D.~L.\ 2017, arXiv:1709.01091

\bibitem[{{Kamionkowski} {et~al.}(1994)}]{Kamionkowskietal1994}
{Kamionkowski}, M., {Ratra}, B., {Spergel}, D.~N., \& {Sugiyama}, N.\ 1994,
\apj, 434, L1 [arXiv:astro-ph/9406069]

\bibitem[{{Lewis} \& {Bridle}(2002)}]{LewisBridle2002}
{Lewis}, A., \& {Bridle}, S.\ 2002, \prd, 66, 103511 [arXiv:astro-ph/0205436]

\bibitem[{{Lewis} {et~al.}(2000)}]{Lewisetal2000}
{Lewis}, A., {Challinor}, A., \& {Lasenby}, A.\ 2000, \apj, 538, 473
[arXiv:astro-ph/9911177]

\bibitem[{{L'Huillier} \& {Shafieloo}(2017)}]{LHuillierShafieloo2017}
{L'Huillier}, B., \& {Shafieloo}, A.\ 2017, JCAP, 1701, 015 [arXiv:1606.06832]

\bibitem[{{Li} {et~al.}(2016)}]{Lietal2016}
{Li}, Z., {Wang}, G.-J., {Liao}, K., \& {Zhu}, Z.-H.\ 2016, 
\apj, 833, 240 [arXiv:1611.00359]

\bibitem[{{Lin} \& {Ishak}(2017)}]{LinIshak2017}
{Lin}, W., \& {Ishak}, M. 2017, \prd, 96, 083532 [arXiv:1708.09813]

\bibitem[{{Lonappan} {et~al.}(2017)}]{Lonappanetal2017}
{Lonappan}, A.~I., {Ruchika}, \& {Sen}, A.~A.\ 2017, arXiv:1705.07336

\bibitem[{{Lucchin} \& {Matarrese}(1985)}]{LucchinMatarrese1985}
{Lucchin}, F., \& {Matarrese}, S. 1985, \prd, 32, 1316

\bibitem[{{Lukovi{\'c}} {et~al.}(2016)}]{Lukovicetal2016}
{Lukovi{\'c}}, V.~V., {D'Agostino}, R., \& {Vittorio}, N.\ 2016,
\aap, 595, A109 [arXiv:1607.05677]

\bibitem[{{Magana} {et~al.}(2017)}]{Maganaetal2017}
{Magana}, J., {Amante}, M.~H., {Garcia-Aspeitia}, M.~A., \& {Motta}, V.\ 2017,
arXiv:1706.09848

\bibitem[{{Martin}(2012)}]{Martin2012}
Martin, J.\ 2012, C.\ R.\ Physique, 13, 566 [arXiv:1205.3365]

\bibitem[{{Mitra} {et~al.}(2018)}]{Mitraetal2018}
{Mitra}, S., {Choudhury}, T.~R., \& {Ratra}, B.\ 2018, \mnras, 479, 4566 
[arXiv:1712.00018]

\bibitem[{{Mitra} {et~al.}(2019)}]{Mitraetal2019}
{Mitra}, S., {Park}, C.-G., {Choudhury}, T.~R., \& {Ratra}, B.\ 2019,  
arXiv:1901.09927

\bibitem[{{Moresco}(2015)}]{Moresco2015}
{Moresco}, M.\ 2015, \mnras, 450, L16
[arXiv:1503.01116]

\bibitem[{{Moresco} {et~al.}(2012)}]{Morescoetal2012}
{Moresco}, M., {Cimatti}, A., {Jimenez}, R., {et~al.} 2012, JCAP, 1208, 006
[arXiv:1201.3609]

\bibitem[{{Moresco} {et~al.}(2016)}]{Morescoetal2016}
{Moresco}, M., {Pozzetti}, L., {Cimatti}, A., {et~al.} 2016, JCAP, 1605, 014
[arXiv:1601:01701]

\bibitem[{{Okumura} {et~al.}(2016)}]{Okumuraetal2016} 
{Okumura}, T., {Hikage}, C., {Totani}, T., {et~al.} 2016, \pasj, 68, 38
[arXiv:1511.08083]

\bibitem[{{Ooba} {et~al.}(2018{\natexlab{a}})}]{Oobaetal2018a}
{Ooba}, J., {Ratra}, B., \& {Sugiyama}, N.\ 2018{\natexlab{a}}, \apj, 864, 80
[arXiv:1707.03452]

\bibitem[{{Ooba} {et~al.}(2018{\natexlab{b}})}]{Oobaetal2018b}
{Ooba}, J., {Ratra}, B., \& {Sugiyama}, N.\ 2018{\natexlab{b}}, \apj, 869, 34 
[arXiv:1710.03271]

\bibitem[{{Ooba} {et~al.}(2018{\natexlab{c}})}]{Oobaetal2018c}
{Ooba}, J., {Ratra}, B., \& {Sugiyama}, N.\ 2018{\natexlab{c}}, \apj, 866, 68
[arXiv:1712.08617]

\bibitem[{{Ooba} {et~al.}(2018d)}]{Oobaetal2018d}
{Ooba}, J., {Ratra}, B., \& {Sugiyama}, N.\ 2018d, arXiv:1802.05571

\bibitem[{{Park} \& {Ratra}(2019)}]{ParkRatra2019}
{Park}, C.-G., \& {Ratra}, B.\ 2019, ApSS, in press [arXiv:1803.05522]

\bibitem[{{Park} \& {Ratra}(2018{\natexlab{a}})}]{ParkRatra2018a}
{Park}, C.-G., \& {Ratra}, B.\ 2018{\natexlab{a}}, \apj, 868, 83 
[arXiv:1807.07421]

\bibitem[{{Park} \& {Ratra}(2018b)}]{ParkRatra2018b}
{Park}, C.-G., \& {Ratra}, B.\ 2018b, arXiv:1809.03598

\bibitem[Pavlov {et~al.}(2013)]{Pavlovetal2013}
{Pavlov}, A., Westmoreland, S., Saaidi, K., \& Ratra, B.\ 2013, 
\prd, 88, 123513 [arXiv:1307.7399]
 
\bibitem[{{Peebles}(1984)}]{Peebles1984}
{Peebles}, P.~J.~E.\ 1984, \apj, 284, 439

\bibitem[{{Peebles} \& {Ratra}(1988)}]{PeeblesRatra1988}
{Peebles}, P.~J.~E., \& {Ratra}, B.\ 1988, \apj, 325, L17

\bibitem[Penton {et~al.}(2018)]{Pentonetal2018}
Penton, J., Peyton, J., Zahoor, A., \& Ratra, B. 2018, \pasp, 130, 114009 
[arXiv:1808.01490]

\bibitem[{{Pezzotta} {et~al.}(2017)}]{Pezzottaetal2017} 
{Pezzotta}, A., {de la Torre}, S., {Bel}, J., {et~al.} 2017, \aap, 604, A33
[arXiv:1612.05645]

\bibitem[{{Planck Collaboration}(2014)}]{PlanckCollaboration2014}
{Planck Collaboration}, {Ade}, P.~A.~R., {Aghanim}, N., {Armitage-Caplan}, C.,
{et~al.} 2014, \aap, 571, A16 [arXiv:1303.5076]

\bibitem[{{Planck Collaboration}(2015)}]{PlanckParameterTables2015}
{Planck Collaboration} 2015, {\it Planck} 2015 Results: Cosmological Parameter Tables
at wiki.cosmos.esa.int/planckpla2015/images/f/f7/Baseline{\_}para{\linebreak}ms{\_}table{\_}2015{\_}limit68.pdf

\bibitem[{{Planck Collaboration}(2016)}]{PlanckCollaboration2016}
{Planck Collaboration}, {Ade}, P.~A.~R., {Aghanim}, N., {Arnaud}, M.,
{et~al.} 2016, \aap, 594, A13 [arXiv:1502.01589]

\bibitem[{{Planck Collaboration}(2017)}]{PlanckCollaboration2017}
{Planck Collaboration}, {Aghanim}, N., {Akrami}, Y., {Ashdown}, M., 
{et~al.} 20167, \aap, 607, A95 [arXiv:1608.02487]

\bibitem[{{Podariu} \& {Ratra}(2001)}]{PodariuRatra2001}
{Podariu}, S., \& {Ratra}, B.\ 2001, \apj, 532, 109
[arXiv:astro-ph/9910527]

\bibitem[{{Rana} {et~al.}(2017)}]{Ranaetal2017}
{Rana}, A., {Jain}, D., {Mahajan}, S., \& {Mukherjee}, A.\ 2017, 
JCAP, 1703, 028 [arXiv:1611.07196]

\bibitem[{{Ratra}(1985)}]{Ratra1985}
{Ratra}, B.\ 1985, \prd, 31, 1931

\bibitem[{{Ratra}(1989)}]{Ratra1989}
{Ratra}, B.\ 1989, \prd, 40, 3939

\bibitem[{{Ratra}(1992)}]{Ratra1992}
{Ratra}, B.\ 1992, \prd, 45, 1913

\bibitem[{{Ratra}(2017)}]{Ratra2017}
{Ratra}, B.\ 2017, \prd, 96, 103534 [arXiv:1707.03439]

\bibitem[{{Ratra} \& {Peebles}(1988)}]{RatraPeebles1988}
{Ratra}, B., \& {Peebles}, P.~J.~E.\ 1988, \prd, 37, 3406

\bibitem[{{Ratra} \& {Peebles}(1994)}]{RatraPeebles1994}
{Ratra}, B., \& {Peebles}, P.~J.~E.\ 1994, \apj, 432, L5

\bibitem[{{Ratra} \& {Peebles}(1995)}]{RatraPeebles1995}
{Ratra}, B., \& {Peebles}, P.~J.~E.\ 1995, \prd, 52, 1837

\bibitem[{{Ratra} \& {Vogeley}(2008)}]{RatraVogeley2008}
{Ratra}, B., \& {Vogeley}, M.\ 2008, \pasp, 120, 235 [arXiv:0706.1565]

\bibitem[{{Ratsimbazafy} {et~al.}(2017)}]{Ratsimbazafyetal2017}
{Ratsimbazafy}, A.~L., {Loubser}, S.~I., {Crawford}, S.~M., {et~al.} 2017,
\mnras, 467, 3239 [arXiv:1702.00418]

\bibitem[{{Rezaei} {et~al.}(2017)}]{Rezaeietal2017}
{Rezaei}, M., {Malekjani}, M., {Basilakos}, S., {Mehrabi}, A., \& {Mota}, 
D.~F.\ 2017, \apj, 843, 65 [arXiv:1706.02537]

\bibitem[{{Rigault} {et~al.}(2015)}]{Rigaultetal2015}
{Rigault}, M., {Aldering}, G., {Kowalski}, M., {et~al.} 2015, \apj,
802, 20 [arXiv:1412.6501]

\bibitem[{{Ross} {et~al.}(2015)}]{Rossetal2015}
{Ross}, A.~J., {Samushia}, L., {Howlett}, C., {et~al.} 2015, \mnras, 449, 835
[arXiv:1409.3242]
 
\bibitem[{{Ryan} {et~al.}(2019)}]{Ryanetal2019}
{Ryan}, J., Chen, Y., \& {Ratra}, B.\ 2019, arXiv:1902.03196

\bibitem[{{Ryan} {et~al.}(2018)}]{Ryanetal2018}
{Ryan}, J., Doshi, S., \& {Ratra}, B.\ 2018, \mnras, 480, 759
[arXiv:1805.06408]

\bibitem[{{Samushia} {et~al.}(2007)}]{Samushiaetal2007}
{Samushia}, L., {Chen}, G., \& {Ratra}, B.\ 2007, arXiv:0706.1963

\bibitem[{{Samushia} \& {Ratra}(2006)}]{SamushiaRatra2006}
{Samushia}, L., \& {Ratra}, B.\ 2006, \apjl, 650, L5 [arXiv:astro-ph/0607301]

\bibitem[{{Sievers} {et~al.}(2013)}]{Sieversetal2013}
{Sievers}, J.~L., {Hlozek}, R.~A., {Nolta}, M.~R., {et~al.} 2013, JCAP,
1310, 060 [arXiv:1301.0824]

\bibitem[{{Simon} {et~al.}(2005)}]{Simonetal2005}
{Simon}, J., {Verde}, L., \& {Jimenez}, R.\ 2005, \prd, 71, 123001
[arXiv:astro-ph/0412269]

\bibitem[{{Simpson} {et~al.}(2016)}]{Simpsonetal2016} 
{Simpson}, F., {Blake}, C., {Peacock}, J.~A., {et~al.} 2016, \prd, 93, 023525
[arXiv:1505.03865]
 
\bibitem[{{Sol\`{a}} {et~al.}(2018)}]{Solaetal2018}
{Sol\`{a}}, J., {G\'{o}mez-Valent}, A., \& {de Cruz P\'{e}rez}, J.\ 2018, 
arXiv:1811.03505

\bibitem[{{Springob} {et~al.}(2016)}]{Springobetal2016} 
{Springob}, C.~M., {Hong}, T., {Staveley-Smith}, L., {et~al.} 2016, 
\mnras, 456, 1886 [arXiv:1511.04849]

\bibitem[{{Stern} {et~al.}(2010)}]{Sternetal2010}
{Stern}, D., {Jimenez}, R., {Verde}, L., {Kamionkowski}, M., \& {Stanford}, 
S.~A.\ 2010, JCAP, 1002, 008
[arXiv:0907.3152]

\bibitem[{{Tripathi} {et~al.}(2017)}]{Tripathietal2017}
{Tripathi}, A., {Sangwan}, A., \& {Jassal}, H.~K.\ 2017, JCAP, 1706, 012
[arXiv:1611.01899]

\bibitem[{Trotta}(2008)]{Trotta2008} 
{Trotta}, R.\ 2008, Contemp.\ Phys.\ 49, 71 [arXiv:0803.4089]

\bibitem[{{Turnbull} {et~al.}(2012)}]{Turnbulletal2012} 
{Turnbull}, S.~J., {Hudson}, M.~J., {Feldman}, H.~A., {et~al.} 2012, 
\mnras, 420, 447 
[arXiv:1111.0631]

\bibitem[{{Wang} {et~al.}(2017)}]{Wangetal2017}
Wang, Y., Xu, L., \& Zhao, G.-B.\ 2017, arXiv:1706.09149

\bibitem[{{Wei} \& {Wu}(2017)}]{WeiWu2017}
Wei, J.-J., \& Wu, X.-F.\ 2017, \apj, 838, 160 [arXiv:1611.00904]

\bibitem[{{Yu} {et~al.}(2018)}]{Yuetal2018}
{Yu}, H., {Ratra}, B., \& {Wang}, F.-Y.\ 2018, \apj, 856, 3 [arXiv:1711.03437]

\bibitem[{{Yu} \& {Wang}(2016)}]{YuWang2016}
{Yu}, H., \& {Wang}, F.~Y.\ 2016, \apj, 828, 85 [arXiv:1605.02483]

\bibitem[{{Zhang} {et~al.}(2017)}]{Zhangetal2017}
{Zhang}, B.~R., {Childress}, M.~J., {Davis}, T.~M., {et~al.} 2017, \mnras,
471, 2254 [arXiv:1706.07573]

\bibitem[{{Zhang} {et~al.}(2014)}]{Zhangetal2014}
{Zhang}, C., {Zhang}, H., {Yuan}, S., {Zhang}, T.-J., \& {Sun}, Y.-C.\ 2014, 
Res.\ Astron.\ Astrophys., 14, 1221 [arXiv:1207.4541]

\end{thebibliography}


\end{document}